\documentclass[acmtog]{acmart}

\AtBeginDocument{%
  }

\acmSubmissionID{1808}

\usepackage{listings}
\usepackage{xcolor}
\usepackage{multirow}
\usepackage{colortbl}
\usepackage{mathtools}
\usepackage{enumitem}
\usepackage{subcaption}
\usepackage{bibunits}

\definecolor{first}{RGB}{220,50,47}     
\definecolor{second}{RGB}{255,140,0}    
\definecolor{third}{RGB}{255,215,0}     

\definecolor{PathBlue}{RGB}{70, 105, 155}
\definecolor{MarchRed}{RGB}{165, 95, 80}

\lstdefinestyle{togpseudo}{
  backgroundcolor=\color{gray!10},
  basicstyle=\ttfamily\small,
  frame=single,
  framerule=0pt,
  xleftmargin=0.2em,
  xrightmargin=0.2em,
  showstringspaces=false,
  columns=fullflexible,
  emph={Path_Forward, Path_Backward},
  emphstyle=\bfseries\color{PathBlue},
  emph={[2]Attributes_Forward, Attributes_Backward},
  emphstyle={[2]\bfseries\color{MarchRed}},
}

\begin{document}

\title{Path-Traced Inverse Rendering with Global Illumination in 3D Gaussian Fields}


\author{Junke Zhu}
\email{junkezhu@mail.ustc.edu.cn}
\affiliation{%
  \institution{University of Science and Technology of China}
  \country{China}
}

\author{Hao Zhang}
\email{zhanghao25@stu.pku.edu.cn}
\affiliation{%
  \institution{Peking University}
  \country{China}
}

\author{Yutian Zhu}
\email{zhuyutian@mail.ustc.edu.cn}
\affiliation{%
  \institution{University of Science and Technology of China}
  \country{China}
}

\author{Ang Li}
\email{psnalaalansp@gmail.com}

\author{Chenxiao Hu}
\email{hineven@pku.edu.cn}

\author{Meng Gai}
\email{gaimeng@pku.edu.cn}

\author{Fei Zhu}
\email{feizhu@pku.edu.cn}
\affiliation{%
  \institution{Peking University}
  \country{China}
}

\author{Zhangjin Huang}
\email{zhuang@ustc.edu.cn}
\authornotemark[1]
\affiliation{%
  \institution{University of Science and Technology of China}
  \country{China}
}

\author{Sheng Li}
\email{lisheng@pku.edu.cn}
\authornote{Sheng Li and Zhangjin Huang are both corresponding authors.}
\affiliation{%
  \institution{Peking University}
  \country{China}
}

\renewcommand{\shortauthors}{Zhu et al.}

\begin{abstract}
Ray tracing enables 3D Gaussian fields to serve as a representation for physically based light transport. Faithful inverse rendering requires forward rendering and backward optimization to be defined within a consistent light-transport pipeline.
Existing Gaussian inverse-rendering methods typically rely on splatting-derived G-buffers and screen-space optimization. The local affine approximation of perspective projection can introduce inconsistencies with path tracing, while simplified rendering formulations often neglect or approximate indirect illumination and visibility.
Therefore, we propose a splatting-free path-traced inverse-rendering framework for 3D Gaussian fields that unifies forward rendering and backward optimization within the same recursive multi-bounce light-transport pipeline. We formulate light transport over overlapping Gaussian primitives directly in path space, enabling Monte Carlo path tracing and pathwise gradient replay over the same recursively sampled transport paths.
The framework jointly optimizes materials and environment light under the full rendering equation, with ray-traced visibility and global illumination explicitly evaluated.
Extensive experiments demonstrate competitive material estimation and improved path-traced rendering quality, producing more plausible shadows, reflections, and relighting under global illumination.
The code is available at \url{https://junkzhu.github.io/project_pages/PTIR}.
\end{abstract}


\setcopyright{cc}
\setcctype{by-nc-nd}
\acmJournal{TOG}
\acmYear{2026} \acmVolume{45} \acmNumber{6} \acmArticle{235}
\acmMonth{12} \acmDOI{10.1145/3842545}


\begin{teaserfigure}
  \centering
  \includegraphics[width=0.98\textwidth]{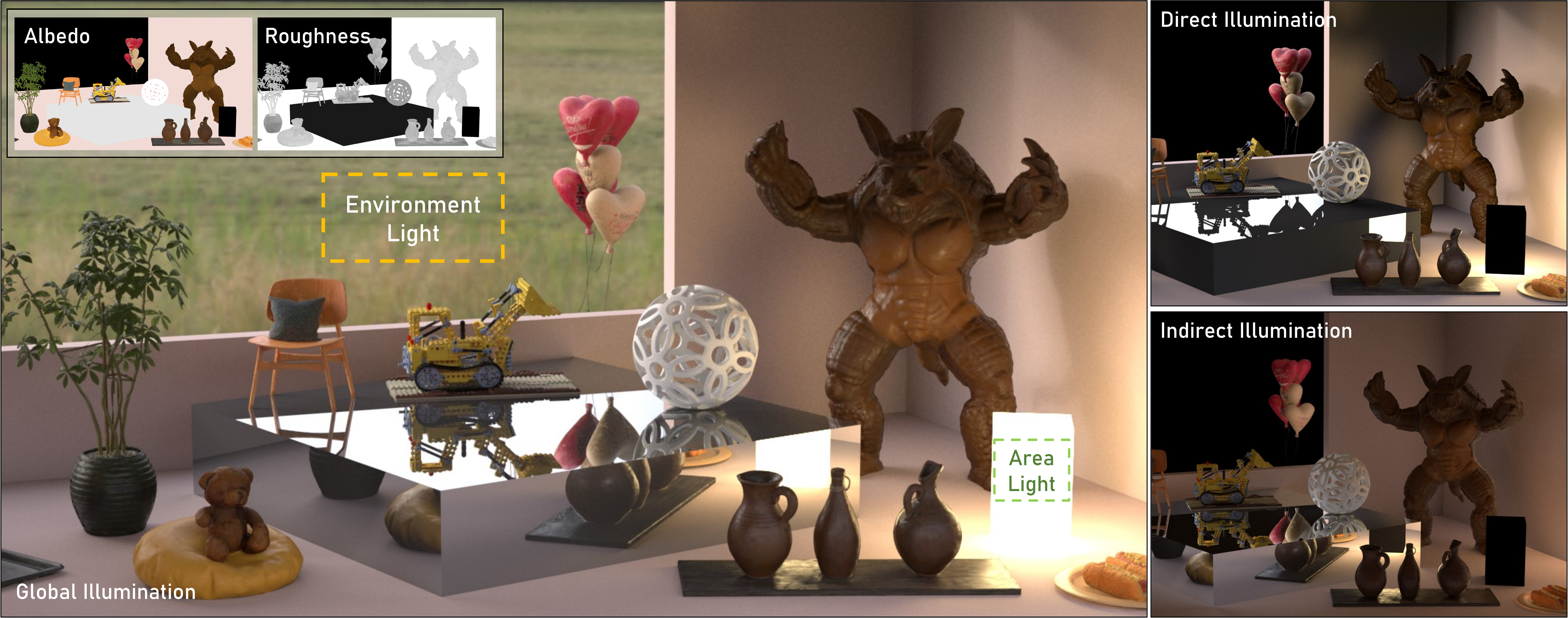}
  \caption{
Our splatting-free framework ensures consistency between forward path-traced
rendering and backward optimization. We demonstrate recovered 3D Gaussian
assets under physically based multi-bounce global illumination in a
heterogeneous scene, where all objects except the light sources, walls, and
metallic box are modeled as 3D Gaussian primitives. This allows the recovered
Gaussian materials to be reused for path-traced rendering and relighting.
}
  \label{fig:teaser}
\end{teaserfigure}

\maketitle
\begin{bibunit}[ACM-Reference-Format]
\section{INTRODUCTION}
\label{sec:Intro}

3D Gaussian Splatting (3DGS) has achieved remarkable success in novel view synthesis~\cite{kerbl3dgaussians}, becoming an important representation for 3D scene perception and understanding. Most 3DGS methods rely on rasterization, projecting Gaussian primitives onto the image plane and compositing them through ordered alpha blending. This formulation is highly effective for
image-space reconstruction, but physically based inverse rendering requires
recovered materials to remain valid under light transport involving ray-space
visibility and indirect illumination. Ray tracing provides a physically based rendering paradigm for 3D Gaussian fields by treating Gaussians as scene primitives in light transport simulation, naturally supporting shadows, reflections, and indirect illumination that remain challenging to model in splatting-based pipelines
~\cite{3dgrt2024,
hu2025realtimeglobalilluminationdynamic, xie2024envgs}.

\begin{figure}
    \centering
    \includegraphics[width=\linewidth]{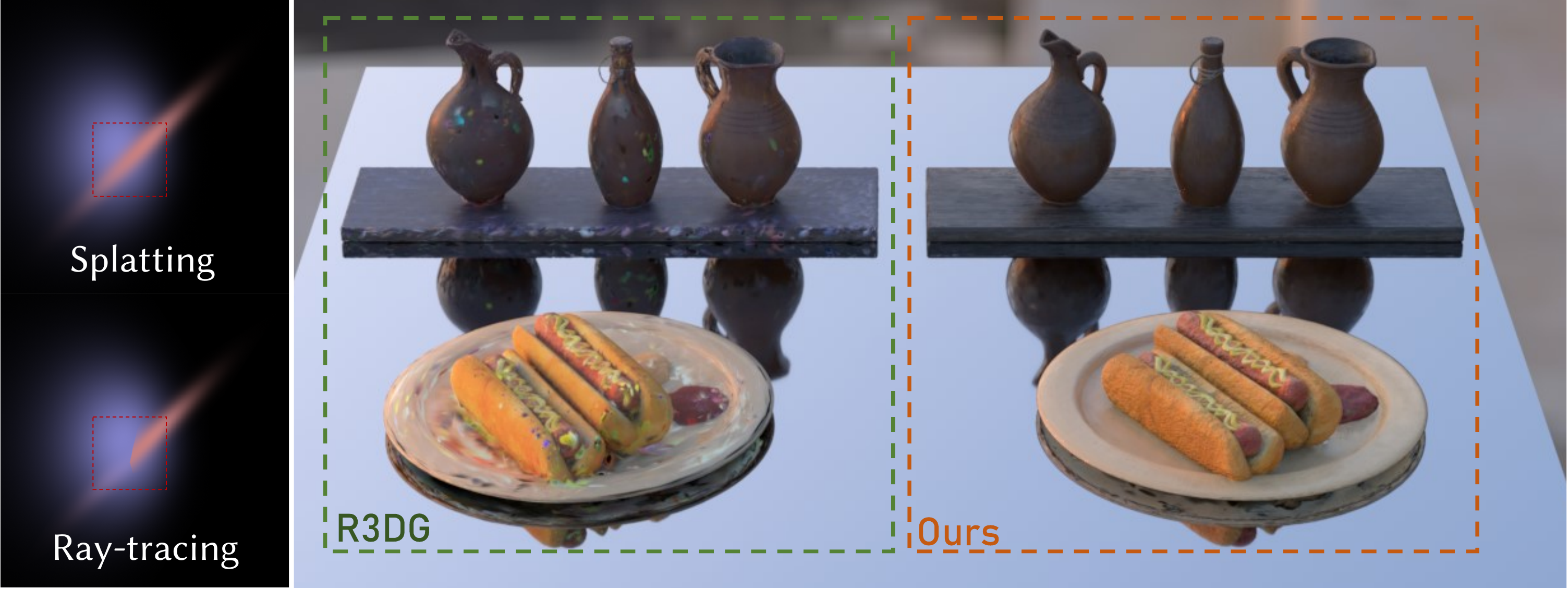}
    \caption{Path-traced evaluation of inverse-rendered 3D Gaussian assets. (Left) Splatting orders Gaussians by their center depths, whereas ray tracing determines effective depths from peak-response locations along each ray. (Right) R3DG~\cite{R3DG} optimizes materials within a splatting-based pipeline, resulting in inconsistencies under path-traced rendering, while our framework produces more consistent results.
}
    \label{fig:motivation}
\end{figure}

As illustrated in Fig.~\ref{fig:motivation}, existing inverse-rendering methods for Gaussian representations typically estimate physically motivated properties from splatting-derived G-buffers optimized with screen-space shading objectives. However, splatting sorts Gaussians according to their center depths~\cite{R3DG}, whereas ray tracing determines interactions from their peak responses along each ray. In addition to the depth inconsistency, projecting a 3D Gaussian into a 2D splat through a local affine approximation can introduce additional errors. Consequently, naively integrating materials inverted with splatting-based methods into a path-traced rendering pipeline can lead to severe artifacts, revealing that the recovered properties remain tied to screen-space optimization rather than path-space transport. Moreover, these methods typically rely on simplified rendering equations. Without multi-bounce path tracing, indirect illumination is often neglected, while visibility effects such as shadows are approximated using precomputed or rasterization-based terms.

We therefore introduce a splatting-free inverse-rendering framework based on a
path-space equivalent interaction model for 3D Gaussian fields. Under this model,
forward Monte Carlo path tracing and
backward optimization are performed over the same ray-traced Gaussian interactions without relying on splatting.
We construct equivalent surface
interactions in ray space, evaluate visibility through ray tracing, and perform
multi-bounce path tracing under the full rendering equation. Gradients are
propagated by replaying the same interactions using path replay
backpropagation (PRB)~\cite{PathReplayBackpropagation}.
Multiple importance
sampling and a compact Spherical-Gaussian environment parameterization are used
for variance reduction and stable lighting optimization.
Experiments demonstrate
competitive material estimation quality together with improved consistency in path-traced
rendering, relighting, and appearance editing.
The main contributions of our work are as follows:
\begin{itemize}[leftmargin=1.5em]
\item We propose a splatting-free path-traced inverse rendering framework for
3D Gaussian fields, where forward rendering and backward optimization are
defined within the same ray-tracing pipeline, avoiding the mismatch between
splatting-based material estimation and path-traced rendering.
\item We formulate path-space light transport for 3D Gaussian fields under the full rendering equation, where equivalent surface interactions define shading states for material and illumination optimization through recursive multi-bounce transport.

\item
We experimentally demonstrate that material and illumination can be recovered in 3D Gaussian fields within a ray-traced light-transport pipeline, achieving competitive inverse-rendering quality while enabling physically based relighting and appearance editing under global illumination.
\end{itemize}

\section{RELATED WORK}
\subsection{Ray Tracing of Gaussian Primitives}
3D Gaussian Splatting (3DGS)~\cite{kerbl3dgaussians} represents radiance fields with anisotropic 3D Gaussian primitives and enables real-time novel-view synthesis through tile-based splatting. However, its screen-space compositing and spherical-harmonic view-dependent appearance limit effects involving incoherent light paths.
HTGS~\cite{hahlbohm2025htgs} performs perspective-correct Gaussian evaluation along per-pixel viewing rays and introduces hybrid transparency to reduce depth-ordering artifacts. 3DGUT~\cite{wu20253dgut} employs an unscented-transform formulation for nonlinear camera models and supports hybrid rendering with traced secondary rays. These methods retain rasterized primary visibility, whereas fully ray-traced formulations process primary and recursively generated rays within a unified ray-tracing pipeline.
For fully ray-traced rendering, 3D Gaussian Ray Tracing (3DGRT)~\cite{3dgrt2024} approximates Gaussians with icosahedral mesh proxies and constructs a scene-level BVH for hardware-accelerated differentiable ray tracing. Similarly, \citet{byrski2025raysplatsraytracingbased} use ellipsoidal proxies for accurate ray--primitive intersections. Several works~\cite{hu2025realtimeglobalilluminationdynamic, sun2025stochasticraytracingtransparent} improve efficiency with probabilistic ray-intersection tests that reduce ray payload overhead and avoid explicit sorting. Another line of research~\cite{DontSplatYourGaussians, zhou2024unifiedgaussianprimitivesscene} reformulates Gaussian light transport for volumetric rendering. EnvGS~\cite{xie2024envgs} models environmental reflections by ray tracing Gaussians visible only through indirect paths. Beyond 3D Gaussians, 2D Gaussian Splatting (2DGS)~\cite{Huang2DGS2024} improves geometry with anisotropic 2D primitives, while IRGS~\cite{IRGS} introduces 2D Gaussian ray tracing for inter-reflections. \citet{edtable-3dgs} use ray tracing for physically based specular reflections.

\subsection{Inverse Rendering of 3D Gaussians}
Reconstructing scene geometry, materials, and illumination from multi-view images via inverse rendering remains highly challenging. Recent advances in 3DGS have inspired several works to integrate traditional shading models into the 3D Gaussian framework to disentangle radiance into explicit material and lighting components~\cite{jiang2024gaussianshader,wu2024deferredgs, guo2024prtgs,SVGIR2025}. For relighting, \citet{R3DG} bake occlusion and indirect illumination via point-based ray tracing, while \citet{GS-IR} adopt a split-sum approximation of the rendering equation and store precomputed occlusion and indirect terms in volumes. In contrast, \citet{bi2024rgs} and \citet{fan25rng} bypass explicit lighting models and instead regress global illumination effects using neural networks, which is particularly effective for challenging structures such as hair. \citet{IRGS} proposed a 2D Gaussian-based ray-tracing approach to efficiently sample incident radiance and compute inter-reflections. \citet{jiang2025radiositygs} extended classical radiosity to Gaussian surfels and formulated a differentiable solution for global light transport; however, due to the low-frequency nature of radiosity, their method cannot faithfully capture high-frequency illumination effects. \citet{edtable-3dgs} optimize materials in 3D Gaussian scenes using path tracing driven by multi-view G-buffers. \citet{EAG-PT} recover diffuse albedo through single-bounce optimization using a reconstructed radiance cache and apply multi-bounce path tracing after scene editing.

Environmental illumination is a key component of inverse rendering and has been
represented by image-based lighting~\cite{paul98ibl}, spherical
harmonics~\cite{ravi01sh}, spherical Gaussians~\cite{wang09sg}, and Haar
wavelets~\cite{ren03haar}. Existing methods either explicitly optimize lighting
representations or implicitly encode illumination in neural
fields~\cite{neural-incident-light-field,zhang2023neilf++,ling24nerfemitter,GS-ID}.
Several works approximate the rendering equation through closed-form SG
integration~\cite{physg2021,zhang2022invrender}, whereas
TensoIR~\cite{Jin2023TensoIR} evaluates environment illumination by ray
sampling, which naturally aligns with our ray-traced formulation.

Our method follows a different formulation from these approaches. Existing
Gaussian inverse-rendering methods typically remain coupled with splatting-based
material buffers, screen-space visibility, precomputed transport terms, or use
ray tracing only as an auxiliary component for selected effects. In contrast, we 
consider a splatting-free, perspective-correct path-tracing setting, where 
material recovery and relighting are performed under multi-bounce global illumination 
without precomputed light-transport approximations.

Concurrent and independent work, SRT-GS~\citep{xu2026Stoch3DGS}, introduces a sorting-free stochastic estimator for differentiable ray-traced 3DGS reconstruction and extends it to a neural relightable representation. It  propagates gradients through Monte Carlo-sampled Gaussian interactions, while our framework uses path replay to optimize explicit material parameters through BRDF evaluations along recursively sampled multi-bounce transport paths. Its stochastic estimator could potentially accelerate the ray--Gaussian evaluation component of our pipeline.

\section{METHOD}
We first introduce the necessary background definitions
(Sec.~\ref{sec:Preliminary}). We then present path-space light transport for
3D Gaussian fields with path-traced global illumination
(Sec.~\ref{sec:forward_light_transport}). Next, we describe the learnable
Spherical-Gaussian environment light
(Sec.~\ref{sec:learnable_environment_lighting}). Finally, we formulate inverse
rendering by replaying the same path-space interactions for gradient propagation
(Sec.~\ref{sec:inverse_rendering}).

\begin{figure*}[t]
  \centering
\includegraphics[width=0.95\linewidth,
        trim=0 0.3in 0 0,
        clip
        ]{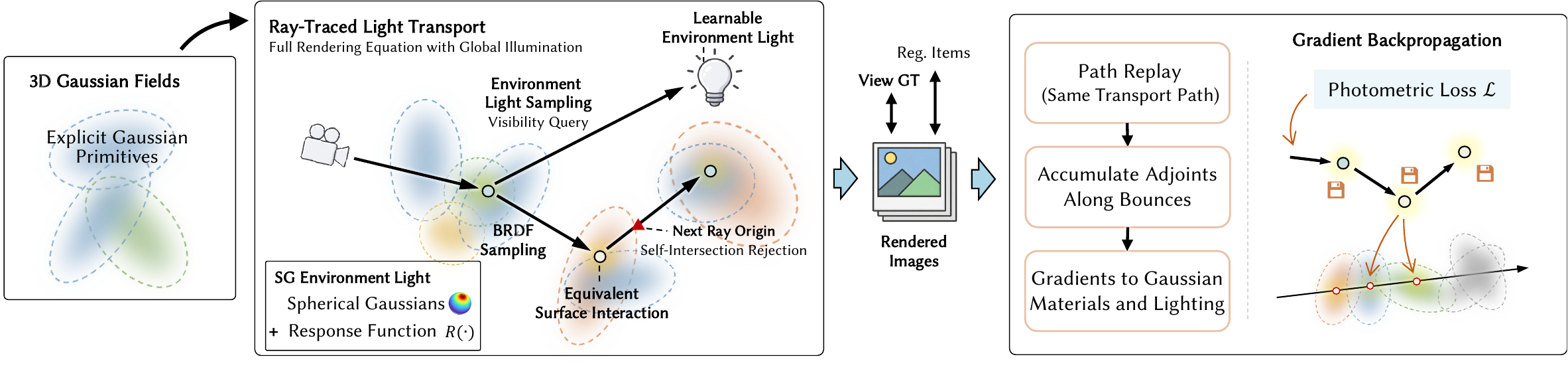}
  \caption{Our framework jointly defines forward rendering and backward optimization for
3D Gaussian fields within a unified ray-tracing pipeline. Given a 3D Gaussian
scene, we construct ray-traced equivalent surface interactions, evaluate
visibility and multi-bounce path-traced light transport under the full rendering equation,
and support optimization of environment illumination using a compact Spherical-Gaussian representation. Multiple
importance sampling reduces the variance of incident radiance estimation.
Gradients are propagated through path replay backpropagation via equivalent
interactions, ensuring consistency between backward optimization and the
forward ray-traced transport paths.}
  \label{fig:pipeline}
\end{figure*}

\subsection{Preliminary}
\label{sec:Preliminary}
\subsubsection{3D Gaussian Primitives.}
Each Gaussian primitive is further associated with material attributes, including shading normal $\mathbf{n}\in\mathbb{R}^3$, albedo $\mathbf{a}\in\mathbb{R}^3$, roughness $r\in\mathbb{R}$, and metallic $m\in\mathbb{R}$, in addition to the vanilla 3D Gaussian parameters~\cite{kerbl3dgaussians}.

For each Gaussian primitive intersected by a ray, we first identify its maximum-response point along the ray $r(t)=\mathbf{o}+t\mathbf{d}$. This point corresponds to the peak of the Gaussian response along the ray and admits a closed-form solution~\cite{3dgrt2024}:
\begin{small}
\begin{equation}
\label{eq:gaussian_peak}
t_{\mathrm{peak}}^{(i)}
=
-\frac{\mathbf{d}^\top \boldsymbol{\Sigma}_i^{-1}(\mathbf{o}-\boldsymbol{\mu}_i)}
       {\mathbf{d}^\top \boldsymbol{\Sigma}_i^{-1}\mathbf{d}},
\quad
\mathbf{p}_{\mathrm{peak}}^{(i)} = r\!\left(t_{\mathrm{peak}}^{(i)}\right),
\end{equation}
\end{small}
where $\boldsymbol{\mu}_i$ denotes the center of the $i$-th Gaussian primitive and $\boldsymbol{\Sigma}_i$ its covariance matrix encoding anisotropic shape. Since Eq.~\eqref{eq:gaussian_peak} is analytic, both $t_{\mathrm{peak}}^{(i)}$ and $\mathbf{p}_{\mathrm{peak}}^{(i)}$ are differentiable with respect to the ray parameters and Gaussian parameters, enabling gradient propagation through the response query. At this point, we evaluate both the Gaussian response and the associated local material attributes of the primitive. The effective opacity is defined as
\begin{small}
\begin{equation}
\alpha_i = \sigma_i \rho\!\left(\mathbf{p}_{\mathrm{peak}}^{(i)}\right),
\end{equation}
\end{small}
where $\sigma_i$ denotes the opacity parameter of the $i$-th Gaussian primitive, and $\rho$ is the Gaussian density evaluated at its maximum-response point. This quantity is later used to determine interaction validity and to weight material aggregation.

\subsubsection{Rendering Equation.}
Consistent with prior Gaussian inverse rendering works~\cite{R3DG, IRGS, SVGIR2025}, we focus on opaque, surface-dominated scenes. Given an equivalent surface interaction at
$\mathbf{x}$ and an outgoing direction $\boldsymbol{\omega}_o$, the outgoing
radiance is described by the classical surface rendering
equation~\citep{kajiya1986rendering}:
\begin{small}
\begin{equation}
\label{eq:renderingequation}
L_o(\mathbf{x},\boldsymbol{\omega}_o)
=
L_e(\mathbf{x},\boldsymbol{\omega}_o)
+
\int_{\Omega}
f_r(\mathbf{x},\boldsymbol{\omega}_i,\boldsymbol{\omega}_o)
L_i(\mathbf{x},\boldsymbol{\omega}_i)
(\mathbf{n}\!\cdot\!\boldsymbol{\omega}_i)
\,d\boldsymbol{\omega}_i,
\end{equation}
\end{small}
where $L_o$, $L_i$, and $L_e$ denote the outgoing, incident, and emitted
radiance, respectively;
$\boldsymbol{\omega}_i$ and $\boldsymbol{\omega}_o$ denote the incident and
outgoing directions; and $\Omega$ denotes the upper hemisphere defined by the
shading normal $\mathbf{n}$. Since we assume non-emissive Gaussians,
$L_e=0$ at Gaussian surface interactions. The function $f_r$ denotes the bidirectional
reflectance distribution function (BRDF). We adopt a simplified Disney BRDF~\citep{burley12disneybrdf}, parameterized by diffuse
albedo $\mathbf{a}$, roughness $r$, and metallic value $m$. The BRDF in Eq.~\eqref{eq:renderingequation} combines a Lambertian diffuse
term $f_d=\frac{\mathbf{a}}{\pi}$ and a Cook--Torrance specular term $f_s$:
\begin{small}
\begin{equation}
f_s(\boldsymbol{\omega}_i,\boldsymbol{\omega}_o)
=
\frac{DFG}
{4(\mathbf{n}\!\cdot\!\boldsymbol{\omega}_i)
(\mathbf{n}\!\cdot\!\boldsymbol{\omega}_o)},
\end{equation}
\end{small}
where $D$, $F$, and $G$ denote the GGX normal distribution function,
the Schlick Fresnel term, and the Schlick--GGX geometry term, respectively.

Applying Eq.~\eqref{eq:renderingequation} to 3D Gaussian fields requires defining both surface interactions and light transport in path space. Instead of querying an explicit surface, each scattering event is constructed from the responses of overlapping Gaussians for BRDF evaluation. The rendering integral is then estimated through stochastic sampling and ray-traced visibility, while recursive ray propagation accounts for multi-bounce indirect illumination. This formulation provides a common physical basis for both path-traced rendering and the subsequent inverse optimization.

\subsection{Ray-Traced Light Transport in 3D Gaussian Fields}
\label{sec:forward_light_transport}
Path tracing naturally models global illumination through multi-bounce light
transport. Ray tracing provides a unified way to evaluate surface interactions,
visibility, and secondary rays in ray space, which remains challenging for
purely rasterization-based pipelines.

We trace rays directly through 3D Gaussian fields in a perspective-correct manner, 
rather than relying on the affine projection used in splatting. Our formulation targets solid objects,
where Gaussian primitives are interpreted as surface-oriented elements rather
than volumetric media such as smoke, as considered by the VPPT integrator
of~\citet{DontSplatYourGaussians}. Under this assumption, the Gaussian field
forms a thin-shell surface representation, allowing ray--surface interactions
to be modeled without treating the field as a participating medium or sampling volumetric scattering events along the ray.
\begin{figure}[h]
  \centering
  \includegraphics[
    width=0.9\linewidth,
        trim=0 0.2in 0 0.1in,
        clip
  ]{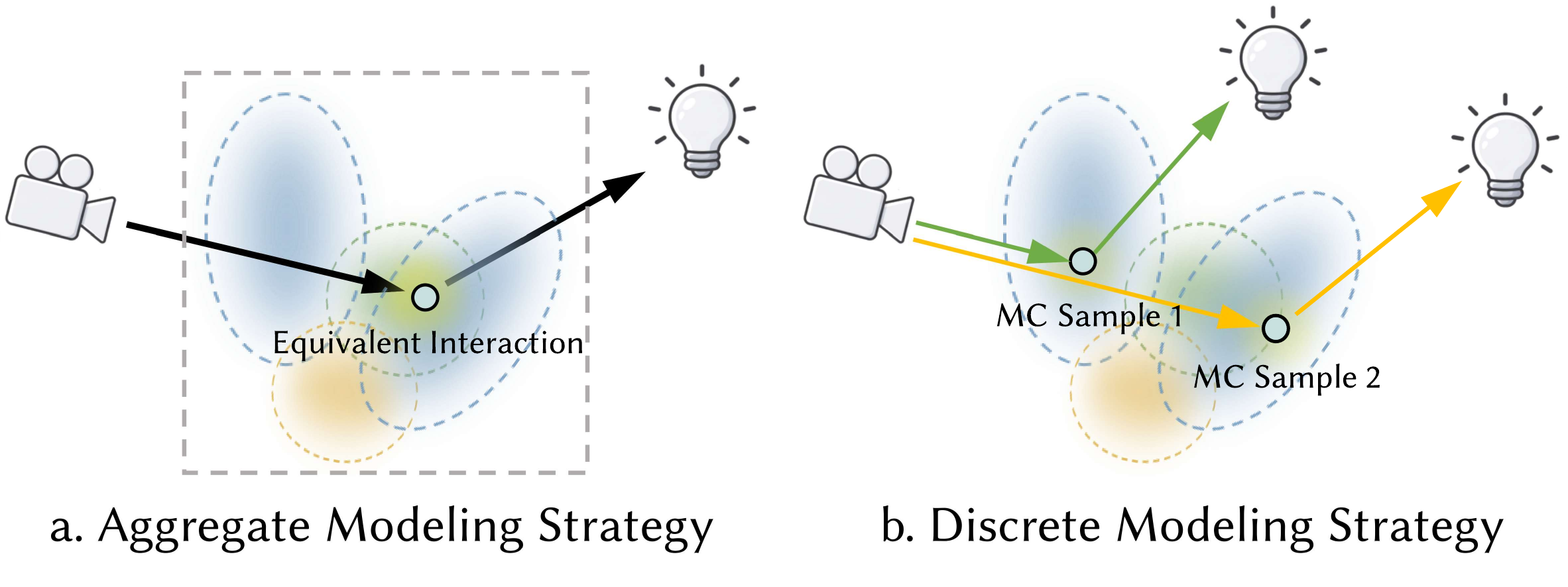}
  \caption{
  Different modeling strategies for ray--Gaussian interactions. (a) The aggregate strategy treats overlapping Gaussians as a single equivalent interaction. (b) The discrete strategy samples one primitive according to the opacity responses and treats it as the interaction.
  }
  \label{fig:modeling_strategies}
\end{figure}

3D Gaussians are semi-explicit primitives: a local surface is typically represented
by multiple overlapping Gaussians, so a ray interaction cannot always be
assigned to a single primitive.
In this work, we adopt an aggregated-surface interpretation (Fig.~\ref{fig:modeling_strategies}(a)), widely used and empirically supported in splatting-based methods, in which nearby contributing Gaussians jointly define an equivalent interaction state.
Ray spawning and shading are then performed from
this representative state, rather than from a stochastically selected primitive.
A discrete micro-surface interpretation (Fig.~\ref{fig:modeling_strategies}(b)) is also compatible with our framework by replacing the aggregation with stochastic selection of a contributing Gaussian primitive, but we found it more sensitive to local geometric inconsistencies in practice (Fig.~\ref{fig:modeling_results}). Path replay can propagate gradients through the selected interaction, so stochastic selection does not introduce additional nondifferentiability.

\subsubsection{Path-Space Equivalent Surface Interaction}
Under the above assumption, our goal is to define a stable path-space
representative interaction for the local aggregated surface encountered by the
ray. This interaction provides a common path-space state for BRDF evaluation,
ray spawning, and gradient replay.

\textit{Interaction Validity. }
Monte Carlo path tracing inherently introduces stochastic variance in the transport estimator. 
In 3D Gaussian fields, however, a ray may additionally accumulate weak responses from Gaussian tails or unstable overlaps that do not correspond to stable local surface interactions. 
If such unreliable responses are directly treated as valid hits, they can lead to unstable interaction depth and material estimates, and this instability further propagates through multi-bounce transport.

Therefore, we enforce a geometric validity criterion based on the accumulated contribution weight along the ray, i.e., the predicted opacity. A candidate path-space interaction is valid
only when this weight exceeds a geometry threshold:
\begin{small}
\begin{equation}
\label{eq:valid_hit}
\chi_{\text{geom}} = \mathbb{I}\!\left( \sum_{i=1}^{N} T_i \, \alpha_i > \tau_g \right),
\end{equation}
\end{small}
where $\chi_{\text{geom}} \in \{0,1\}$ denotes a binary indicator of geometric validity, $T_i$ is the accumulated transmittance up to the $i$-th Gaussian along the ray, $\alpha_i$ is the effective opacity of the $i$-th Gaussian evaluated at the ray interaction location, and $\tau_g$ is set to 0.5 by default.
Since the Gaussian geometry is constructed using the same response-based accumulation mechanism during training and is further regularized by a mask entropy loss that encourages foreground opacity toward one, this criterion does not introduce noticeable discontinuities during inverse rendering in practice.

\textit{Interaction State. }
Conditioned on a valid path-space interaction $\mathbf{p}$, we then define an equivalent interaction state $\mathbf{s}_{\mathbf{p}}$ that represents the aggregate contribution of the Gaussian primitives.
Since this state is intended to represent the local surface configuration rather than the accumulated response magnitude itself, we use normalized contribution weights to aggregate the contributing primitives.
Accordingly, the representative interaction depth is defined as
\begin{small}
\begin{equation}
\mathcal{D}
=
\mathbb{E}[d \mid \chi_{\mathrm{geom}}=1]
=
\sum_{i=1}^{N} w_i \, t_{\mathrm{peak}}^{(i)},
\quad
w_i = \frac{T_i \, \alpha_i}{\sum_{j=1}^{N} T_j \, \alpha_j}.
\end{equation}
\end{small}
Under the valid-hit condition, ${w_i}_{i=1}^{N}$ defines a conditional distribution over the contributing Gaussian primitives. Accordingly, $\mathcal{D}$ serves as the representative depth of the aggregated interaction.

Using the same normalized contribution weights, we further aggregate the local material attributes to obtain the effective interaction state. Specifically, the effective shading normal $\mathcal{N}\in\mathbb{R}^3$, albedo $\mathcal{A}\in\mathbb{R}^3$, roughness $\mathcal{R}\in\mathbb{R}$, and metallic $\mathcal{M}\in\mathbb{R}$ are defined as
\begin{small}
\begin{equation}
\label{eq:attribute}
\{\mathcal{N}, \mathcal{A}, \mathcal{R}, \mathcal{M}\}
=
\sum_{i=1}^{N} w_i \, \{\mathbf{n}_i, \mathbf{a}_i, r_i, m_i\},
\quad
w_i = \frac{T_i \alpha_i}{\sum_{j=1}^{N} T_j \alpha_j}.
\end{equation}
\end{small}
The aggregated normal $\mathcal{N}$ is normalized before BRDF evaluation.
The contribution-weighted attributes define the path-space shading
state of our equivalent-surface model, under which the induced light-transport integral is evaluated using Monte Carlo path tracing.

\subsubsection{Ray Spawning and Visibility Test}
\label{sec:ray_spawning}
We use multiple importance sampling (MIS)~\cite{Veach1997RobustMC} to estimate
incident radiance at each equivalent surface interaction, which reduces
variance by combining environment-light sampling and BRDF importance sampling.
At the $i$-th bounce, the incident radiance term in Eq.~\eqref{eq:rendering_mis} can be compactly expressed as
\begin{small}
\begin{equation}
\hat{L}_i
=
\underbrace{
w_{\mathrm{env}}^{(i)}
\frac{
V(\mathbf{p}_i,\boldsymbol{\omega}_{\mathrm{env}}^{(i)})
\,L_{\mathrm{env}}(\boldsymbol{\ell},\boldsymbol{\omega}_{\mathrm{env}}^{(i)})
}{
p_{\mathrm{env}}(\boldsymbol{\omega}_{\mathrm{env}}^{(i)})
}
}_{\mathclap{\text{Environment-sampled Query}}}
+
\underbrace{
w_{\mathrm{brdf}}^{(i)}
\frac{
\hat{L}_{i+1}\!\left(\mathbf{r}_{i+1}\right)
}{
p_{\mathrm{brdf}}(\boldsymbol{\omega}_{\mathrm{brdf}}^{(i)} \mid \mathbf{s}_{\mathbf{p}_i})
}
}_{\mathclap{\text{BRDF-sampled Query}}},
\label{eq:rendering_mis}
\end{equation}
\end{small}
where $w_{\mathrm{env}}^{(i)}$ and $w_{\mathrm{brdf}}^{(i)}$ are the MIS weights, $V(\cdot)$ denotes the visibility estimated by the shadow-ray test, and $\hat{L}_{i+1}(\mathbf{r}_{i+1})$ is the radiance carried by the next-bounce ray $\mathbf{r}_{i+1}$.

\emph{Shadow Ray Test:}
We estimate shadow-ray visibility $V(\cdot)$ using a stochastic transmittance
test in the any-hit shader, following~\citet{hu2025realtimeglobalilluminationdynamic}.
This yields a reliable estimate of transmittance while allowing early
termination in high-opacity regions for improved efficiency.

\emph{Next-Ray Event Determination:} 
The BRDF-sampled term in Eq.~\eqref{eq:rendering_mis} is evaluated by tracing
the spawned ray $\mathbf{r}_{i+1}$ and determining the next transport event.
For continuation rays, the stochastic transmittance test used for shadow rays
must be coupled with a geometric validity check, since weak Gaussian-tail
responses should not be treated as valid secondary surface interactions.
Accordingly, we augment stochastic transmittance
continuation with an additional geometric validity criterion:
\begin{small}
\begin{equation}
\label{eq:geom_gate}
\chi_{\mathrm{pass}}
=
\mathbb{I}\!\left(\chi_{\mathrm{geom}}=0\right)\,
\mathbb{I}\!\left(\xi < T_{\mathrm{ray}}\right),
\quad
\xi \sim \mathcal{U}(0,1),
\end{equation}
\end{small}
where $\chi_{\mathrm{pass}}\in\{0,1\}$ indicates whether the spawned ray 
passes through the Gaussian field when no valid subsequent interaction is formed,
$\chi_{\mathrm{geom}}$ is the geometric validity indicator defined in
Eq.~\eqref{eq:valid_hit}, and $T_{\mathrm{ray}}$ denotes the accumulated
transmittance along the ray.

\textit{Self-Intersection Rejection:}
A direct consequence of representing a surface by overlapping Gaussian
primitives is that rays spawned from an equivalent interaction may immediately
encounter residual responses from primitives that already contributed to the
same local surface. Such responses do not correspond to a new transport event,
but to repeated intersections with the same aggregated thin-shell surface, and
can therefore introduce spurious self-occlusion for both shadow rays and
continuation rays.

A fixed origin offset treats self-intersection as a purely numerical issue and
does not generalize well across models, since the appropriate offset magnitude
depends on object scale and local Gaussian support thickness. We therefore
introduce a backface-aware origin-offset strategy that classifies local
back-facing responses along the spawned ray as self-intersections and advances
the ray origin past the corresponding Gaussian peak:
\begin{small}
\begin{equation}
\label{eq:self_occ}
\hat{\mathbf{o}}
=
\mathbf{o}
+
\chi_{\mathrm{self}}\,(t_{\mathrm{peak}}+\epsilon)\,\mathbf{d},
\quad
\chi_{\mathrm{self}}
=
\mathbb{I}\!\left(t_{\mathrm{peak}}>0\right)
\mathbb{I}\!\left(\mathbf{d}^{\top}\mathbf{n}>0\right).
\end{equation}
\end{small}
Here $\hat{\mathbf{o}}$ denotes the updated ray origin,
$t_{\mathrm{peak}}$ is defined in Eq.~\eqref{eq:gaussian_peak},
$\epsilon$ is a small positive constant for numerical robustness,
and $\mathbf{n}$ denotes the shading normal of the candidate Gaussian primitive.
The condition $\mathbf{d}^{\top}\mathbf{n}>0$ identifies a back-facing response
relative to the spawned ray, which is treated as a repeated hit of the current
aggregated surface rather than a valid occluding interaction.

\subsection{Learnable Environment Light}
\label{sec:learnable_environment_lighting}

Incident directions sampled from the environment illumination directly participate in BRDF evaluation and can affect gradient variance. In addition to conventional 2D environment maps or cube maps, we optionally parameterize environment illumination using a compact set of $N_{\mathrm{sg}}$ Spherical Gaussians (SGs). This continuous and differentiable representation is compatible with Monte Carlo ray sampling and multiple importance sampling (MIS), while its compact parameter space yields smoother optimization and improved stability under low-SPP training.

The SG lobe directions $\{\boldsymbol{\eta}_i\}_{i=1}^{N_{\text{sg}}}$ are
initialized uniformly over the unit sphere via Fibonacci sampling~\cite{gonz09fib},
where $\boldsymbol{\eta}_i \in \mathbb{S}^2$ denotes the unit direction of the
$i$-th SG lobe. The radiance along direction $\mathbf{v}$ is modeled as
\begin{small}
\begin{equation}
L(\mathbf{v}) =
\sum_{i=1}^{N_{\text{sg}}}
\mathbf{c}_i
\exp\!\big(\lambda_i(\boldsymbol{\eta}_i \cdot \mathbf{v} - 1)\big),
\quad
\mathbf{c}_i = R(\mathbf{x}_i;\gamma),
\end{equation}
\end{small}
where $\mathbf{x}_i \in \mathbb{R}^3$ denotes the unconstrained RGB amplitude
parameter of the $i$-th SG lobe, $\lambda_i$ controls its sharpness, and
$R(\mathbf{x};\gamma)=\exp(\operatorname{softplus}(\mathbf{x})/\gamma)-1$
maps raw parameters to non-negative HDR radiance values. This compact
parameterization preserves high-dynamic-range illumination, avoids invalid
negative radiance, and stabilizes environment-light optimization in our
ray-traced inverse rendering framework.

\subsection{Inverse Rendering}
\label{sec:inverse_rendering}
\subsubsection{Problem Formulation}
Building upon the path-tracing framework defined above, we formulate physically based rendering with multiple importance sampling (MIS) over 3D Gaussian fields. 
A camera ray is denoted as $\mathbf{r}(\mathbf{o},\mathbf{d})$, where $\mathbf{o}$ and $\mathbf{d}$ denote the ray origin and direction, respectively. Given a camera ray $\mathbf{r}$, the rendered radiance is expressed as
\begin{equation}
\mathcal{T}(\mathbf{r})
=
\sum_{i=0}^{K}
\beta_i \,
f_{\mathrm{brdf}}(\mathbf{s}_{\mathbf{p}_i})\,
L_i(\boldsymbol{\ell}, \mathbf{d}_i),
\quad
\mathbf{s}_{\mathbf{p}_i}
=
\Phi\!\left(
\mathbf{r}_i,\,
\sum_{j=1}^{N_i} w_j^{(i)} \boldsymbol{\Theta}_j
\right),
\end{equation}
where $\beta_i$ is the path throughput up to the $i$-th bounce, $f_{\mathrm{brdf}}(\mathbf{s}_{\mathbf{p}_i})$ is the BRDF evaluated at the equivalent interaction state $\mathbf{s}_{\mathbf{p}_i}$, and $L_i(\boldsymbol{\ell}, \mathbf{d}_i)$ denotes the incident radiance arriving from direction $\mathbf{d}_i$ under lighting $\boldsymbol{\ell}$. Each path-space interaction $\mathbf{p}_i$ is associated with a local shading state $\mathbf{s}_{\mathbf{p}_i}$ formed by aggregating the attributes of the Gaussians contributing to that interaction, where $\Phi(\cdot)$ denotes the mapping from the current ray and aggregated Gaussian attributes to the local shading state.

Our goal is to jointly recover material attributes $\boldsymbol{\Theta}$ and lighting $\boldsymbol{\ell}$ by minimizing the discrepancy between rendered images and ground-truth images $I_{\text{gt}}$:
\begin{small}
\begin{equation}
\label{eq:inverse_problem}
\min_{\boldsymbol{\Theta},\,\boldsymbol{\ell}}
\;
\mathbb{E}_{\mathbf{r}\sim \pi(\mathbf{r})}
\left[
\mathcal{L}\!\left(
\mathcal{T}(\mathbf{r};\boldsymbol{\Theta},\boldsymbol{\ell}),
I_{\text{gt}}(\mathbf{r})
\right)
\right],
\end{equation}
\end{small}
where $\pi(\mathbf{r})$ denotes the camera-ray sampling distribution over image pixels, and $p(\mathbf{r})=(\mathbf{p}_0,\mathbf{p}_1,\dots,\mathbf{p}_K)$ denotes the multi-bounce light transport path induced by the camera ray $\mathbf{r}$.

\subsubsection{Gradient Propagation over Equivalent Interactions}
Direct differentiation of Eq.~\eqref{eq:inverse_problem} with respect to
the learnable parameters is impractical for Monte Carlo path tracing: it requires
retaining a large intermediate computation graph across many sampled paths, while Gaussian aggregation further requires tracking the contributing primitives and their interaction states, leading to substantial memory overhead.
Existing inverse rendering methods in 3D Gaussian fields~\cite{R3DG}
typically address this by caching intermediate screen-space states via splatting
(e.g., G-buffers). In a fully ray-traced pipeline, however, explicitly caching per-path states becomes increasingly expensive as the sample count and the number of bounces increase.

We instantiate path replay backpropagation (PRB)~\cite{PathReplayBackpropagation}
over the same path-space equivalent interactions used by the forward estimator.
During replay, equivalent interactions are reconstructed from replayed rays and
random numbers, yielding a pathwise gradient conditioned on the replayed sampling
decisions. The replayed contribution uses the same weighted estimator form as
the forward pass, with PDFs, MIS weights, visibility indicators, and validity
decisions evaluated on the replayed path and held fixed during differentiation.
Gradients propagate only through smooth pathwise terms, including BRDF evaluation,
equivalent interaction attributes, incident radiance, and ray-direction reparameterization. The
resulting pathwise replay gradient can be expressed as
\begin{small}
\begin{equation}
\frac{\partial \mathcal{L}}{\partial \boldsymbol{\Theta}_j}
=
\sum_i
\frac{\partial \mathcal{L}}{\partial \mathcal{T}_i}
\Bigg(
\underbrace{
\frac{\partial \mathcal{T}_i}{\partial f_{\mathrm{brdf}}}
\frac{\partial f_{\mathrm{brdf}}}{\partial \mathbf{s}_{\mathbf{p}_i}}
}_{\mathclap{\text{BRDF gradient}}}
+
\underbrace{
\frac{\partial \mathcal{T}_i}{\partial L_i}
\frac{\partial L_i}{\partial \mathbf{d}_i}
\frac{\partial \mathbf{d}_i}{\partial \mathbf{s}_{\mathbf{p}_i}}
}_{\mathclap{\text{Ray-induced gradient}}}
\Bigg)
\frac{\partial \mathbf{s}_{\mathbf{p}_i}}{\partial \boldsymbol{\Theta}_j}.
\label{eq:gs_gradient_decomp}
\end{equation}
\end{small}
Since the local BRDF value enters the path throughput multiplicatively at the
current interaction, its derivative is accumulated through the replayed
throughput along the sampled path. For the BRDF parameterization in this work,
albedo $\mathbf{a}$ only affects the local BRDF value but not the sampled
direction and therefore receives gradients only through the BRDF branch, while
roughness $r$ affects both the local BRDF value and the reparameterized outgoing
direction, and therefore additionally receives gradients through the ray-induced
branch.

Since each equivalent interaction is explicitly aggregated from contributing
3D Gaussian primitives, the contribution weight of each participating primitive
is known. For local material attributes, this allows gradients of the
equivalent interaction state to be directly redistributed to the contributing
Gaussians:
\begin{small}
\begin{equation}
\frac{\partial \mathbf{s}_{\mathbf p_i}}{\partial \mathbf g_j}
=
w_j^{(i)}\mathbf I,
\quad
\bar{\mathbf g}_j
\mathrel{+}
=
w_j^{(i)}
\bar{\mathbf{s}}_{\mathbf p_i},
\label{eq:interaction_grad}
\end{equation}
\end{small}
where $\mathbf{g}_j$ denotes the local material-attribute vector of the
$j$-th Gaussian primitive, and $w_j^{(i)}$ is its contribution weight at the
$i$-th replayed interaction. This weight-sharing between primal aggregation and
gradient redistribution is what keeps the backward pass aligned with the
forward ray-traced estimator.

The gradient of the environment light parameters can be expressed as
\begin{small}
\begin{equation}
\frac{\partial \mathcal{L}}{\partial \boldsymbol{\ell}}
=
\sum_i
\frac{\partial \mathcal{L}}{\partial \mathcal{T}_i}
\frac{\partial \mathcal{T}_i}{\partial L_i}
\frac{\partial L_i}{\partial \boldsymbol{\ell}}
=
\sum_i
\frac{\partial \mathcal{L}}{\partial \mathcal{T}_i}
\,\beta_i\,
f_{\mathrm{brdf}}
\frac{\partial L_i}{\partial \boldsymbol{\ell}},
\label{eq:lighting_gradient}
\end{equation}
\end{small}
where the gradient flows from the rendering loss to the environment parameters
through the path throughput and the incident radiance evaluated along replayed
paths.

\subsubsection{Training Objective}
\label{sec:training_loss}
We optimize the material attributes $\Theta$ and the light parameters $\ell$ by minimizing an objective composed of an appearance term $\mathcal{L}_{\text{view}}$ and regularization terms $\mathcal{L}_{\text{reg}}$. Following~\citet{kerbl3dgaussians}, the appearance term combines an $\ell_1$ term with a D-SSIM term:
\begin{small}
\begin{equation}
\mathcal{L}_{\text{view}}
=
(1 - \lambda_s)\,\mathcal{L}_{1}
+
\lambda_s\,\mathcal{L}_{\text{D-SSIM}}.
\end{equation}
\end{small}

For regularization $\mathcal{L}_{\mathrm{reg}}$, we apply an edge-aware total
variation penalty to the predicted material attribute maps. Let $\mathcal{I}$
denote the set of regularized attribute maps. The smoothness term is defined as
\begin{small}
\begin{equation}
\label{eq:tv_loss}
\mathcal{L}_{\mathrm{tv}}
=
\sum_{I \in \mathcal{I}}
\sum_{\mathbf p}
\|\nabla I(\mathbf p)\|_{2}^{2}
\exp\!\left(-\|\nabla I_{\mathrm{gt}}(\mathbf p)\|\right),
\end{equation}
\end{small}
where $I_{\mathrm{gt}}$ is the reference RGB image used to compute the
edge-aware weights.

Jointly estimating material properties and environment light can be ambiguous.
Following~\citet{GS-ID}, we introduce an early-stage material prior to alleviate
this coupling without requiring per-scene parameter tuning. Specifically, we
employ a monocular diffusion-based estimator~\citep{rgb2x} to provide material
priors and regularize the predicted material maps during the early optimization
stage:
\begin{small}
\begin{equation}
\mathcal{L}_{\text{reg}}
=
\lambda_{\mathrm{tv}}\,\mathcal{L}_{\mathrm{tv}}
+
\lambda_{\text{pri}}
\sum_{I\in\mathcal I}
\mathcal{L}_{2}(I, I_{\text{pri}}),
\end{equation}
\end{small}
where $I_{\text{pri}}$ denotes the corresponding values predicted by the monocular prior. The prior weight $\lambda_{\text{pri}}$ is initialized to $0.5$, linearly decayed to $0.005$ over the first 400 iterations, and kept fixed thereafter. The prior is assigned a relatively small weight and mainly provides weak guidance for material initialization during early optimization, with the RGB reconstruction loss remaining the primary supervision.
\begin{figure*}[!t]
    \centering
    \includegraphics[
        width=0.8\linewidth,
        trim=0 0.3in 0 0.2in,
        clip
    ]{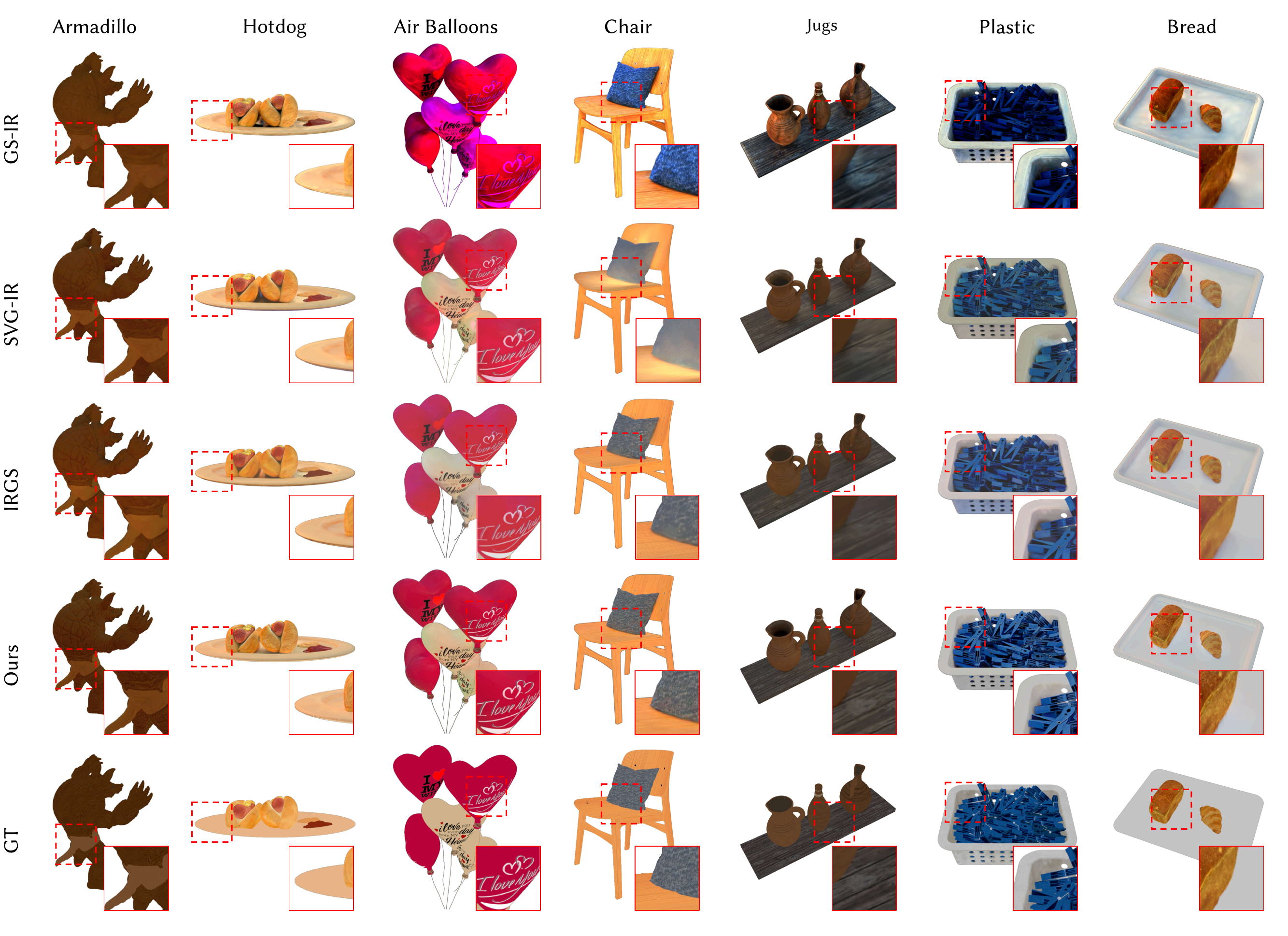}
    \caption{Albedo recovery results on benchmark datasets, demonstrating that our framework achieves superior performance on albedo recovery.
}
 \label{fig:albedo}
\end{figure*}

\section{EXPERIMENTS AND ABLATIONS}
\subsection{Implementation Details}
We implement our method in an NVIDIA OptiX-based ray-tracing framework~\cite{optix} and run all experiments on an NVIDIA RTX~3090 GPU. We adopt a two-stage optimization pipeline. In the first stage, Gaussian normals are supervised by depth-derived pseudo-normals using an angular normal-consistency loss, together with a normal prior from the same diffusion-based estimator described in Sec.~\ref{sec:training_loss}. Gaussian densification via splitting and cloning is enabled in this stage.
In the inverse-rendering stage, we fix the geometry and optimize the model for 800 iterations. Optimization is performed with 32 samples per pixel (SPP) and a maximum of 2 bounces, while final rendering uses 256 SPP. The D-SSIM mixing weight is set to $\lambda_{\mathrm{s}}=0.2$, and the TV regularization weight is set to $\lambda_{\mathrm{tv}}=0.095$.
For the spherical Gaussian representation, we set $N_{\mathrm{sg}}=24$ and $\gamma=0.3$, and use a learning rate of $0.035$ for both material and light parameters.
A single filtering pass with a fixed \(3\times3\) normalized binomial kernel is applied to the rendered image during optimization to suppress Monte Carlo noise.

\begin{table}[!htbp]
  \centering
  \caption{Quantitative evaluation on TensoIR dataset.}
  \label{tab:TensoIR}
  \setlength{\tabcolsep}{5pt}
  \renewcommand{\arraystretch}{1.1}
  \resizebox{0.98\linewidth}{!}{
  \begin{tabular}{lcccccc}
    \toprule
    \multirow{2}{*}{Method}
    & Albedo
    & NVS
    & Relight
    & Roughness
    & Normal
    & Training\\
    
    & PSNR$\uparrow$
    & PSNR$\uparrow$
    & PSNR$\uparrow$
    & MSE$\downarrow$
    & MAE$\downarrow$
    & Hours$\downarrow$\\
    \midrule

    TensoIR~\citep{Jin2023TensoIR}
      & 29.27
      & 35.09
      & 28.58
      & \cellcolor{first!25}{0.013} 
      & \cellcolor{second!25}4.100
      & 4.0\\

    GS-IR~\citep{GS-IR}
      & 29.94
      & 35.33
      & 24.37
      & 0.027
      & 4.948
      & \cellcolor{second!25}0.5\\

    R3DG~\citep{R3DG}
      & 29.27
      & 33.35
      & 27.37
      & \cellcolor{third!25}{0.016}
      & 5.927
      & 1.1\\

    SVG-IR~\citep{SVGIR2025}
      & 30.48
      & \cellcolor{first!25}{36.71}
      & \cellcolor{third!25}{31.10}
      & 0.033
      & 4.358
      & 1.1\\

    IRGS~\citep{IRGS}
      & \cellcolor{third!25}{30.62}
      & 35.43
      & 29.91
      & 0.054 
      & \cellcolor{third!25}4.209
      & \cellcolor{third!25}0.9\\

    \midrule
    Ours w/o Prior
      & \cellcolor{second!25}{31.04}
      & \cellcolor{second!25}{36.50} 
      & \cellcolor{second!25}{31.31}
      & 0.023
      & \cellcolor{first!25}
      & \cellcolor{first!25}\\
    
    Ours
      & \cellcolor{first!25}{32.12}
      & \cellcolor{third!25}{36.17} 
      & \cellcolor{first!25}{31.84}
      & \cellcolor{second!25}{0.015}
      & \cellcolor{first!25}\multirow{-2}{*}{4.028}
      & \cellcolor{first!25}\multirow{-2}{*}{0.4}\\
    \bottomrule
  \end{tabular}
  }
\end{table}
\begin{table}[!htbp]
  \centering
  \caption{Quantitative evaluation on Synthetic4Relight dataset.}
  \label{tab:Synthetic4Relight}
  \setlength{\tabcolsep}{5pt}
  \renewcommand{\arraystretch}{1.1}
  \resizebox{0.98\linewidth}{!}{
  \begin{tabular}{lccccc}
    \toprule
    \multirow{2}{*}{Method}
    & Albedo 
    & NVS
    & Relight 
    & Roughness
    & Training \\
    
    & PSNR$\uparrow$
    & PSNR$\uparrow$
    & PSNR$\uparrow$
    & MSE$\downarrow$ 
    & Hours$\downarrow$\\
    \midrule

    GS-IR~\citep{GS-IR}
      & 19.48
      & 33.95
      & 25.40
      & 0.011 
      & \cellcolor{first!25}0.4\\
    
    R3DG~\citep{R3DG}
      & 28.65
      & {34.10}
      & {33.12}
      & {0.010} 
      & 0.9\\

    SVG-IR~\citep{SVGIR2025}
      & 29.06
      & \cellcolor{third!25}36.10
      & 32.59
      & 0.009
      & 1.0 \\
    
    IRGS~\citep{IRGS}
      & \cellcolor{third!25}{30.50}
      & 34.42
      & \cellcolor{second!25}{34.35}
      & \cellcolor{first!25}{0.008} 
      & \cellcolor{third!25}0.7 \\

    \midrule    
    Ours w/o Prior
      & \cellcolor{second!25}30.80
      & \cellcolor{first!25}36.84
      & \cellcolor{third!25}33.82
      & \cellcolor{first!25}0.008
      & \cellcolor{second!25}\\
      
    Ours
      & \cellcolor{first!25}{31.09}
      & \cellcolor{second!25}{36.66}
      & \cellcolor{first!25}{34.59}
      & \cellcolor{first!25}{0.008}
      & \cellcolor{second!25}\multirow{-2}{*}{0.6} \\
    \bottomrule
  \end{tabular}
  }
\end{table}
\begin{table}[!htbp]
  \centering
  \caption{Quantitative evaluation on RT4Relight dataset.}
  \label{tab:RT4Relight}
  \setlength{\tabcolsep}{5pt}
  \renewcommand{\arraystretch}{1.1}
  \resizebox{0.98\linewidth}{!}{
  \begin{tabular}{lccccc}
    \toprule
    \multirow{2}{*}{Method}
    & Albedo
    & NVS
    & Relight
    & Roughness
    & Training\\
    
    & PSNR$\uparrow$
    & PSNR$\uparrow$
    & PSNR$\uparrow$
    & MSE$\downarrow$
    & Hours$\downarrow$\\
    \midrule

    GS-IR~\cite{GS-IR}
      & 23.89
      & 33.87
      & 24.91
      & 0.083 
      & \cellcolor{first!25}{0.5}\\

    R3DG~\cite{R3DG}
      & 25.80
      & 34.61
      & 24.60
      & {0.025}
      & 0.9\\

    SVG-IR~\cite{SVGIR2025}
      & 26.71
      & \cellcolor{first!25}{36.63}
      & {27.77}
      & \cellcolor{third!25}{0.022} 
      & 1.4\\

    IRGS~\cite{IRGS}
      & \cellcolor{third!25}{29.28}
      & \cellcolor{third!25}{36.35}
      & \cellcolor{third!25}{29.56}
      & 0.065 
      & \cellcolor{third!25}{0.7}\\

    \midrule
    Ours w/o Prior
      & \cellcolor{second!25}{29.34}        
      & \cellcolor{second!25}{36.49}
      & \cellcolor{second!25}{29.81}
      & \cellcolor{second!25}{0.018}
      & \cellcolor{second!25}\\
    Ours
      & \cellcolor{first!25}{29.57}
      & {35.99}
      & \cellcolor{first!25}{30.76}
      & \cellcolor{first!25}{0.016} 
      & \cellcolor{second!25}\multirow{-2}{*}{0.6}\\
    \bottomrule
  \end{tabular}
  }
\end{table}

\subsection{Results}
\label{sec: results}
We compare our method with state-of-the-art splatting-based inverse-rendering
approaches~\cite{GS-IR,R3DG,SVGIR2025,IRGS}. IRGS~\cite{IRGS} and
SVG-IR~\cite{SVGIR2025} incorporate indirect illumination through
screen-space approximations. To align with these baselines, we fix the metallic parameter to zero and use a 128×64 2D environment map for benchmark evaluations. Given sufficient optimization, this higher-capacity representation offers flexibility for fitting complex
illumination and can achieve stronger benchmark performance. Separately, for our standalone real-world experiments, we adopt the compact spherical Gaussian illumination representation, which enables faster convergence and smoother optimization gradients.

\begin{figure}[!htbp]
    \centering
    \includegraphics[
        width=0.95\linewidth,
        trim=0.9in 1.1in 0.6in 0,
        clip
    ]{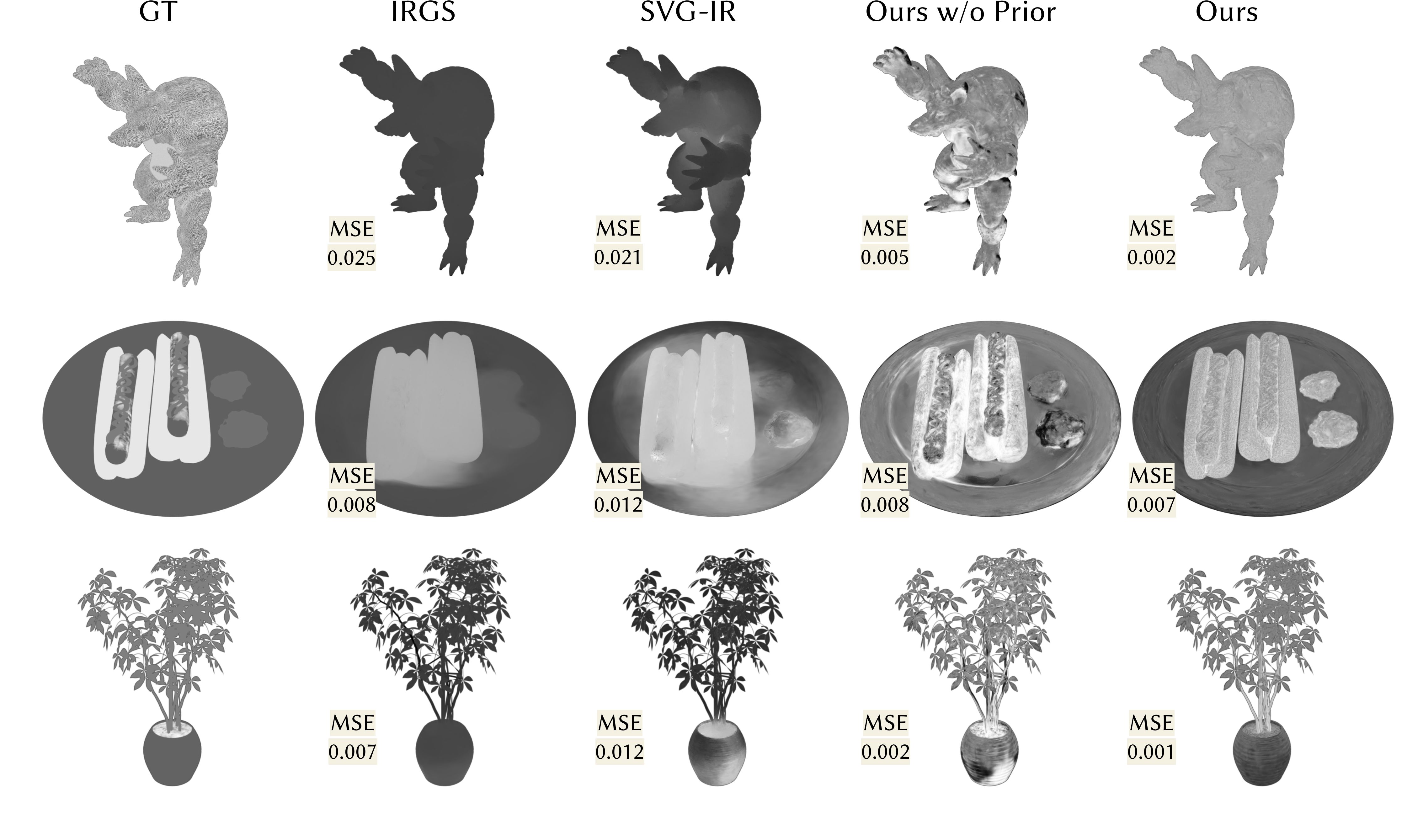}
    \caption{Roughness recovery results. Our recovered roughness is closer to the
ground truth, while the diffusion prior helps regularize early optimization and
suppress degenerate low-roughness artifacts.}
    \label{fig:roughness}
\end{figure}

\textit{Results on Benchmark Datasets.}
We evaluate on TensoIR~\cite{Jin2023TensoIR}, Synthetic4Relight~\cite{zhang2022invrender}, and RT4Relight. These synthetic datasets provide ground-truth materials and relighting images, enabling quantitative evaluation of material decomposition and relighting quality. 
Quantitative results in
Tabs.~\ref{tab:TensoIR},~\ref{tab:Synthetic4Relight}, and~\ref{tab:RT4Relight}
show that our method achieves competitive and stable performance across most metrics while maintaining competitive training efficiency. Training time is averaged across all scenes.
Qualitative albedo and roughness results are shown in
Figs.~\ref{fig:albedo} and~\ref{fig:roughness}.

\begin{figure}[!htbp]
    \centering
    \includegraphics[
        width=0.95\linewidth,
        trim=0 0.1in 0 0.2in,
        clip
    ]{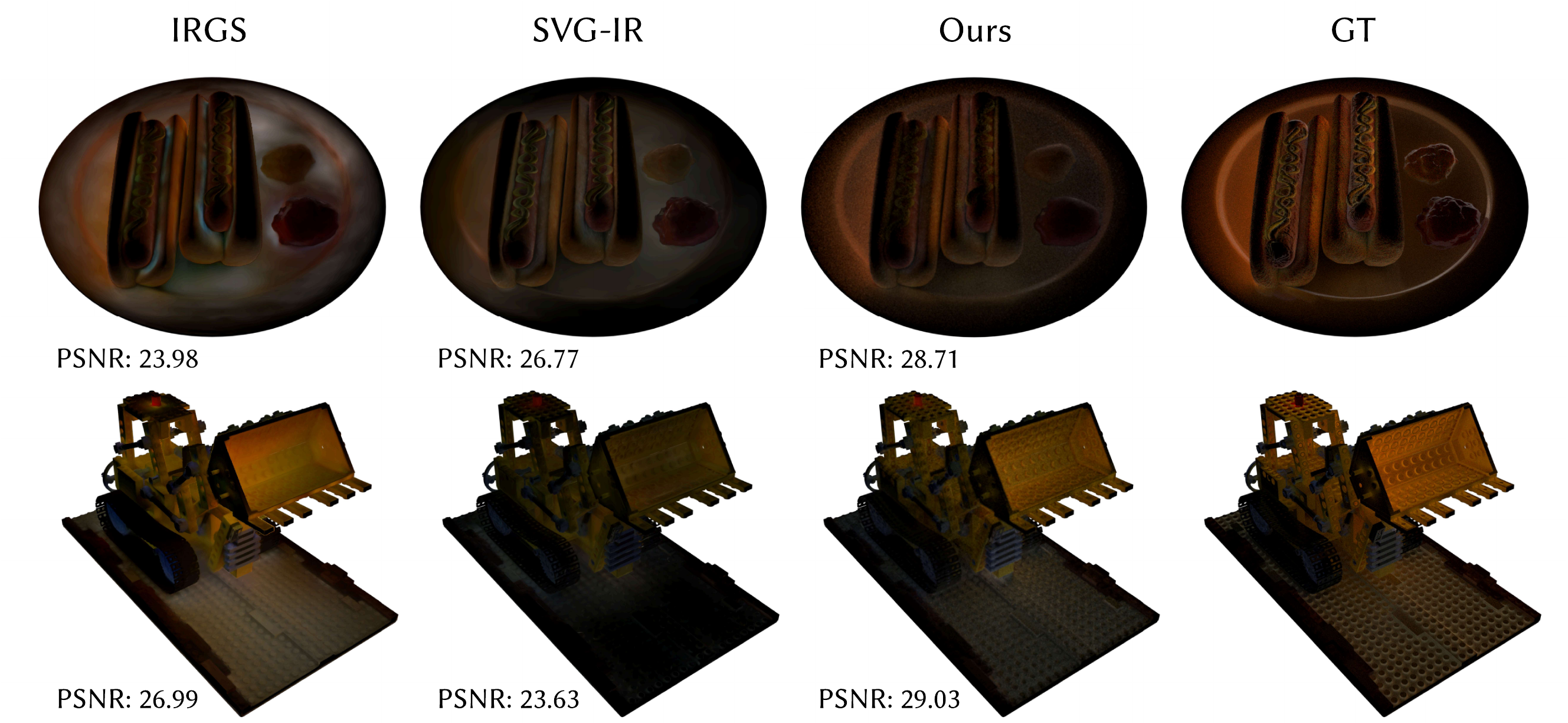}
    \caption{Indirect-only rendering results from inverse-rendered scenes. Our result is obtained through path tracing under the full
    rendering equation, yielding indirect-light estimates that are closer to the
    ground truth.}
    \label{fig:indirect_light}
\end{figure}

\begin{figure}[!htbp]
    \centering
    \includegraphics[width=\linewidth]{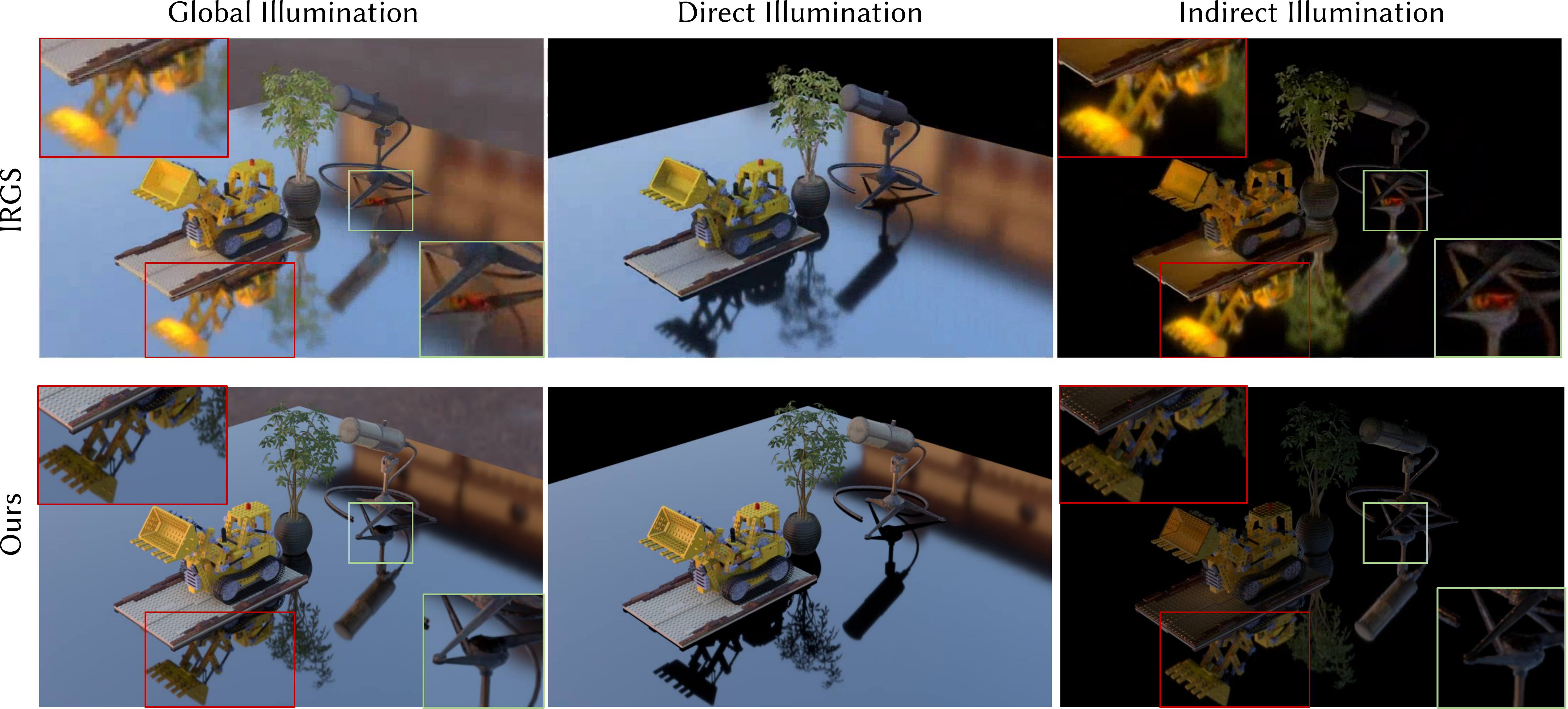}
    \caption{Multi-bounce light transport result. Compared with IRGS~\cite{IRGS},
our path-traced framework explicitly traces repeated light bounces under
the full rendering equation, enabling more consistent direct illumination,
reflections, and indirect radiance.}
    \label{fig:irgs}
\end{figure}

\begin{figure}[!b]
    \centering
    \includegraphics[
        width=\linewidth,
        trim=0 0 0 0in,
        clip
    ]{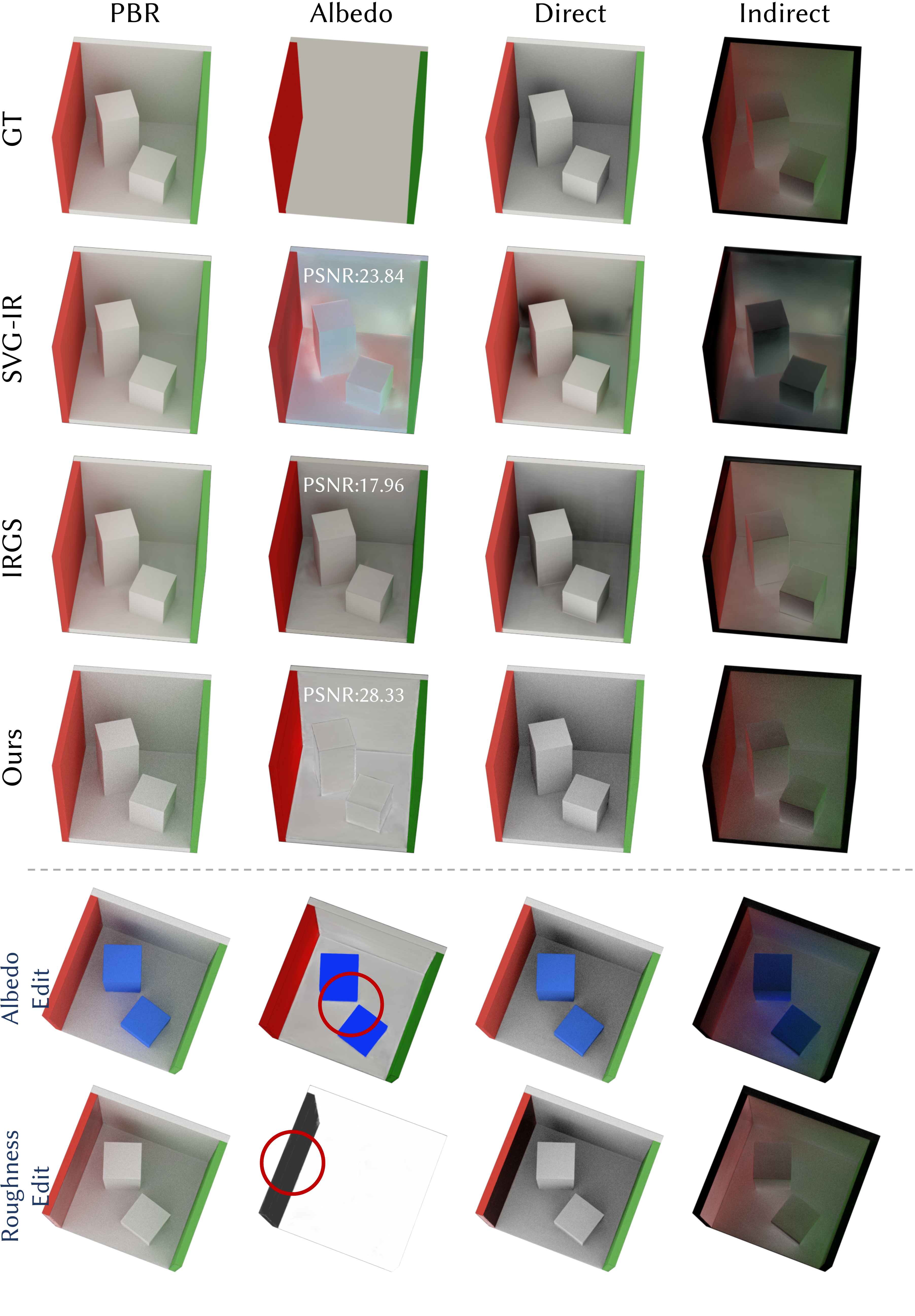}
    \caption{Cornell box inverse rendering and material editing with a maximum of 4 bounces. 
    Material edits in our framework, including albedo and roughness changes, 
    affect not only the directly visible appearance but also global illumination 
    effects such as indirect radiance and color bleeding.}
    \label{fig:cornell_box}
\end{figure}

\clearpage
\begin{figure*}[t]
    \centering
    \includegraphics[width=0.99\linewidth]{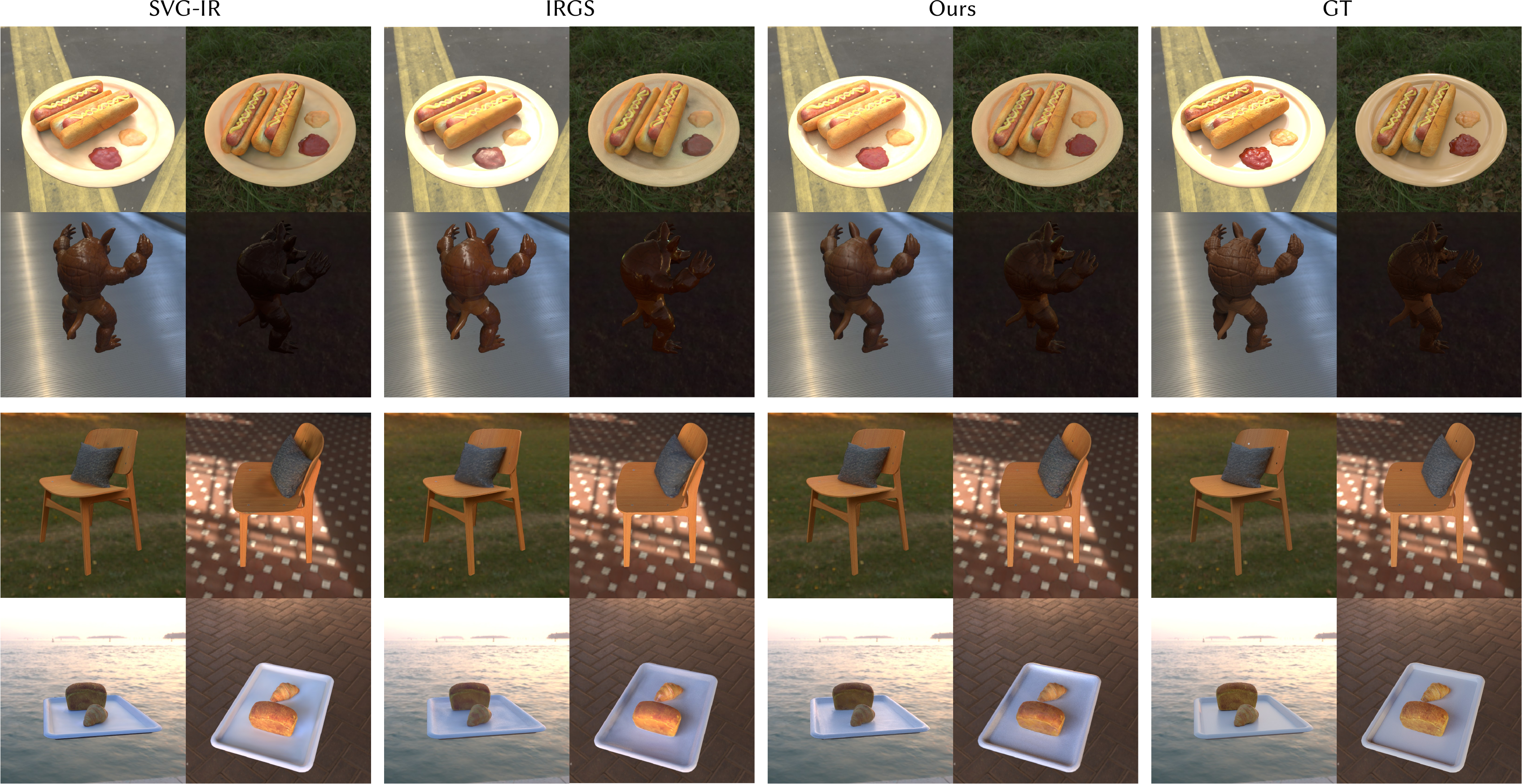}
    \caption{Relighting comparison on benchmark datasets. Our framework achieves more natural relighting with soft shadows and realistic reflections.}
    \label{fig:object_relight}
\end{figure*}

\begin{figure*}[t]
  \centering
  \includegraphics[width=0.99\linewidth]{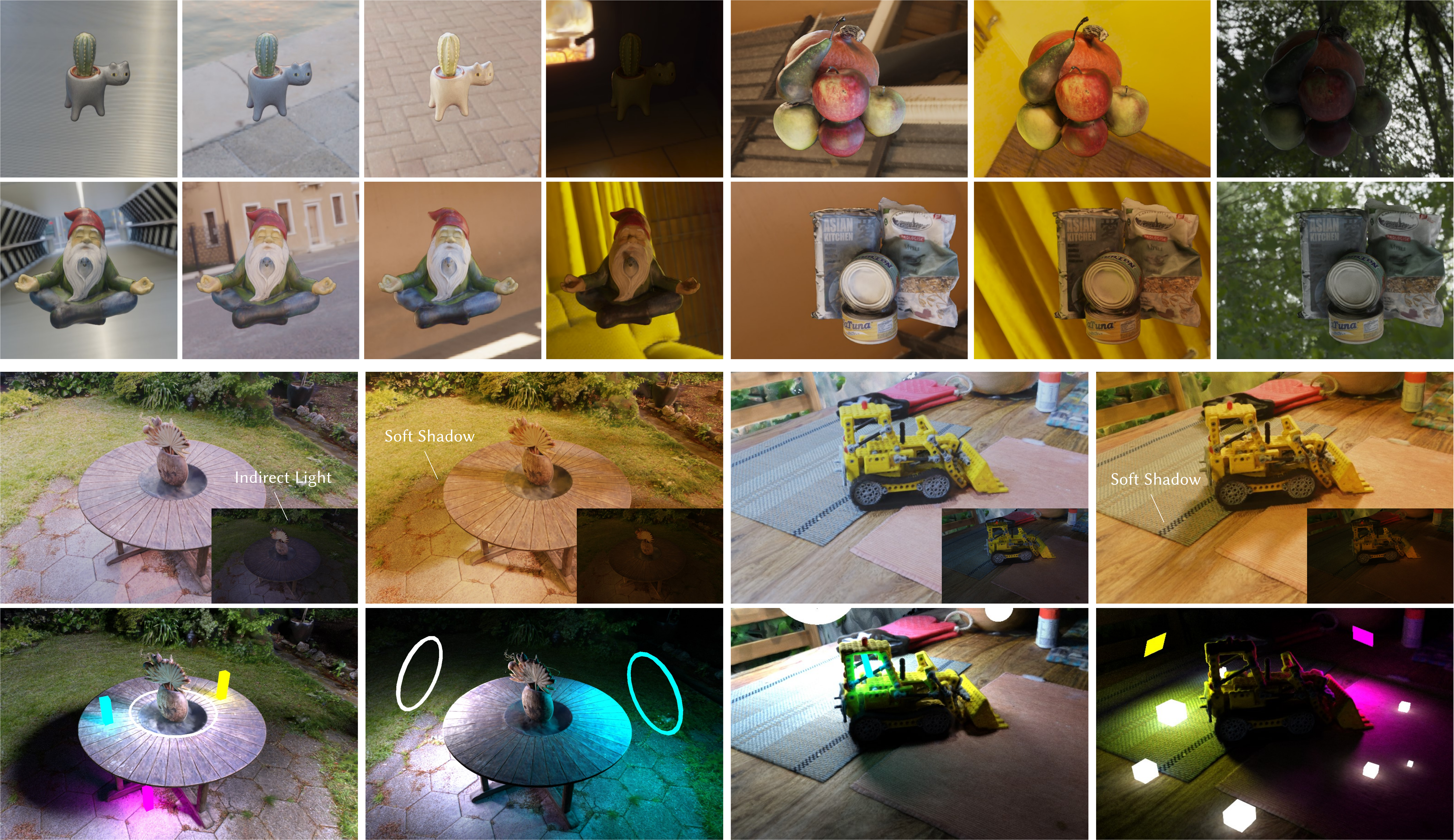}
  \caption{Relighting results from our method on
  Stanford-ORB~\cite{kuang2023stanfordorb}, DTU~\cite{dtu}, and
  Mip-NeRF 360~\cite{barron2022mipnerf360} scenes. We demonstrate
  relighting with environment maps and near-field light sources, producing
  consistent illumination and soft shadows under the target illumination.}
  \label{fig:real_relight}
\end{figure*}
\clearpage

\textit{Results on Global Illumination.}
We compare our method with IRGS~\cite{IRGS} and SVG-IR~\cite{SVGIR2025} in
terms of indirect illumination (Fig.~\ref{fig:indirect_light}). When the
recovered objects are rendered in a forward path-tracing pipeline, our method
produces more plausible indirect illumination (Fig.~\ref{fig:irgs}). IRGS
models indirect illumination using spherical-harmonic values attached to
Gaussians, which can lead to inaccurate indirect-light intensity when evaluated
under multi-bounce global illumination. Experiments on the Cornell-box scene
(Fig.~\ref{fig:cornell_box}) suggest that multi-bounce global illumination helps
disentangle surface albedo from color bleeding. Moreover, due to the consistency
between forward rendering and backward optimization, material edits
propagate to the resulting global illumination.

\textit{Relighting Results on Synthetic and Real Datasets.}
We first compare relighting on synthetic benchmarks
(Fig.~\ref{fig:object_relight}). For real objects without ground-truth
relighting, we show single-object results on Stanford-ORB and DTU, and
scene-level results on Mip-NeRF 360 scenes (Fig.~\ref{fig:real_relight}) under
environment maps and near-field lights.
Our path-traced framework naturally accounts for visibility under the target
illumination. Relighting is evaluated through Monte Carlo path tracing without
relying on screen-space visibility or cached transport. This enables realistic
soft shadows and illumination changes, especially under near-field lights,
supporting physically consistent relighting and appearance editing.

\subsection{Ablation Study}
We conduct ablation studies on the TensoIR dataset without prior regularization to validate the effectiveness of key components (Tab.~\ref{tab:ablation_studies}) and analyze the effects of different implementation settings (Tab.~\ref{tab:ablation_spp}).

\begin{figure}[!htbp]
    \centering
    \includegraphics[
        width=\linewidth,
        trim=0 0in 0 0in,
        clip
    ]{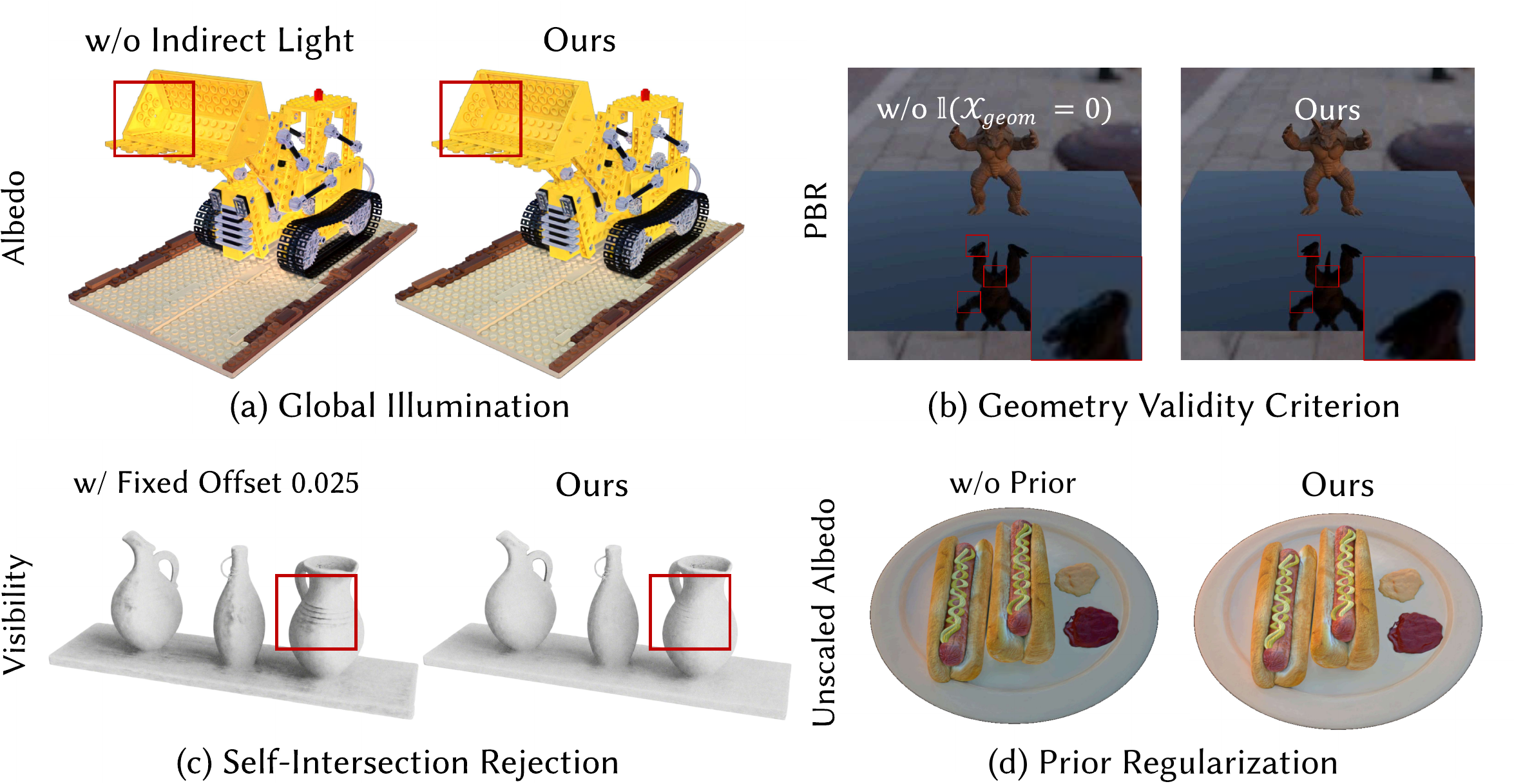}
    \caption{
    Ablation studies on various components of our framework.
    }
    \label{fig:ablation_studies}
\end{figure}

\textit{Global Illumination.}
Modeling global illumination with the full rendering equation via multi-bounce transport improves material recovery and relighting quality over direct illumination alone, highlighting the importance of indirect light transport for inverse rendering
(Fig.~\ref{fig:ablation_studies}(a) and the 1-bounce setting in Tab.~\ref{tab:ablation_spp}).

\begin{table}[!htbp]
  \centering
  \caption{Ablation study of key components on TensoIR dataset (w/o Prior), demonstrating the effectiveness of each component.}
  \label{tab:ablation_studies}
  \setlength{\tabcolsep}{4pt}
  \renewcommand{\arraystretch}{1.2}
  \resizebox{1.0\columnwidth}{!}{
  \begin{tabular}{lccc}
    \toprule
    Method
    & Albedo PSNR$\uparrow$
    & NVS PSNR$\uparrow$
    & Relight PSNR$\uparrow$\\
    \midrule

    w/o Geom. Criterion
      & 30.96
      & 36.19
      & 31.13 \\

    w/o Backface-aware Offset, $\epsilon_o=0.025$
      & 30.89
      & 36.16
      & 31.13 \\

    w/o Backface-aware Offset, $\epsilon_o=0.05$
      & 30.94
      & 36.38
      & 31.24 \\

    Discrete Interaction
      & 30.71
      & 36.15
      & 30.65 \\

    \midrule
    Full
      & \textbf{31.04}
      & \textbf{36.50}
      & \textbf{31.31} \\
    \bottomrule
  \end{tabular}
  }
\end{table}

\textit{Geometric Validity Criterion.}
The geometric-validity criterion for ray continuation in Eq.~\eqref{eq:geom_gate} introduces a
stricter condition for secondary interactions during multi-bounce transport,
improving their reliability and avoiding light-leaking artifacts
(Fig.~\ref{fig:ablation_studies}(b)).

\textit{Self-Intersection Rejection.}
We evaluate the backface-aware origin-offset rule in
Eq.~\eqref{eq:self_occ}. It suppresses spurious self-occlusion by rejecting
local back-facing responses from overlapping Gaussians that represent the same
aggregated surface. Compared with a fixed ray-origin offset, our rule reduces
self-occlusion artifacts without relying on a manually selected offset magnitude
(Fig.~\ref{fig:ablation_studies}(c)).

\textit{Prior Regularization.}
The diffusion prior provides a coarse perceptual anchor during the early stage
of path-traced inverse rendering, guiding the material parameters toward a
stable and spatially consistent solution. In particular, it yields more
plausible albedo intensity before scale alignment and reduces the sensitivity
of material optimization to scene-specific hyperparameter tuning
(Figs.~\ref{fig:ablation_studies}(d) and~\ref{fig:roughness}).

\begin{figure}[!htbp]
    \centering
    \includegraphics[
        width=\linewidth,
        trim=1in 0 0 0,
        clip
    ]{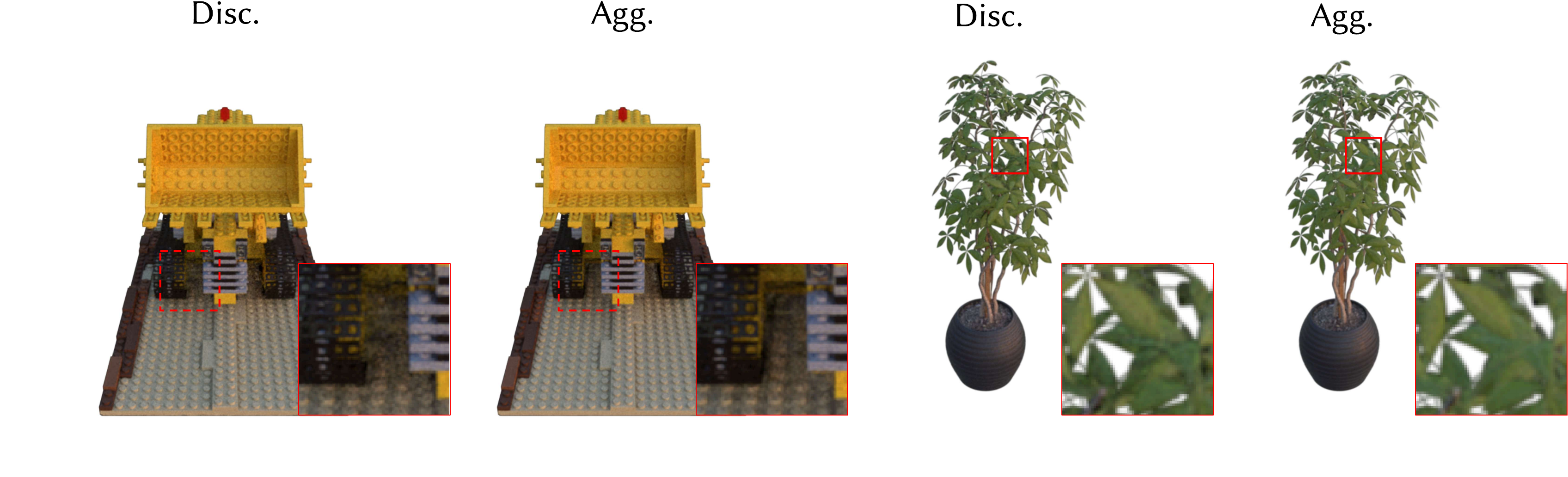}
    \caption{
    Ablation study of ray--Gaussian interaction models during material optimization with 32 SPP. Aggregate interaction modeling typically yields more reliable results in our implementation.
    }
    \label{fig:modeling_results}
\end{figure}

\textit{Discrete Interaction Modeling.}
Discrete modeling can also be optimized through differentiable path tracing, but in our implementation, it exhibits greater sensitivity to geometric consistency, as each interaction depends on a single sampled Gaussian rather than an aggregated surface state (Fig.~\ref{fig:modeling_results}). A possible explanation is the mismatch between the aggregate interaction model used for geometry reconstruction and the discrete interaction model adopted during inverse rendering. \citet{xu2026Stoch3DGS} introduce a sorting-free stochastic reconstruction estimator that may produce geometry better aligned with the discrete interaction model used during inversion.

\textit{SPP and Bounces.}
We ablate the number of samples per pixel (SPP) and the maximum number of bounces during inversion in Tab.~\ref{tab:ablation_spp}. Increasing either setting generally improves inversion quality while increasing optimization time, with higher SPP additionally raising peak GPU memory usage. The slightly higher Albedo PSNR with the 1-bounce setting is mainly due to the Armadillo scene (+0.67 dB), where the missing indirect illumination can be partially absorbed into the recovered albedo scale and subsequently favored by scale alignment during evaluation. The maximum number of bounces has little effect on peak GPU memory because path traversal is performed iteratively, updating a fixed-size per-path payload in place rather than allocating depth-indexed buffers. 
For comparison, the average peak GPU memory usage is 12.81 GB for IRGS~\cite{IRGS} and 22.65 GB for SVG-IR~\cite{SVGIR2025} under their default settings.
Our default implementation settings (32 SPP and $2$ Bounces) provide a favorable quality-time trade-off.

\begin{table}[h]
  \centering
  \caption{Ablation of SPP and bounce count on the TensoIR dataset (w/o Prior). Each factor is varied independently from the default optimization setting (32 SPP and a maximum of 2 bounces), with the other fixed. The default setting provides a favorable quality--time trade-off.}
  \label{tab:ablation_spp}
  \setlength{\tabcolsep}{4pt}
  \renewcommand{\arraystretch}{1.2}
  \resizebox{1.0\columnwidth}{!}{
  \begin{tabular}{lccccc}
    \toprule
    \multirow{2}{*}{Setting}
    & Albedo
    & NVS
    & Relight
    & Avg. Peak GPU mem. 
    & Training \\

    &PSNR$\uparrow$
    &PSNR$\uparrow$
    &PSNR$\uparrow$
    &GB$\downarrow$
    &Hours$\downarrow$\\
    
    \midrule
    1 SPP
    & 28.12
    & 34.89
    & 30.00
    & \textbf{1.51}
    & \textbf{0.04}\\

    16 SPP
    & 30.53
    & 36.45
    & 31.30
    & 3.91
    & 0.21\\
    
    64 SPP
    & \textbf{31.45}
    & \textbf{36.63}
    & \textbf{31.75}
    & 8.83
    & 0.77\\
    
    \midrule

    1 Bounce
    & \textbf{31.18}
    & 34.91
    & 31.17
    & \textbf{5.67}
    & \textbf{0.32}\\

    3 Bounces
    & 31.06
    & 36.52
    & 31.32
    & 5.70
    & 0.42    \\

    4 Bounces
    & 31.06
    & \textbf{36.53}
    & 31.36
    & 5.71
    & 0.44\\
    
    9 Bounces
    & 31.07
    & \textbf{36.53}
    & \textbf{31.44}
    & 5.71
    & 0.47\\

    \midrule
    Default
      & 31.04
      & 36.50
      & 31.31
      & 5.70
      & 0.39\\
    \bottomrule
  \end{tabular}
  }
\end{table}
\section{DISCUSSION AND LIMITATIONS}

\textit{Computational Overhead.}
Our goal is not real-time novel-view synthesis, but physically consistent
inverse rendering of Gaussian materials for path-traced rendering, relighting,
and appearance editing. This requires optimizing materials under the same
ray-traced transport used for final rendering. Path replay backpropagation
avoids storing intermediate path states, but adds a replay pass that repeats
Gaussian--ray intersections, shadow tests, and MIS evaluations. With $32$ SPP
for both primal and replay passes, the first-interaction cost is about $32\times$
higher than 3DGRT~\cite{3dgrt2024}. Runtime, training time, and peak GPU memory
are reported in Tab.~\ref{tab:runtime}. For comparison, under a 256-SPP setting on the TensoIR dataset, our method takes 3.85 s per frame on average, compared with 1.24 s for IRGS~\cite{IRGS}, requiring approximately $3.1\times$ the rendering time. Reducing this cost remains an important
future direction; adaptive sampling, transport caching, and neural
radiosity-style approximations may help amortize multi-bounce transport.

\begin{table}[!htbp]
  \centering
  \caption{Runtime and memory statistics across different Gaussian-count scales under $32$ SPP and a maximum of $2$ bounces. Training time is estimated for 800 optimization iterations, where the $32$ samples per pixel are packed and processed as one ray chunk.}
  \label{tab:runtime}
  \setlength{\tabcolsep}{5pt}
  \renewcommand{\arraystretch}{1.15}
  \resizebox{\linewidth}{!}{
  \begin{tabular}{lccccc}
    \toprule
    Gaussian-count scale
    & Res.
    & Inversion
    & Peak GPU mem.
    & Train time
    & Post-opt. render \\
    \midrule
    158K (TensoIR)
      & 800
      & 1.66 s/iter
      & 5.70 GB
      & 0.39 h
      & 0.59 s/frame \\
    179K (Synthetic4Relight)
      & 800
      & 2.63 s/iter
      & 5.66 GB
      & 0.60 h
      & 0.95 s/frame \\
    106K (Stanford-ORB)
      & 1200
      & 8.64 s/iter
      & 9.29 GB
      & 2.01 h
      & 3.07 s/frame \\
    2.5M (Mip-NeRF 360)
      & 1237
      & 24.92 s/iter
      & 13.68 GB
      & 5.65 h
      & 9.04 s/frame \\
    \bottomrule
  \end{tabular}
  }
\end{table}

\textit{Opaque Surface Scenes.}
Our formulation targets opaque, surface-dominated scenes. Instead of treating each Gaussian as an independent volumetric medium, we aggregate overlapping Gaussian responses into an equivalent surface interaction and perform PBR-based recursive multi-bounce light transport over this representation. Accordingly, our light-transport formulation and self-intersection rejection mechanism assume surface-like interactions and are not intended for genuinely volumetric or transmissive phenomena, such as smoke, hair, fur, or refractive materials.

\textit{Discontinuities and Non-differentiability Considerations.}
Our path-traced inversion involves discrete decisions in geometry-validity
evaluation and shadow-ray visibility testing. The hard geometry-validity
threshold is applied after the Gaussian geometry and opacity have been
reconstructed, and both remain fixed throughout the subsequent inversion.
Following prior Gaussian inverse-rendering methods~\cite{IRGS, SVGIR2025}, the reconstruction stage
employs a mask entropy loss that encourages foreground opacity toward one,
thereby reducing ambiguous responses near the threshold. Since the Gaussian geometry remains fixed during inversion,
the validity decision for a given ray is independent of the optimized
material and illumination parameters. We observed no noticeable threshold-induced artifacts
in our experiments, although regions with uncertain geometry may be more
sensitive to this transition. Shadow-ray visibility decisions likewise depend
only on the fixed scene geometry. Path replay reproduces the forward visibility
decisions and propagates gradients through the subsequent continuous shading
evaluation. Accordingly, these discrete decisions do not obstruct gradient
propagation with respect to the material and illumination parameters optimized
in our setting.

\textit{Alternative Interaction Models.}
Our framework is compatible with different formulations of ray--Gaussian interactions. As described in Sec.~\ref{sec:forward_light_transport}, the aggregate model constructs an equivalent local shading state, whereas the discrete model replaces this aggregation step with Monte Carlo sampling of a single Gaussian primitive at each interaction. The two formulations share the same path-tracing and path-replay pipeline. During backward optimization, path replay reproduces the stochastic selection made in the forward pass and propagates gradients through the corresponding interaction. Therefore, the sampling decision need not be differentiated through and does not prevent gradient propagation through the subsequent shading evaluation.
Due to nonlinear shading evaluation during path tracing, using a consistent interaction formulation for optimization and rendering is more appropriate.
This modular formulation can also accommodate alternative Gaussian representations such as 2DGS~\citep{Huang2DGS2024}. Such representations can be incorporated by replacing the current ray--Gaussian evaluation and local interaction construction with their 2D counterparts while retaining the subsequent light-transport and optimization pipeline. Surface-aligned primitives may improve interaction consistency around thin structures and geometric boundaries, although gaps, overlaps, and grazing-angle intersections between planar primitives require careful handling during recursive light transport.

\textit{Non-Smooth Gradients.}
In our implementation, ray-traced Gaussian optimization may yield less smooth
material gradients than splatting-based optimization, as neighboring rays can
interact with different primitives and follow different transport paths under
limited sampling. Splatting-based methods, by contrast, introduce implicit
spatial smoothing through screen-space accumulation. This effect can be more
pronounced in the absence of explicit smoothing regularization. Structure- or
cluster-aware regularization over spatial or semantic neighborhoods could further
mitigate this artifact.

\begin{figure}[!htbp]
    \centering
    \includegraphics[
        width=0.9\linewidth,
        trim=0 2in 0 1.0in,
        clip
    ]{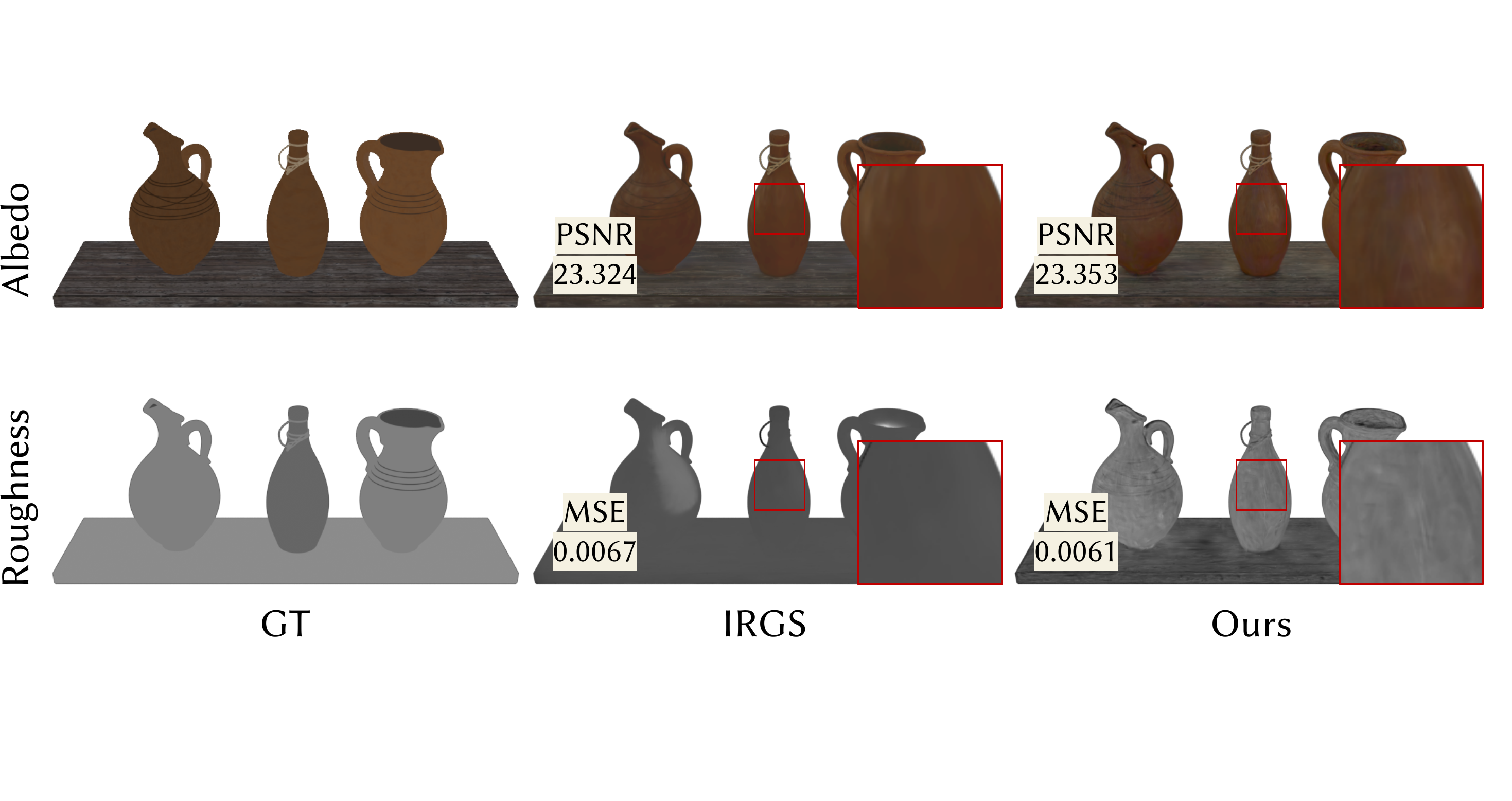}
    \caption{Non-smooth material artifact. Although the quantitative metrics are
    comparable, ray-traced optimization may yield less smooth material maps than
    splatting-based methods under limited sampling.}
    \label{fig:failure}
\end{figure}

\section{CONCLUSION}
We have explored inverse rendering in ray-traced 3D Gaussian scenes and
introduced a splatting-free framework that defines forward rendering and
backward optimization within the same ray-based light transport pipeline. This
consistency allows material estimation and illumination optimization to be performed under a unified physical process, rather than relying
on rasterization-derived buffers or screen-space approximations. Our results
demonstrate the feasibility of using 3D Gaussian fields as optimizable scene
representations for ray-traced inverse rendering, showing that Gaussian
primitives can be integrated into physically based light transport beyond
view-synthesis-oriented splatting.

By accounting for multi-bounce transport, our framework optimizes material and
illumination parameters with the same path-traced process used for relighting,
appearance editing, and physically based evaluation. This enables 3D Gaussian
assets to be used in ray-traced pipelines, where visibility, shadows,
reflections, and indirect illumination are evaluated within a unified path-space
transport process. It supports scenes where Gaussians coexist with meshes,
environment illumination, and analytic light sources. We hope this work encourages further exploration of efficient and robust ray-traced Gaussian representations, including adaptive sampling, transport caching, neural approximations, structure- or cluster-aware regularization to encourage coherent optimization within Gaussian groups, and extensions to a broader range of Gaussian representations.

\begin{acks}
We thank the anonymous reviewers for valuable comments. We also thank NVIDIA for their contributions to ray tracing and EPFL for their work on path replay backpropagation. This work was supported in part by the National Key Research and Development Program of China (No. 2023YFF0905103).

\end{acks}

\putbib[chapters/reference]

\end{bibunit}



\begin{thebibliography}{45}


\ifx \showCODEN    \undefined \def \showCODEN     #1{\unskip}     \fi
\ifx \showISBNx    \undefined \def \showISBNx     #1{\unskip}     \fi
\ifx \showISBNxiii \undefined \def \showISBNxiii  #1{\unskip}     \fi
\ifx \showISSN     \undefined \def \showISSN      #1{\unskip}     \fi
\ifx \showLCCN     \undefined \def \showLCCN      #1{\unskip}     \fi
\ifx \shownote     \undefined \def \shownote      #1{#1}          \fi
\ifx \showarticletitle \undefined \def \showarticletitle #1{#1}   \fi
\ifx \showURL      \undefined \def \showURL       {\relax}        \fi
\providecommand\bibfield[2]{#2}
\providecommand\bibinfo[2]{#2}
\providecommand\natexlab[1]{#1}
\providecommand\showeprint[2][]{arXiv:#2}

\bibitem[Aan{\ae}s et~al\mbox{.}(2016)]%
        {dtu}
\bibfield{author}{\bibinfo{person}{Henrik Aan{\ae}s}, \bibinfo{person}{Rasmus~Ramsb{\o}l Jensen}, \bibinfo{person}{George Vogiatzis}, \bibinfo{person}{Engin Tola}, {and} \bibinfo{person}{Anders~Bjorholm Dahl}.} \bibinfo{year}{2016}\natexlab{}.
\newblock \showarticletitle{Large-Scale Data for Multiple-View Stereopsis}.
\newblock \bibinfo{journal}{\emph{International Journal of Computer Vision}} (\bibinfo{year}{2016}), \bibinfo{pages}{153–168}.
\newblock


\bibitem[Barron et~al\mbox{.}(2022)]%
        {barron2022mipnerf360}
\bibfield{author}{\bibinfo{person}{Jonathan~T. Barron}, \bibinfo{person}{Ben Mildenhall}, \bibinfo{person}{Dor Verbin}, \bibinfo{person}{Pratul~P. Srinivasan}, {and} \bibinfo{person}{Peter Hedman}.} \bibinfo{year}{2022}\natexlab{}.
\newblock \showarticletitle{Mip-NeRF 360: Unbounded Anti-Aliased Neural Radiance Fields}.
\newblock \bibinfo{journal}{\emph{CVPR}} (\bibinfo{year}{2022}).
\newblock


\bibitem[Bi et~al\mbox{.}(2024)]%
        {bi2024rgs}
\bibfield{author}{\bibinfo{person}{Zoubin Bi}, \bibinfo{person}{Yixin Zeng}, \bibinfo{person}{Chong Zeng}, \bibinfo{person}{Fan Pei}, \bibinfo{person}{Xiang Feng}, \bibinfo{person}{Kun Zhou}, {and} \bibinfo{person}{Hongzhi Wu}.} \bibinfo{year}{2024}\natexlab{}.
\newblock \showarticletitle{GS\textsuperscript{3}: Efficient Relighting with Triple Gaussian Splatting}. In \bibinfo{booktitle}{\emph{SIGGRAPH Asia 2024 Conference Papers}}.
\newblock


\bibitem[Burley(2012)]%
        {burley12disneybrdf}
\bibfield{author}{\bibinfo{person}{Brent Burley}.} \bibinfo{year}{2012}\natexlab{}.
\newblock \showarticletitle{Physically-Based Shading at {Disney}}. In \bibinfo{booktitle}{\emph{{ACM SIGGRAPH} Course Notes}}. \bibinfo{pages}{1--7}.
\newblock


\bibitem[Byrski et~al\mbox{.}(2025)]%
        {byrski2025raysplatsraytracingbased}
\bibfield{author}{\bibinfo{person}{Krzysztof Byrski}, \bibinfo{person}{Marcin Mazur}, \bibinfo{person}{Jacek Tabor}, \bibinfo{person}{Tadeusz Dziarmaga}, \bibinfo{person}{Marcin Kądziołka}, \bibinfo{person}{Dawid Baran}, {and} \bibinfo{person}{Przemysław Spurek}.} \bibinfo{year}{2025}\natexlab{}.
\newblock \bibinfo{title}{RaySplats: Ray Tracing based Gaussian Splatting}.
\newblock
\showeprint[arxiv]{2501.19196}~[cs.CV]
\urldef\tempurl%
\url{https://arxiv.org/abs/2501.19196}
\showURL{%
\tempurl}


\bibitem[Condor et~al\mbox{.}(2025)]%
        {DontSplatYourGaussians}
\bibfield{author}{\bibinfo{person}{Jorge Condor}, \bibinfo{person}{Sebastien Speierer}, \bibinfo{person}{Lukas Bode}, \bibinfo{person}{Aljaz Bozic}, \bibinfo{person}{Simon Green}, \bibinfo{person}{Piotr Didyk}, {and} \bibinfo{person}{Adrian Jarabo}.} \bibinfo{year}{2025}\natexlab{}.
\newblock \showarticletitle{{Don't Splat your Gaussians: Volumetric Ray-Traced Primitives for Modeling and Rendering Scattering and Emissive Media}}.
\newblock \bibinfo{journal}{\emph{ACM Trans. Graph.}} (\bibinfo{date}{Jan.} \bibinfo{year}{2025}).
\newblock
\showISSN{0730-0301}
\href{https://doi.org/10.1145/3711853}{doi:\nolinkurl{10.1145/3711853}}


\bibitem[Debevec(1998)]%
        {paul98ibl}
\bibfield{author}{\bibinfo{person}{Paul Debevec}.} \bibinfo{year}{1998}\natexlab{}.
\newblock \showarticletitle{Rendering synthetic objects into real scenes: bridging traditional and image-based graphics with global illumination and high dynamic range photography}. In \bibinfo{booktitle}{\emph{Proceedings of the 25th Annual Conference on Computer Graphics and Interactive Techniques}} \emph{(\bibinfo{series}{SIGGRAPH '98})}. \bibinfo{publisher}{Association for Computing Machinery}, \bibinfo{address}{New York, NY, USA}, \bibinfo{pages}{189–198}.
\newblock
\showISBNx{0897919998}
\href{https://doi.org/10.1145/280814.280864}{doi:\nolinkurl{10.1145/280814.280864}}


\bibitem[Du et~al\mbox{.}(2024)]%
        {GS-ID}
\bibfield{author}{\bibinfo{person}{Kang Du}, \bibinfo{person}{Zhihao Liang}, {and} \bibinfo{person}{Zeyu Wang}.} \bibinfo{year}{2024}\natexlab{}.
\newblock \bibinfo{title}{GS-ID: Illumination Decomposition on Gaussian Splatting via Diffusion Prior and Parametric Light Source Optimization}.
\newblock
\showeprint[arxiv]{2408.08524}~[cs.CV]
\urldef\tempurl%
\url{https://arxiv.org/abs/2408.08524}
\showURL{%
\tempurl}


\bibitem[Fan et~al\mbox{.}(2025)]%
        {fan25rng}
\bibfield{author}{\bibinfo{person}{Jiahui Fan}, \bibinfo{person}{Fujun Luan}, \bibinfo{person}{Jian Yang}, \bibinfo{person}{Milos Hasan}, {and} \bibinfo{person}{Beibei Wang}.} \bibinfo{year}{2025}\natexlab{}.
\newblock \showarticletitle{RNG: Relightable Neural Gaussians}.
\newblock \bibinfo{journal}{\emph{Proceedings of CVPR 2025}} (\bibinfo{year}{2025}).
\newblock


\bibitem[Gao et~al\mbox{.}(2024)]%
        {R3DG}
\bibfield{author}{\bibinfo{person}{Jian Gao}, \bibinfo{person}{Chun Gu}, \bibinfo{person}{Youtian Lin}, \bibinfo{person}{Zhihao Li}, \bibinfo{person}{Hao Zhu}, \bibinfo{person}{Xun Cao}, \bibinfo{person}{Li Zhang}, {and} \bibinfo{person}{Yao Yao}.} \bibinfo{year}{2024}\natexlab{}.
\newblock \showarticletitle{Relightable 3D Gaussians: Realistic Point Cloud Relighting with BRDF Decomposition and Ray Tracing}. In \bibinfo{booktitle}{\emph{Computer Vision – ECCV 2024: 18th European Conference, Milan, Italy, September 29–October 4, 2024, Proceedings, Part XLV}} (Milan, Italy). \bibinfo{publisher}{Springer-Verlag}, \bibinfo{address}{Berlin, Heidelberg}, \bibinfo{pages}{73–89}.
\newblock
\showISBNx{978-3-031-72994-2}
\href{https://doi.org/10.1007/978-3-031-72995-9_5}{doi:\nolinkurl{10.1007/978-3-031-72995-9_5}}


\bibitem[González(2009)]%
        {gonz09fib}
\bibfield{author}{\bibinfo{person}{Álvaro González}.} \bibinfo{year}{2009}\natexlab{}.
\newblock \showarticletitle{Measurement of Areas on a Sphere Using Fibonacci and Latitude–Longitude Lattices}.
\newblock \bibinfo{journal}{\emph{Mathematical Geosciences}} \bibinfo{volume}{42}, \bibinfo{number}{1} (\bibinfo{date}{Nov.} \bibinfo{year}{2009}), \bibinfo{pages}{49–64}.
\newblock
\showISSN{1874-8953}
\href{https://doi.org/10.1007/s11004-009-9257-x}{doi:\nolinkurl{10.1007/s11004-009-9257-x}}


\bibitem[Gu et~al\mbox{.}(2025)]%
        {IRGS}
\bibfield{author}{\bibinfo{person}{Chun Gu}, \bibinfo{person}{Xiaofei Wei}, \bibinfo{person}{Zixuan Zeng}, \bibinfo{person}{Yuxuan Yao}, {and} \bibinfo{person}{Li Zhang}.} \bibinfo{year}{2025}\natexlab{}.
\newblock \showarticletitle{IRGS: Inter-Reflective Gaussian Splatting with 2D Gaussian Ray Tracing}. In \bibinfo{booktitle}{\emph{CVPR}}.
\newblock


\bibitem[Guo et~al\mbox{.}(2024)]%
        {guo2024prtgs}
\bibfield{author}{\bibinfo{person}{Yijia Guo}, \bibinfo{person}{Yuanxi Bai}, \bibinfo{person}{Liwen Hu}, \bibinfo{person}{Ziyi Guo}, \bibinfo{person}{Mianzhi Liu}, \bibinfo{person}{Yu Cai}, \bibinfo{person}{Tiejun Huang}, {and} \bibinfo{person}{Lei Ma}.} \bibinfo{year}{2024}\natexlab{}.
\newblock \showarticletitle{PRTGS: Precomputed Radiance Transfer of Gaussian Splats for Real-Time High-Quality Relighting} \emph{(\bibinfo{series}{MM '24})}. \bibinfo{publisher}{Association for Computing Machinery}, \bibinfo{address}{New York, NY, USA}, \bibinfo{pages}{5112–5120}.
\newblock
\showISBNx{9798400706868}
\href{https://doi.org/10.1145/3664647.3680893}{doi:\nolinkurl{10.1145/3664647.3680893}}


\bibitem[Hahlbohm et~al\mbox{.}(2025)]%
        {hahlbohm2025htgs}
\bibfield{author}{\bibinfo{person}{Florian Hahlbohm}, \bibinfo{person}{Fabian Friederichs}, \bibinfo{person}{Tim Weyrich}, \bibinfo{person}{Linus Franke}, \bibinfo{person}{Moritz Kappel}, \bibinfo{person}{Susana Castillo}, \bibinfo{person}{Marc Stamminger}, \bibinfo{person}{Martin Eisemann}, {and} \bibinfo{person}{Marcus Magnor}.} \bibinfo{year}{2025}\natexlab{}.
\newblock \showarticletitle{Efficient Perspective-Correct 3D Gaussian Splatting Using Hybrid Transparency}.
\newblock \bibinfo{journal}{\emph{Computer Graphics Forum}} \bibinfo{volume}{44}, \bibinfo{number}{2} (\bibinfo{year}{2025}).
\newblock
\href{https://doi.org/10.1111/cgf.70014}{doi:\nolinkurl{10.1111/cgf.70014}}


\bibitem[Hu et~al\mbox{.}(2026)]%
        {hu2025realtimeglobalilluminationdynamic}
\bibfield{author}{\bibinfo{person}{Chenxiao Hu}, \bibinfo{person}{Meng Gai}, \bibinfo{person}{Guoping Wang}, {and} \bibinfo{person}{Sheng Li}.} \bibinfo{year}{2026}\natexlab{}.
\newblock \showarticletitle{Dynamic Global Illumination for Interactive Gaussian Splatting Scenes in Real Time}.
\newblock \bibinfo{journal}{\emph{IEEE Transactions on Visualization and Computer Graphics}} \bibinfo{volume}{32}, \bibinfo{number}{7} (\bibinfo{year}{2026}), \bibinfo{pages}{6919--6931}.
\newblock
\href{https://doi.org/10.1109/TVCG.2026.3689199}{doi:\nolinkurl{10.1109/TVCG.2026.3689199}}


\bibitem[Huang et~al\mbox{.}(2024)]%
        {Huang2DGS2024}
\bibfield{author}{\bibinfo{person}{Binbin Huang}, \bibinfo{person}{Zehao Yu}, \bibinfo{person}{Anpei Chen}, \bibinfo{person}{Andreas Geiger}, {and} \bibinfo{person}{Shenghua Gao}.} \bibinfo{year}{2024}\natexlab{}.
\newblock \showarticletitle{2D Gaussian Splatting for Geometrically Accurate Radiance Fields}. In \bibinfo{booktitle}{\emph{SIGGRAPH 2024 Conference Papers}}. \bibinfo{publisher}{Association for Computing Machinery}.
\newblock
\href{https://doi.org/10.1145/3641519.3657428}{doi:\nolinkurl{10.1145/3641519.3657428}}


\bibitem[Jiang et~al\mbox{.}(2025)]%
        {jiang2025radiositygs}
\bibfield{author}{\bibinfo{person}{Kaiwen Jiang}, \bibinfo{person}{Jia-Mu Sun}, \bibinfo{person}{Zilu Li}, \bibinfo{person}{Dan Wang}, \bibinfo{person}{Tzu-Mao Li}, {and} \bibinfo{person}{Ravi Ramamoorthi}.} \bibinfo{year}{2025}\natexlab{}.
\newblock \showarticletitle{Differentiable Light Transport with Gaussian Surfels via Adapted Radiosity for Efficient Relighting and Geometry Reconstruction}.
\newblock \bibinfo{journal}{\emph{ACM Transactions on Graphics (TOG)}} \bibinfo{volume}{44}, \bibinfo{number}{6} (\bibinfo{date}{December} \bibinfo{year}{2025}).
\newblock


\bibitem[Jiang et~al\mbox{.}(2024)]%
        {jiang2024gaussianshader}
\bibfield{author}{\bibinfo{person}{Yingwenqi Jiang}, \bibinfo{person}{Jiadong Tu}, \bibinfo{person}{Yuan Liu}, \bibinfo{person}{Xifeng Gao}, \bibinfo{person}{Xiaoxiao Long}, \bibinfo{person}{Wenping Wang}, {and} \bibinfo{person}{Yuexin Ma}.} \bibinfo{year}{2024}\natexlab{}.
\newblock \showarticletitle{{GaussianShader}: {3D} gaussian splatting with shading functions for reflective surfaces}. In \bibinfo{booktitle}{\emph{Proceedings of the IEEE/CVF Conference on Computer Vision and Pattern Recognition}}. \bibinfo{pages}{5322--5332}.
\newblock


\bibitem[Jin et~al\mbox{.}(2023)]%
        {Jin2023TensoIR}
\bibfield{author}{\bibinfo{person}{Haian Jin}, \bibinfo{person}{Isabella Liu}, \bibinfo{person}{Peijia Xu}, \bibinfo{person}{Xiaoshuai Zhang}, \bibinfo{person}{Songfang Han}, \bibinfo{person}{Sai Bi}, \bibinfo{person}{Xiaowei Zhou}, \bibinfo{person}{Zexiang Xu}, {and} \bibinfo{person}{Hao Su}.} \bibinfo{year}{2023}\natexlab{}.
\newblock \showarticletitle{TensoIR: Tensorial Inverse Rendering}. In \bibinfo{booktitle}{\emph{Proceedings of the IEEE/CVF Conference on Computer Vision and Pattern Recognition (CVPR)}}.
\newblock


\bibitem[Kajiya(1986)]%
        {kajiya1986rendering}
\bibfield{author}{\bibinfo{person}{James~T. Kajiya}.} \bibinfo{year}{1986}\natexlab{}.
\newblock \showarticletitle{The rendering equation}.
\newblock \bibinfo{journal}{\emph{SIGGRAPH Comput. Graph.}} \bibinfo{volume}{20}, \bibinfo{number}{4} (\bibinfo{date}{Aug.} \bibinfo{year}{1986}), \bibinfo{pages}{143–150}.
\newblock
\showISSN{0097-8930}
\href{https://doi.org/10.1145/15886.15902}{doi:\nolinkurl{10.1145/15886.15902}}


\bibitem[Kerbl et~al\mbox{.}(2023)]%
        {kerbl3dgaussians}
\bibfield{author}{\bibinfo{person}{Bernhard Kerbl}, \bibinfo{person}{Georgios Kopanas}, \bibinfo{person}{Thomas Leimk{\"u}hler}, {and} \bibinfo{person}{George Drettakis}.} \bibinfo{year}{2023}\natexlab{}.
\newblock \showarticletitle{3D Gaussian Splatting for Real-Time Radiance Field Rendering}.
\newblock \bibinfo{journal}{\emph{ACM Transactions on Graphics}} \bibinfo{volume}{42}, \bibinfo{number}{4} (\bibinfo{date}{July} \bibinfo{year}{2023}).
\newblock


\bibitem[Kuang et~al\mbox{.}(2023)]%
        {kuang2023stanfordorb}
\bibfield{author}{\bibinfo{person}{Zhengfei Kuang}, \bibinfo{person}{Yunzhi Zhang}, \bibinfo{person}{Hong-Xing Yu}, \bibinfo{person}{Samir Agarwala}, \bibinfo{person}{Shangzhe Wu}, \bibinfo{person}{Jiajun Wu}, {et~al\mbox{.}}} \bibinfo{year}{2023}\natexlab{}.
\newblock \showarticletitle{Stanford-ORB: a real-world 3D object inverse rendering benchmark}.
\newblock \bibinfo{journal}{\emph{Advances in Neural Information Processing Systems Datasets and Benchmarks Track}}.
\newblock


\bibitem[Liang et~al\mbox{.}(2024)]%
        {GS-IR}
\bibfield{author}{\bibinfo{person}{Zhihao Liang}, \bibinfo{person}{Qi Zhang}, \bibinfo{person}{Ying Feng}, \bibinfo{person}{Ying Shan}, {and} \bibinfo{person}{Kui Jia}.} \bibinfo{year}{2024}\natexlab{}.
\newblock \showarticletitle{GS-IR: 3D Gaussian Splatting for Inverse Rendering}. In \bibinfo{booktitle}{\emph{Proceedings of the IEEE/CVF Conference on Computer Vision and Pattern Recognition (CVPR)}}. \bibinfo{pages}{21644--21653}.
\newblock


\bibitem[Ling et~al\mbox{.}(2024)]%
        {ling24nerfemitter}
\bibfield{author}{\bibinfo{person}{Jingwang Ling}, \bibinfo{person}{Ruihan Yu}, \bibinfo{person}{Feng Xu}, \bibinfo{person}{Chun Du}, {and} \bibinfo{person}{Shuang Zhao}.} \bibinfo{year}{2024}\natexlab{}.
\newblock \showarticletitle{NeRF as a Non-Distant Environment Emitter in Physics-based Inverse Rendering}. In \bibinfo{booktitle}{\emph{ACM SIGGRAPH 2024 Conference Papers}} (Denver, CO, USA) \emph{(\bibinfo{series}{SIGGRAPH '24})}. \bibinfo{publisher}{Association for Computing Machinery}, \bibinfo{address}{New York, NY, USA}, Article \bibinfo{articleno}{39}, \bibinfo{numpages}{12}~pages.
\newblock
\showISBNx{9798400705250}
\href{https://doi.org/10.1145/3641519.3657404}{doi:\nolinkurl{10.1145/3641519.3657404}}


\bibitem[Moenne-Loccoz et~al\mbox{.}(2024)]%
        {3dgrt2024}
\bibfield{author}{\bibinfo{person}{Nicolas Moenne-Loccoz}, \bibinfo{person}{Ashkan Mirzaei}, \bibinfo{person}{Or Perel}, \bibinfo{person}{Riccardo de Lutio}, \bibinfo{person}{Janick Martinez~Esturo}, \bibinfo{person}{Gavriel State}, \bibinfo{person}{Sanja Fidler}, \bibinfo{person}{Nicholas Sharp}, {and} \bibinfo{person}{Zan Gojcic}.} \bibinfo{year}{2024}\natexlab{}.
\newblock \showarticletitle{{3D} Gaussian Ray Tracing: Fast Tracing of Particle Scenes}.
\newblock \bibinfo{journal}{\emph{ACM Transactions on Graphics}} \bibinfo{volume}{43}, \bibinfo{number}{6}, Article \bibinfo{articleno}{232} (\bibinfo{date}{Nov.} \bibinfo{year}{2024}), \bibinfo{numpages}{19}~pages.
\newblock
\href{https://doi.org/10.1145/3687934}{doi:\nolinkurl{10.1145/3687934}}


\bibitem[Ng et~al\mbox{.}(2003)]%
        {ren03haar}
\bibfield{author}{\bibinfo{person}{Ren Ng}, \bibinfo{person}{Ravi Ramamoorthi}, {and} \bibinfo{person}{Pat Hanrahan}.} \bibinfo{year}{2003}\natexlab{}.
\newblock \showarticletitle{All-Frequency Shadows Using Non-Linear Wavelet Lighting Approximation}.
\newblock \bibinfo{journal}{\emph{ACM Transactions on Graphics}} \bibinfo{volume}{22}, \bibinfo{number}{3} (\bibinfo{date}{July} \bibinfo{year}{2003}), \bibinfo{pages}{376--381}.
\newblock
\href{https://doi.org/10.1145/882262.882280}{doi:\nolinkurl{10.1145/882262.882280}}


\bibitem[Parker et~al\mbox{.}(2010)]%
        {optix}
\bibfield{author}{\bibinfo{person}{Steven~G. Parker}, \bibinfo{person}{James Bigler}, \bibinfo{person}{Andreas Dietrich}, \bibinfo{person}{Heiko Friedrich}, \bibinfo{person}{Jared Hoberock}, \bibinfo{person}{David Luebke}, \bibinfo{person}{David McAllister}, \bibinfo{person}{Morgan McGuire}, \bibinfo{person}{Keith Morley}, \bibinfo{person}{Austin Robison}, {and} \bibinfo{person}{Martin Stich}.} \bibinfo{year}{2010}\natexlab{}.
\newblock \showarticletitle{OptiX: a general purpose ray tracing engine}.
\newblock \bibinfo{journal}{\emph{ACM Trans. Graph.}} \bibinfo{volume}{29}, \bibinfo{number}{4}, Article \bibinfo{articleno}{66} (\bibinfo{date}{July} \bibinfo{year}{2010}), \bibinfo{numpages}{13}~pages.
\newblock
\showISSN{0730-0301}
\href{https://doi.org/10.1145/1778765.1778803}{doi:\nolinkurl{10.1145/1778765.1778803}}


\bibitem[Poirier-Ginter et~al\mbox{.}(2025)]%
        {edtable-3dgs}
\bibfield{author}{\bibinfo{person}{Yohan Poirier-Ginter}, \bibinfo{person}{Jeffrey Hu}, \bibinfo{person}{Jean-Fran{\c c}ois Lalonde}, {and} \bibinfo{person}{George Drettakis}.} \bibinfo{year}{2025}\natexlab{}.
\newblock \showarticletitle{{Editable Physically-based Reflections in Raytraced Gaussian Radiance Fields}}. In \bibinfo{booktitle}{\emph{{SIGGRAPH Asia 2025 - 18th ACM SIGGRAPH Conference and Exhibition on Computer Graphics and Interactive Techniques in Asia}}}. \bibinfo{address}{Hong Kong, Hong Kong SAR China}.
\newblock
\href{https://doi.org/10.1145/3757377.3763971}{doi:\nolinkurl{10.1145/3757377.3763971}}


\bibitem[Ramamoorthi and Hanrahan(2001)]%
        {ravi01sh}
\bibfield{author}{\bibinfo{person}{Ravi Ramamoorthi} {and} \bibinfo{person}{Pat Hanrahan}.} \bibinfo{year}{2001}\natexlab{}.
\newblock \showarticletitle{An efficient representation for irradiance environment maps}. In \bibinfo{booktitle}{\emph{Proceedings of the 28th Annual Conference on Computer Graphics and Interactive Techniques}} \emph{(\bibinfo{series}{SIGGRAPH '01})}. \bibinfo{publisher}{Association for Computing Machinery}, \bibinfo{address}{New York, NY, USA}, \bibinfo{pages}{497–500}.
\newblock
\showISBNx{158113374X}
\href{https://doi.org/10.1145/383259.383317}{doi:\nolinkurl{10.1145/383259.383317}}


\bibitem[Sun et~al\mbox{.}(2025a)]%
        {SVGIR2025}
\bibfield{author}{\bibinfo{person}{Hanxiao Sun}, \bibinfo{person}{Yupeng Gao}, \bibinfo{person}{Jin Xie}, \bibinfo{person}{Jian Yang}, {and} \bibinfo{person}{Beibei Wang}.} \bibinfo{year}{2025}\natexlab{a}.
\newblock \showarticletitle{SVG-IR: Spatially-Varying Gaussian Splatting for Inverse Rendering}. In \bibinfo{booktitle}{\emph{Proceedings of the IEEE/CVF Conference on Computer Vision and Pattern Recognition (CVPR)}}. \bibinfo{pages}{16143--16152}.
\newblock


\bibitem[Sun et~al\mbox{.}(2025b)]%
        {sun2025stochasticraytracingtransparent}
\bibfield{author}{\bibinfo{person}{Xin Sun}, \bibinfo{person}{Iliyan Georgiev}, \bibinfo{person}{Yun Fei}, {and} \bibinfo{person}{Miloš Hašan}.} \bibinfo{year}{2025}\natexlab{b}.
\newblock \showarticletitle{Stochastic Ray Tracing of 3D Transparent Gaussians}.
\newblock \bibinfo{journal}{\emph{EGSR conference proceedings}} (\bibinfo{year}{2025}).
\newblock


\bibitem[Veach(1997)]%
        {Veach1997RobustMC}
\bibfield{author}{\bibinfo{person}{Eric Veach}.} \bibinfo{year}{1997}\natexlab{}.
\newblock \emph{\bibinfo{title}{Robust {Monte Carlo} Methods for Light Transport Simulation}}.
\newblock \bibinfo{thesistype}{Ph.\,D. Dissertation}. \bibinfo{school}{Stanford University}, \bibinfo{address}{Stanford, CA, USA}.
\newblock
\href{https://doi.org/10.5555/927297}{doi:\nolinkurl{10.5555/927297}}


\bibitem[Vicini et~al\mbox{.}(2021)]%
        {PathReplayBackpropagation}
\bibfield{author}{\bibinfo{person}{Delio Vicini}, \bibinfo{person}{Sébastien Speierer}, {and} \bibinfo{person}{Wenzel Jakob}.} \bibinfo{year}{2021}\natexlab{}.
\newblock \showarticletitle{Path Replay Backpropagation: Differentiating Light Paths using Constant Memory and Linear Time}.
\newblock \bibinfo{journal}{\emph{Transactions on Graphics (Proceedings of SIGGRAPH)}} \bibinfo{volume}{40}, \bibinfo{number}{4} (\bibinfo{date}{Aug.} \bibinfo{year}{2021}), \bibinfo{pages}{108:1--108:14}.
\newblock


\bibitem[Wang et~al\mbox{.}(2009)]%
        {wang09sg}
\bibfield{author}{\bibinfo{person}{Jiaping Wang}, \bibinfo{person}{Peiran Ren}, \bibinfo{person}{Minmin Gong}, \bibinfo{person}{John Snyder}, {and} \bibinfo{person}{Baining Guo}.} \bibinfo{year}{2009}\natexlab{}.
\newblock \showarticletitle{All-frequency rendering of dynamic, spatially-varying reflectance}.
\newblock \bibinfo{journal}{\emph{ACM Trans. Graph.}} \bibinfo{volume}{28}, \bibinfo{number}{5} (\bibinfo{date}{Dec.} \bibinfo{year}{2009}), \bibinfo{pages}{1–10}.
\newblock
\showISSN{0730-0301}
\href{https://doi.org/10.1145/1618452.1618479}{doi:\nolinkurl{10.1145/1618452.1618479}}


\bibitem[Wu et~al\mbox{.}(2025)]%
        {wu20253dgut}
\bibfield{author}{\bibinfo{person}{Qi Wu}, \bibinfo{person}{Janick Martinez~Esturo}, \bibinfo{person}{Ashkan Mirzaei}, \bibinfo{person}{Nicolas Moenne-Loccoz}, {and} \bibinfo{person}{Zan Gojcic}.} \bibinfo{year}{2025}\natexlab{}.
\newblock \showarticletitle{3DGUT: Enabling Distorted Cameras and Secondary Rays in Gaussian Splatting}.
\newblock \bibinfo{journal}{\emph{Conference on Computer Vision and Pattern Recognition (CVPR)}} (\bibinfo{year}{2025}).
\newblock


\bibitem[Wu et~al\mbox{.}(2024)]%
        {wu2024deferredgs}
\bibfield{author}{\bibinfo{person}{Tong Wu}, \bibinfo{person}{Jia-Mu Sun}, \bibinfo{person}{Yu-Kun Lai}, \bibinfo{person}{Yuewen Ma}, \bibinfo{person}{Leif Kobbelt}, {and} \bibinfo{person}{Lin Gao}.} \bibinfo{year}{2024}\natexlab{}.
\newblock \showarticletitle{DeferredGS: Decoupled and Editable Gaussian Splatting with Deferred Shading}.
\newblock \bibinfo{journal}{\emph{arXiv preprint arXiv:2404.09412}} (\bibinfo{year}{2024}).
\newblock


\bibitem[Xie et~al\mbox{.}(2024)]%
        {xie2024envgs}
\bibfield{author}{\bibinfo{person}{Tao Xie}, \bibinfo{person}{Xi Chen}, \bibinfo{person}{Zhen Xu}, \bibinfo{person}{Yiman Xie}, \bibinfo{person}{Yudong Jin}, \bibinfo{person}{Yujun Shen}, \bibinfo{person}{Sida Peng}, \bibinfo{person}{Hujun Bao}, {and} \bibinfo{person}{Xiaowei Zhou}.} \bibinfo{year}{2024}\natexlab{}.
\newblock \showarticletitle{EnvGS: Modeling View-Dependent Appearance with Environment Gaussian}.
\newblock \bibinfo{journal}{\emph{arXiv preprint arXiv:2412.15215}} (\bibinfo{year}{2024}).
\newblock


\bibitem[Xu et~al\mbox{.}(2026)]%
        {xu2026Stoch3DGS}
\bibfield{author}{\bibinfo{person}{Peiyu Xu}, \bibinfo{person}{Xin Sun}, \bibinfo{person}{Krishna Mullia}, \bibinfo{person}{Raymond Fei}, \bibinfo{person}{Iliyan Georgiev}, {and} \bibinfo{person}{Shuang Zhao}.} \bibinfo{year}{2026}\natexlab{}.
\newblock \bibinfo{title}{Stochastic Ray Tracing for the Reconstruction of 3D Gaussian Splatting}.
\newblock
\showeprint[arxiv]{2603.23637}~[cs.CV]
\urldef\tempurl%
\url{https://arxiv.org/abs/2603.23637}
\showURL{%
\tempurl}


\bibitem[Yang et~al\mbox{.}(2026)]%
        {EAG-PT}
\bibfield{author}{\bibinfo{person}{Xijie Yang}, \bibinfo{person}{Mulin Yu}, \bibinfo{person}{Changjian Jiang}, \bibinfo{person}{Kerui Ren}, \bibinfo{person}{Tao Lu}, \bibinfo{person}{Jiangmiao Pang}, \bibinfo{person}{Dahua Lin}, \bibinfo{person}{Bo Dai}, {and} \bibinfo{person}{Linning Xu}.} \bibinfo{year}{2026}\natexlab{}.
\newblock \showarticletitle{EAG-PT: Emission-Aware Gaussians and Path Tracing for Diffuse Indoor Scene Reconstruction and Editing}. In \bibinfo{booktitle}{\emph{Proceedings of the Special Interest Group on Computer Graphics and Interactive Techniques Conference Conference Papers}} \emph{(\bibinfo{series}{SIGGRAPH Conference Papers '26})}. \bibinfo{publisher}{Association for Computing Machinery}, \bibinfo{address}{New York, NY, USA}, Article \bibinfo{articleno}{149}, \bibinfo{numpages}{12}~pages.
\newblock
\showISBNx{9798400725548}
\href{https://doi.org/10.1145/3799902.3811054}{doi:\nolinkurl{10.1145/3799902.3811054}}


\bibitem[Yao et~al\mbox{.}(2022)]%
        {neural-incident-light-field}
\bibfield{author}{\bibinfo{person}{Yao Yao}, \bibinfo{person}{Jingyang Zhang}, \bibinfo{person}{Jingbo Liu}, \bibinfo{person}{Yihang Qu}, \bibinfo{person}{Tian Fang}, \bibinfo{person}{David McKinnon}, \bibinfo{person}{Yanghai Tsin}, {and} \bibinfo{person}{Long Quan}.} \bibinfo{year}{2022}\natexlab{}.
\newblock \showarticletitle{NeILF: Neural Incident Light Field for Material and Lighting Estimation}. In \bibinfo{booktitle}{\emph{ECCV}}.
\newblock
\urldef\tempurl%
\url{https://arxiv.org/abs/2203.07182}
\showURL{%
\tempurl}


\bibitem[Zeng et~al\mbox{.}(2024)]%
        {rgb2x}
\bibfield{author}{\bibinfo{person}{Zheng Zeng}, \bibinfo{person}{Valentin Deschaintre}, \bibinfo{person}{Iliyan Georgiev}, \bibinfo{person}{Yannick Hold-Geoffroy}, \bibinfo{person}{Yiwei Hu}, \bibinfo{person}{Fujun Luan}, \bibinfo{person}{Ling-Qi Yan}, {and} \bibinfo{person}{Milo\v{s} Ha\v{s}an}.} \bibinfo{year}{2024}\natexlab{}.
\newblock \showarticletitle{RGB$\leftrightarrow$X: Image decomposition and synthesis using material- and lighting-aware diffusion models}. In \bibinfo{booktitle}{\emph{ACM SIGGRAPH 2024 Conference Papers}} (Denver, CO, USA) \emph{(\bibinfo{series}{SIGGRAPH '24})}. \bibinfo{publisher}{Association for Computing Machinery}, \bibinfo{address}{New York, NY, USA}, Article \bibinfo{articleno}{75}, \bibinfo{numpages}{11}~pages.
\newblock
\showISBNx{9798400705250}
\href{https://doi.org/10.1145/3641519.3657445}{doi:\nolinkurl{10.1145/3641519.3657445}}


\bibitem[Zhang et~al\mbox{.}(2023)]%
        {zhang2023neilf++}
\bibfield{author}{\bibinfo{person}{Jingyang Zhang}, \bibinfo{person}{Yao Yao}, \bibinfo{person}{Shiwei Li}, \bibinfo{person}{Jingbo Liu}, \bibinfo{person}{Tian Fang}, \bibinfo{person}{David McKinnon}, \bibinfo{person}{Yanghai Tsin}, {and} \bibinfo{person}{Long Quan}.} \bibinfo{year}{2023}\natexlab{}.
\newblock \showarticletitle{NeILF++: Inter-reflectable Light Fields for Geometry and Material Estimation}.
\newblock \bibinfo{journal}{\emph{International Conference on Computer Vision (ICCV)}} (\bibinfo{year}{2023}).
\newblock


\bibitem[Zhang et~al\mbox{.}(2021)]%
        {physg2021}
\bibfield{author}{\bibinfo{person}{Kai Zhang}, \bibinfo{person}{Fujun Luan}, \bibinfo{person}{Qianqian Wang}, \bibinfo{person}{Kavita Bala}, {and} \bibinfo{person}{Noah Snavely}.} \bibinfo{year}{2021}\natexlab{}.
\newblock \showarticletitle{{PhySG}: {I}nverse Rendering with Spherical Gaussians for Physics-based Material Editing and Relighting}. In \bibinfo{booktitle}{\emph{The IEEE/CVF Conference on Computer Vision and Pattern Recognition (CVPR)}}.
\newblock


\bibitem[Zhang et~al\mbox{.}(2022)]%
        {zhang2022invrender}
\bibfield{author}{\bibinfo{person}{Yuanqing Zhang}, \bibinfo{person}{Jiaming Sun}, \bibinfo{person}{Xingyi He}, \bibinfo{person}{Huan Fu}, \bibinfo{person}{Rongfei Jia}, {and} \bibinfo{person}{Xiaowei Zhou}.} \bibinfo{year}{2022}\natexlab{}.
\newblock \showarticletitle{Modeling Indirect Illumination for Inverse Rendering}. In \bibinfo{booktitle}{\emph{CVPR}}.
\newblock


\bibitem[Zhou et~al\mbox{.}(2024)]%
        {zhou2024unifiedgaussianprimitivesscene}
\bibfield{author}{\bibinfo{person}{Yang Zhou}, \bibinfo{person}{Songyin Wu}, {and} \bibinfo{person}{Ling-Qi Yan}.} \bibinfo{year}{2024}\natexlab{}.
\newblock \bibinfo{title}{Unified Gaussian Primitives for Scene Representation and Rendering}.
\newblock
\showeprint[arxiv]{2406.09733}~[cs.GR]
\urldef\tempurl%
\url{https://arxiv.org/abs/2406.09733}
\showURL{%
\tempurl}


\end{thebibliography}



\begin{thebibliography}{10}


\ifx \showCODEN    \undefined \def \showCODEN     #1{\unskip}     \fi
\ifx \showISBNx    \undefined \def \showISBNx     #1{\unskip}     \fi
\ifx \showISBNxiii \undefined \def \showISBNxiii  #1{\unskip}     \fi
\ifx \showISSN     \undefined \def \showISSN      #1{\unskip}     \fi
\ifx \showLCCN     \undefined \def \showLCCN      #1{\unskip}     \fi
\ifx \shownote     \undefined \def \shownote      #1{#1}          \fi
\ifx \showarticletitle \undefined \def \showarticletitle #1{#1}   \fi
\ifx \showURL      \undefined \def \showURL       {\relax}        \fi
\providecommand\bibfield[2]{#2}
\providecommand\bibinfo[2]{#2}
\providecommand\natexlab[1]{#1}
\providecommand\showeprint[2][]{arXiv:#2}

\bibitem[Burley(2012)]%
        {burley12disneybrdf}
\bibfield{author}{\bibinfo{person}{Brent Burley}.} \bibinfo{year}{2012}\natexlab{}.
\newblock \showarticletitle{Physically-Based Shading at {Disney}}. In \bibinfo{booktitle}{\emph{{ACM SIGGRAPH} Course Notes}}. \bibinfo{pages}{1--7}.
\newblock


\bibitem[Du et~al\mbox{.}(2024)]%
        {GS-ID}
\bibfield{author}{\bibinfo{person}{Kang Du}, \bibinfo{person}{Zhihao Liang}, {and} \bibinfo{person}{Zeyu Wang}.} \bibinfo{year}{2024}\natexlab{}.
\newblock \bibinfo{title}{GS-ID: Illumination Decomposition on Gaussian Splatting via Diffusion Prior and Parametric Light Source Optimization}.
\newblock
\showeprint[arxiv]{2408.08524}~[cs.CV]
\urldef\tempurl%
\url{https://arxiv.org/abs/2408.08524}
\showURL{%
\tempurl}


\bibitem[Gao et~al\mbox{.}(2024)]%
        {R3DG}
\bibfield{author}{\bibinfo{person}{Jian Gao}, \bibinfo{person}{Chun Gu}, \bibinfo{person}{Youtian Lin}, \bibinfo{person}{Zhihao Li}, \bibinfo{person}{Hao Zhu}, \bibinfo{person}{Xun Cao}, \bibinfo{person}{Li Zhang}, {and} \bibinfo{person}{Yao Yao}.} \bibinfo{year}{2024}\natexlab{}.
\newblock \showarticletitle{Relightable 3D Gaussians: Realistic Point Cloud Relighting with BRDF Decomposition and Ray Tracing}. In \bibinfo{booktitle}{\emph{Computer Vision – ECCV 2024: 18th European Conference, Milan, Italy, September 29–October 4, 2024, Proceedings, Part XLV}} (Milan, Italy). \bibinfo{publisher}{Springer-Verlag}, \bibinfo{address}{Berlin, Heidelberg}, \bibinfo{pages}{73–89}.
\newblock
\showISBNx{978-3-031-72994-2}
\href{https://doi.org/10.1007/978-3-031-72995-9_5}{doi:\nolinkurl{10.1007/978-3-031-72995-9_5}}


\bibitem[Gu et~al\mbox{.}(2025)]%
        {IRGS}
\bibfield{author}{\bibinfo{person}{Chun Gu}, \bibinfo{person}{Xiaofei Wei}, \bibinfo{person}{Zixuan Zeng}, \bibinfo{person}{Yuxuan Yao}, {and} \bibinfo{person}{Li Zhang}.} \bibinfo{year}{2025}\natexlab{}.
\newblock \showarticletitle{IRGS: Inter-Reflective Gaussian Splatting with 2D Gaussian Ray Tracing}. In \bibinfo{booktitle}{\emph{CVPR}}.
\newblock


\bibitem[Kajiya(1986)]%
        {kajiya1986rendering}
\bibfield{author}{\bibinfo{person}{James~T. Kajiya}.} \bibinfo{year}{1986}\natexlab{}.
\newblock \showarticletitle{The rendering equation}.
\newblock \bibinfo{journal}{\emph{SIGGRAPH Comput. Graph.}} \bibinfo{volume}{20}, \bibinfo{number}{4} (\bibinfo{date}{Aug.} \bibinfo{year}{1986}), \bibinfo{pages}{143–150}.
\newblock
\showISSN{0097-8930}
\href{https://doi.org/10.1145/15886.15902}{doi:\nolinkurl{10.1145/15886.15902}}


\bibitem[Kerbl et~al\mbox{.}(2023)]%
        {kerbl3dgaussians}
\bibfield{author}{\bibinfo{person}{Bernhard Kerbl}, \bibinfo{person}{Georgios Kopanas}, \bibinfo{person}{Thomas Leimk{\"u}hler}, {and} \bibinfo{person}{George Drettakis}.} \bibinfo{year}{2023}\natexlab{}.
\newblock \showarticletitle{3D Gaussian Splatting for Real-Time Radiance Field Rendering}.
\newblock \bibinfo{journal}{\emph{ACM Transactions on Graphics}} \bibinfo{volume}{42}, \bibinfo{number}{4} (\bibinfo{date}{July} \bibinfo{year}{2023}).
\newblock


\bibitem[Sun et~al\mbox{.}(2025)]%
        {SVGIR2025}
\bibfield{author}{\bibinfo{person}{Hanxiao Sun}, \bibinfo{person}{Yupeng Gao}, \bibinfo{person}{Jin Xie}, \bibinfo{person}{Jian Yang}, {and} \bibinfo{person}{Beibei Wang}.} \bibinfo{year}{2025}\natexlab{}.
\newblock \showarticletitle{SVG-IR: Spatially-Varying Gaussian Splatting for Inverse Rendering}. In \bibinfo{booktitle}{\emph{Proceedings of the IEEE/CVF Conference on Computer Vision and Pattern Recognition (CVPR)}}. \bibinfo{pages}{16143--16152}.
\newblock


\bibitem[Vicini et~al\mbox{.}(2021)]%
        {PathReplayBackpropagation}
\bibfield{author}{\bibinfo{person}{Delio Vicini}, \bibinfo{person}{Sébastien Speierer}, {and} \bibinfo{person}{Wenzel Jakob}.} \bibinfo{year}{2021}\natexlab{}.
\newblock \showarticletitle{Path Replay Backpropagation: Differentiating Light Paths using Constant Memory and Linear Time}.
\newblock \bibinfo{journal}{\emph{Transactions on Graphics (Proceedings of SIGGRAPH)}} \bibinfo{volume}{40}, \bibinfo{number}{4} (\bibinfo{date}{Aug.} \bibinfo{year}{2021}), \bibinfo{pages}{108:1--108:14}.
\newblock


\bibitem[Zeng et~al\mbox{.}(2024)]%
        {rgb2x}
\bibfield{author}{\bibinfo{person}{Zheng Zeng}, \bibinfo{person}{Valentin Deschaintre}, \bibinfo{person}{Iliyan Georgiev}, \bibinfo{person}{Yannick Hold-Geoffroy}, \bibinfo{person}{Yiwei Hu}, \bibinfo{person}{Fujun Luan}, \bibinfo{person}{Ling-Qi Yan}, {and} \bibinfo{person}{Milo\v{s} Ha\v{s}an}.} \bibinfo{year}{2024}\natexlab{}.
\newblock \showarticletitle{RGB$\leftrightarrow$X: Image decomposition and synthesis using material- and lighting-aware diffusion models}. In \bibinfo{booktitle}{\emph{ACM SIGGRAPH 2024 Conference Papers}} (Denver, CO, USA) \emph{(\bibinfo{series}{SIGGRAPH '24})}. \bibinfo{publisher}{Association for Computing Machinery}, \bibinfo{address}{New York, NY, USA}, Article \bibinfo{articleno}{75}, \bibinfo{numpages}{11}~pages.
\newblock
\showISBNx{9798400705250}
\href{https://doi.org/10.1145/3641519.3657445}{doi:\nolinkurl{10.1145/3641519.3657445}}


\bibitem[Zhang et~al\mbox{.}(2026)]%
        {RaDe-GS}
\bibfield{author}{\bibinfo{person}{Baowen Zhang}, \bibinfo{person}{Chuan Fang}, \bibinfo{person}{Rakesh Shrestha}, \bibinfo{person}{Yixun Liang}, \bibinfo{person}{Xiao-Xiao Long}, {and} \bibinfo{person}{Ping Tan}.} \bibinfo{year}{2026}\natexlab{}.
\newblock \showarticletitle{RaDe-GS: Rasterizing Depth in Gaussian Splatting}.
\newblock \bibinfo{journal}{\emph{ACM Trans. Graph.}} \bibinfo{volume}{45}, \bibinfo{number}{2}, Article \bibinfo{articleno}{19} (\bibinfo{date}{March} \bibinfo{year}{2026}), \bibinfo{numpages}{14}~pages.
\newblock
\showISSN{0730-0301}
\href{https://doi.org/10.1145/3789201}{doi:\nolinkurl{10.1145/3789201}}


\end{thebibliography}

\clearpage
\begin{bibunit}[ACM-Reference-Format]
\appendix
\setcounter{page}{1}
\setcounter{section}{0}
\setcounter{figure}{0}
\setcounter{table}{0}
\setcounter{equation}{0}

\section*{Appendix}

\section{Physically Based Rendering Details}
\label{sec:supp_pbr_model}

Our framework is built on physically based path tracing. For a non-emissive
surface point $\mathbf{x} \in \mathbb{R}^3$, the outgoing radiance
$L_o \in \mathbb{R}^3$ along direction $\boldsymbol{\omega}_o$ is governed by
the classical rendering equation~\cite{kajiya1986rendering}:
\begin{small}
\begin{equation}
\label{eq:rendering_equation}
L_o(\mathbf{x}, \boldsymbol{\omega}_o)
=
\int_{\Omega}
f_r(\mathbf{x}, \boldsymbol{\omega}_i, \boldsymbol{\omega}_o)\,
L_i(\mathbf{x}, \boldsymbol{\omega}_i)\,
(\mathbf{n}\!\cdot\!\boldsymbol{\omega}_i)\,
d\boldsymbol{\omega}_i,
\end{equation}
\end{small}
where $\boldsymbol{\omega}_i \in \Omega$ denotes the incident direction,
$f_r(\cdot)$ is the bidirectional reflectance distribution function (BRDF),
$L_i \in \mathbb{R}^3$ is the incident radiance,
$\mathbf{n} \in \mathbb{R}^3$ is the surface normal, and $\Omega$ denotes the
upper hemisphere aligned with $\mathbf{n}$.
The incident radiance $L_i$ recursively encodes light transport from the entire
scene, including direct illumination, visibility, and multi-bounce scattering.

We adopt a simplified Disney BRDF~\cite{burley12disneybrdf} parameterized by
albedo $\mathbf{a} \in \mathbb{R}^3$, roughness $r \in \mathbb{R}$, and
metallic $m \in \mathbb{R}$. The metallic parameter controls the relative diffuse and specular contributions, and the BRDF is modeled as a combination of a diffuse term and a GGX microfacet specular term:
\begin{small}
\begin{equation}
\label{eq:supp_disney_brdf}
\begin{gathered}
f_r(\boldsymbol{\omega}_i,\boldsymbol{\omega}_o)
=
(1-m)\frac{\mathbf{a}}{\pi}(\textbf{1}-\mathbf{F})
+
\frac{
D(\mathbf{n},\mathbf{h})\,
\mathbf{F}(\boldsymbol{\omega}_o,\mathbf{h})\,
G(\boldsymbol{\omega}_i,\boldsymbol{\omega}_o)
}{
4(\mathbf{n}\!\cdot\!\boldsymbol{\omega}_i)
 (\mathbf{n}\!\cdot\!\boldsymbol{\omega}_o)
}, \\[1mm]
\mathbf{h}
=
\frac{\boldsymbol{\omega}_i+\boldsymbol{\omega}_o}
{\|\boldsymbol{\omega}_i+\boldsymbol{\omega}_o\|}, \\[1mm]
\mathbf{F}_0
=
(1-m)\cdot 0.04 + m\mathbf{a}, \\[1mm]
\mathbf{F}(\boldsymbol{\omega}_o,\mathbf{h})
=
\mathbf{F}_0
+
(1-\mathbf{F}_0)
\left(1-\max(\boldsymbol{\omega}_o\!\cdot\!\mathbf{h},0)\right)^5,
\end{gathered}
\end{equation}
\end{small}
where $\mathbf{h}$ is the half vector and $\mathbf{F}$ is the Fresnel term.
The GGX normal distribution function and the Schlick--GGX approximation of the geometry term are defined as
\begin{small}
\begin{equation}
\label{eq:supp_ggx_terms}
\begin{gathered}
D(\mathbf{n},\mathbf{h}) 
= 
\frac{\alpha^2}
{ \pi \left[ (\mathbf{n}\!\cdot\!\mathbf{h})^2(\alpha^2-1)+1 \right]^2 },
\\[1mm]
G(\boldsymbol{\omega}_i,\boldsymbol{\omega}_o)
= 
G_1(\mathbf{n},\boldsymbol{\omega}_i)\, G_1(\mathbf{n},\boldsymbol{\omega}_o), 
\\[1mm] 
G_1(\mathbf{n},\boldsymbol{\omega}) = \frac{\mathbf{n}\!\cdot\!\boldsymbol{\omega}} { (\mathbf{n}\!\cdot\!\boldsymbol{\omega})(1-k)+k }, \qquad k=\frac{\alpha}{2}, \quad \alpha=r^2 . \end{gathered} 
\end{equation} 
\end{small}

In implementation, we evaluate the cosine-weighted BRDF term
\begin{small}
\begin{equation}
\tilde{f}_r
=
f_r(\boldsymbol{\omega}_i,\boldsymbol{\omega}_o)
\max(\mathbf{n}\!\cdot\!\boldsymbol{\omega}_i,0),
\end{equation}
\end{small}
which directly corresponds to the BRDF--cosine factor in the integrand of
Eq.~(\ref{eq:rendering_equation}).

\section{Inverse Rendering Algorithm Details}
\label{sec:supp_inverse_rendering_detail}

In Sec.~\ref{sec:inverse_rendering} of the main
paper, the gradient of the material parameters is
decomposed into a local BRDF branch and a ray-induced branch:
\begin{small}
\begin{equation}
\frac{\partial \mathcal{L}}{\partial \boldsymbol{\Theta}_j}
=
\sum_i
\frac{\partial \mathcal{L}}{\partial \mathcal{T}_i}
\left(
\frac{\partial \mathcal{T}_i}{\partial f_{\mathrm{brdf}}}
\frac{\partial f_{\mathrm{brdf}}}{\partial \mathbf{s}_{\mathbf{p}_i}}
+
\frac{\partial \mathcal{T}_i}{\partial L_i}
\frac{\partial L_i}{\partial \mathbf{d}_i}
\frac{\partial \mathbf{d}_i}{\partial \mathbf{s}_{\mathbf{p}_i}}
\right)
\frac{\partial \mathbf{s}_{\mathbf{p}_i}}{\partial \boldsymbol{\Theta}_j}.
\label{eq:supp_gs_gradient_decomp}
\end{equation}
\end{small}

Under multiple importance sampling (MIS), each interaction evaluates one environment-sampled query and
one BRDF-sampled continuation query. We write the contribution of all replayed
paths passing through the \(i\)-th interaction as
\begin{small}
\begin{equation}
\begin{aligned}
\mathcal T_i
=
\beta_i
\Big(
&
f_i^{\mathrm{env}} L_i
+
f_i^{\mathrm{brdf}} f_{i+1}^{\mathrm{env}} L_{i+1}
+
f_i^{\mathrm{brdf}} f_{i+1}^{\mathrm{brdf}}
f_{i+2}^{\mathrm{env}} L_{i+2}
+\cdots
\Big),
\end{aligned}
\label{eq:supp_mis_expansion}
\end{equation}
\end{small}
where \(\beta_i\) denotes the path throughput before the \(i\)-th interaction and \(L_i\) denotes the incident radiance along the sampled environment direction.
The factors \(f_i^{\mathrm{env}}\) and \(f_i^{\mathrm{brdf}}\) are the evaluated
scattering terms for the environment-sampling and BRDF-continuation branches,
including the BRDF value, cosine term, PDF correction, and MIS weight. During
replay, PDF corrections and MIS weights are treated as fixed estimator weights and are not differentiated. With fixed geometry and path replay~\cite{PathReplayBackpropagation}, this treatment does not introduce additional non-differentiable operations into the replayed optimization.

Under this convention, gradients through the scattering factors below are propagated only through the evaluated BRDF--cosine terms. From Eq.~\eqref{eq:supp_mis_expansion}, the derivatives with respect to the two local scattering factors are
\begin{small}
\begin{equation}
\frac{\partial \mathcal T_i}{\partial f_i^{\mathrm{env}}}
=
\beta_i L_i,
\quad
\frac{\partial \mathcal T_i}{\partial f_i^{\mathrm{brdf}}}
=
\frac{
\mathcal T_i-\beta_i f_i^{\mathrm{env}}L_i
}{
f_i^{\mathrm{brdf}}
}.
\label{eq:supp_mis_scattering_grad}
\end{equation}
\end{small}
These derivatives provide the adjoints propagated to the equivalent surface interaction through path replay.
We denote the local cosine-weighted BRDF term by \(\mathbf f_{\mathrm{brdf}}\) and derive its material derivatives with respect to albedo,
roughness, and metallic, with the BRDF normal detached. Omitting the interaction
index \(i\), we define
\[
\mathrm{denom}
=
4(\mathbf n\!\cdot\!\boldsymbol{\omega}_o)
(\mathbf n\!\cdot\!\boldsymbol{\omega}_i)+\epsilon,
\quad
\kappa=\frac{DG}{\mathrm{denom}},
\quad
c=(1-\boldsymbol{\omega}_o\!\cdot\!\mathbf h)^5 ,
\]
where \(\epsilon\) is a small constant for numerical stability.
The derivatives with respect to albedo, roughness, and metallic are
\begin{small}
\begin{equation}
\begin{aligned}
\frac{\partial \mathbf f_{\mathrm{brdf}}}{\partial a_k}
&=
(\mathbf n\!\cdot\!\boldsymbol{\omega}_i)
\left[
\kappa\,m(1-c)\mathbf e_k
+
\frac{1-m}{\pi}
\left(
(1-F_k)\mathbf e_k
-
\mathbf a\,
\frac{\partial \mathbf F}{\partial a_k}
\right)
\right],
\\[1.5mm]
\frac{\partial \mathbf f_{\mathrm{brdf}}}{\partial r}
&=
(\mathbf n\!\cdot\!\boldsymbol{\omega}_i)\,
\mathbf F
\frac{
G\frac{\partial D}{\partial r}
+
D\frac{\partial G}{\partial r}
}
{\mathrm{denom}},
\\[1.5mm]
\frac{\partial \mathbf f_{\mathrm{brdf}}}{\partial m}
&=
(\mathbf n\!\cdot\!\boldsymbol{\omega}_i)
\left[
\kappa(1-c)(\mathbf a-0.04\mathbf 1)
-
\frac{\mathbf a}{\pi}
\left(
\mathbf 1-\mathbf F
+
(1-m)
\frac{\partial \mathbf F}{\partial m}
\right)
\right],
\end{aligned}
\label{eq:supp_brdf_material_derivatives}
\end{equation}
\end{small}
where \(a_k\) and \(F_k\) denote the \(k\)-th channels of
\(\mathbf a\) and \(\mathbf F\), respectively; \(\mathbf e_k\) is the
corresponding RGB basis vector, \(\mathbf 1\) denotes the RGB all-ones vector.

Following Eq.~\eqref{eq:interaction_grad} of the main paper, the resulting
adjoints are propagated to each contributing Gaussian as
\begin{small}
\begin{equation}
\begin{aligned}
\bar{\mathbf a}_j
&\mathrel{+}=
w_j^{(i)}
\boldsymbol{\eta}_i^{\top}
\frac{\partial \mathbf f_i}{\partial \mathbf a_i},
\\[1mm]
\bar r_j
&\mathrel{+}=
w_j^{(i)}
\left(
\boldsymbol{\eta}_i^{\top}
\frac{\partial \mathbf f_i}{\partial r_i}
+
\boldsymbol{\gamma}_i^{\top}
\frac{\partial \mathbf d_i}{\partial r_i}
\right),
\\[1mm]
\bar m_j
&\mathrel{+}=
w_j^{(i)}
\boldsymbol{\eta}_i^{\top}
\frac{\partial \mathbf f_i}{\partial m_i}
,
\end{aligned}
\label{eq:supp_final_gs_attr_grad}
\end{equation}
\end{small}
where \(\boldsymbol{\eta}_i\) and \(\boldsymbol{\gamma}_i\) denote the adjoints of
the cosine-weighted BRDF term and replayed ray direction, respectively. 

For completeness, the normalized aggregation also induces an adjoint with respect to the
unnormalized contribution \(q_j^{(i)}\). With
\(w_j^{(i)}=q_j^{(i)}/Q_i\) and
\(Q_i=\sum_{k\in\mathcal C_i}q_k^{(i)}\), the corresponding adjoint is
\begin{small}
\begin{equation}
\bar q_j^{(i)}
=
\frac{
\bar{\mathbf{s}}_{\mathbf p_i}^{\top}
\left(
\mathbf g_j-\mathbf{s}_{\mathbf p_i}
\right)
}{
Q_i
}.
\label{eq:supp_q_adjoint}
\end{equation}
\end{small}
Material attributes are updated using
Eq.~\eqref{eq:supp_final_gs_attr_grad}, while Gaussian opacity and geometry
remain fixed throughout the inverse-rendering stage and therefore receive no
gradient updates.


\section{Geometry Optimization}
\label{sec:supp_normal}
Geometry plays an important role in inverse rendering, as accurate geometric information is essential for reliable surface interaction, visibility estimation, and material recovery. Following existing methods~\citep{R3DG, IRGS}, we adopt a lightweight strategy that uses pseudo normals estimated from rendered depth to supervise the optimized Gaussian normals.

Given the predicted distance map and the corresponding camera ray directions,
we first reconstruct a pseudo surface and estimate a pseudo normal target
$\hat{\mathbf N}$. The predicted normal map $\mathbf N$ is then supervised by a
cosine loss on valid foreground pixels, with a linearly increasing weight during
training:
\begin{small}
\begin{equation}
    \mathcal L_n
    =
    \frac{1}{|\Omega|}
    \sum_{\mathbf u \in \Omega}
    \left(
    1 -
    \left\langle
    \mathbf N(\mathbf u),
    \hat{\mathbf N}(\mathbf u)
    \right\rangle
    \right),
    \quad
    \lambda_n(t)
    =
    \lambda_n
    \min\left(\frac{t}{T}, 1\right),
\end{equation}
\end{small}
where $\Omega$ denotes the set of valid foreground pixels, $t$ is the current
training iteration, and $T$ is the total number of iterations. This warm-up
avoids enforcing noisy pseudo-normal targets too strongly at the early stage of
optimization.

Following GS-ID~\citep{GS-ID}, we further regularize the optimized normals using
dataset-provided normal priors~\citep{rgb2x} and an image-space smoothness term.
The provided normal maps are loaded and transformed into world space to obtain
$\mathbf N^p$. We use an $\ell_2$ loss for the normal prior, and adopt the
edge-aware total variation loss defined in Eq.~\eqref{eq:tv_loss} for normal
smoothness, using the ground-truth RGB image as the edge reference:
\begin{small}
\begin{equation}
    \mathcal L_p
    =
    \frac{1}{|\Omega|}
    \sum_{\mathbf u \in \Omega}
    \left\|
    \mathbf N(\mathbf u) -
    \mathbf N^p(\mathbf u)
    \right\|_2^2,
    \quad
    \mathcal L_{tv}
    =
    \mathcal L_{\mathrm{TV}}^{\mathrm{edge}}
    \left(
    \mathbf N,
    \mathbf I
    \right),
\end{equation}
\end{small}
where $\mathbf I$ is the ground-truth RGB image. The foreground mask is used
when computing both regularization terms. The normal prior is applied
throughout training without warm-up, while the normal smoothness term is only
used during the geometry and normal initialization stage and is disabled in the
subsequent PBR inverse-rendering stage.

In addition, we adopt a mask binary cross-entropy loss to encourage foreground opacity toward 1 and background opacity toward 0:
\begin{scriptsize}
\begin{equation}
\mathcal{L}_{\mathrm{mask}}
=
-\frac{1}{|\mathcal{P}|}
\sum_{\mathbf{p}\in\mathcal{P}}
\left[
M(\mathbf{p})\log \hat{A}(\mathbf{p})
+
\bigl(1-M(\mathbf{p})\bigr)
\log\bigl(1-\hat{A}(\mathbf{p})\bigr)
\right],
\end{equation}
\end{scriptsize}
where $M(\mathbf{p})\in\{0,1\}$ denotes the foreground mask and
$\hat{A}(\mathbf{p})$ is the rendered opacity at pixel $\mathbf{p}$.
This suppresses floating Gaussians and keeps foreground opacity away from the
hard-threshold boundary.

\begin{figure}[t]
    \centering
    \includegraphics[width=0.9\linewidth]{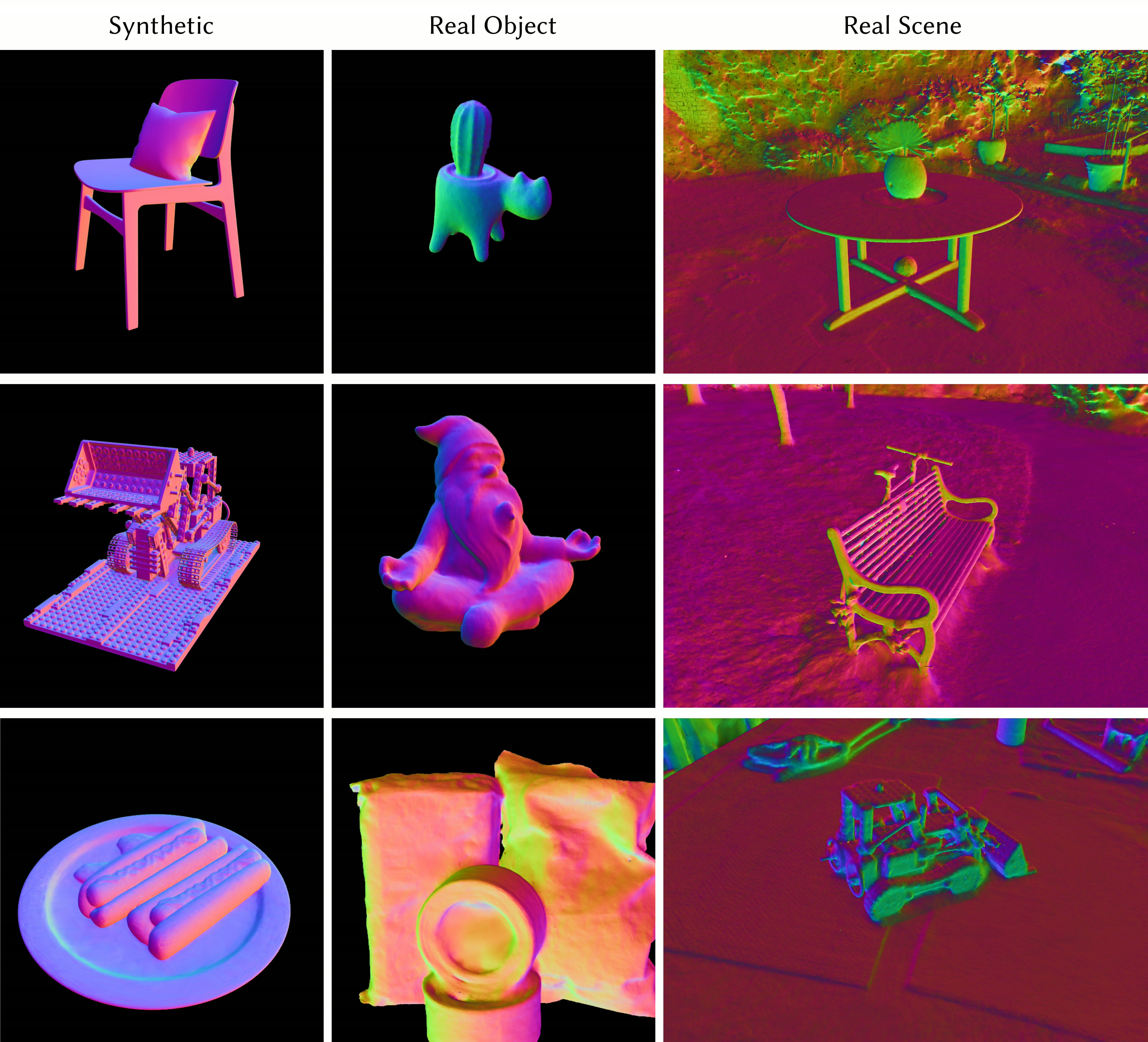}
    \caption{
    Normal results on synthetic objects, real objects, and real scenes.
    }
    \label{fig:normal_results}
\end{figure}

The overall normal-related objective is therefore:
\begingroup
\small
\begin{equation}
    \mathcal L_N
    =
    \lambda_n(t)\mathcal L_n
    +
    \lambda_p\mathcal L_p
    +
    \lambda_{tv}\mathcal L_{tv}
    +
    \lambda_{mask}\mathcal{L}_{\mathrm{mask}}.
\end{equation}
\endgroup
We set $\lambda_n=0.01$, $\lambda_{tv}=0.01$, and $\lambda_{mask}=0.1$ in all experiments. The prior term is enabled only for the Hotdog scene and the Synthetic4Relight scenes, with $\lambda_p=0.01$; it is disabled for all other scenes. Since normal supervision can also affect geometry-related Gaussian parameters through rendering weights and visibility, we reuse the densification strategy of vanilla 3D Gaussian Splatting~\citep{kerbl3dgaussians}, including its clone and split operations. The optimized geometry and normal results are visualized in Fig.~\ref{fig:normal_results} and Figs.~\ref{fig:armadillo}--\ref{fig:jugs}. 

\begin{figure}[!htbp]
    \centering
    \includegraphics[width=\linewidth]{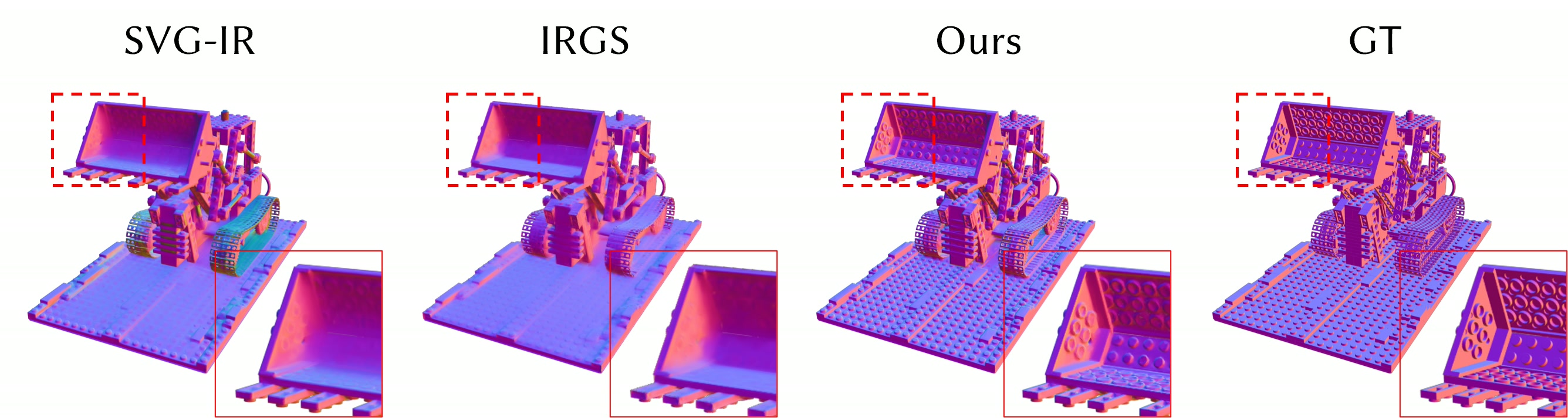}
    \vspace{-0.1in}
    \caption{Lego Normal Qualitative Comparisons(w/o Prior), showing that our framework recovers finer normal details.}
    \label{fig:lego_normal}
    \vspace{-0.15in}
\end{figure}

Ray tracing determines Gaussian interactions according to their peak responses along each ray rather than center depths~\citep{RaDe-GS}, providing more accurate geometric cues for normal estimation. Moreover, the volumetric flexibility of 3D Gaussians can better preserve complex local geometric details in challenging regions. As shown in Fig.~\ref{fig:lego_normal}, our framework recovers finer normal details on the Lego scene. 

\section{Extensive Results}
\subsection{Additional Cornell Box Results}
\label{sec:supp_cornell_box_analysis}
\begin{figure}[!htbp]
    \centering
    \includegraphics[width=0.6\linewidth]{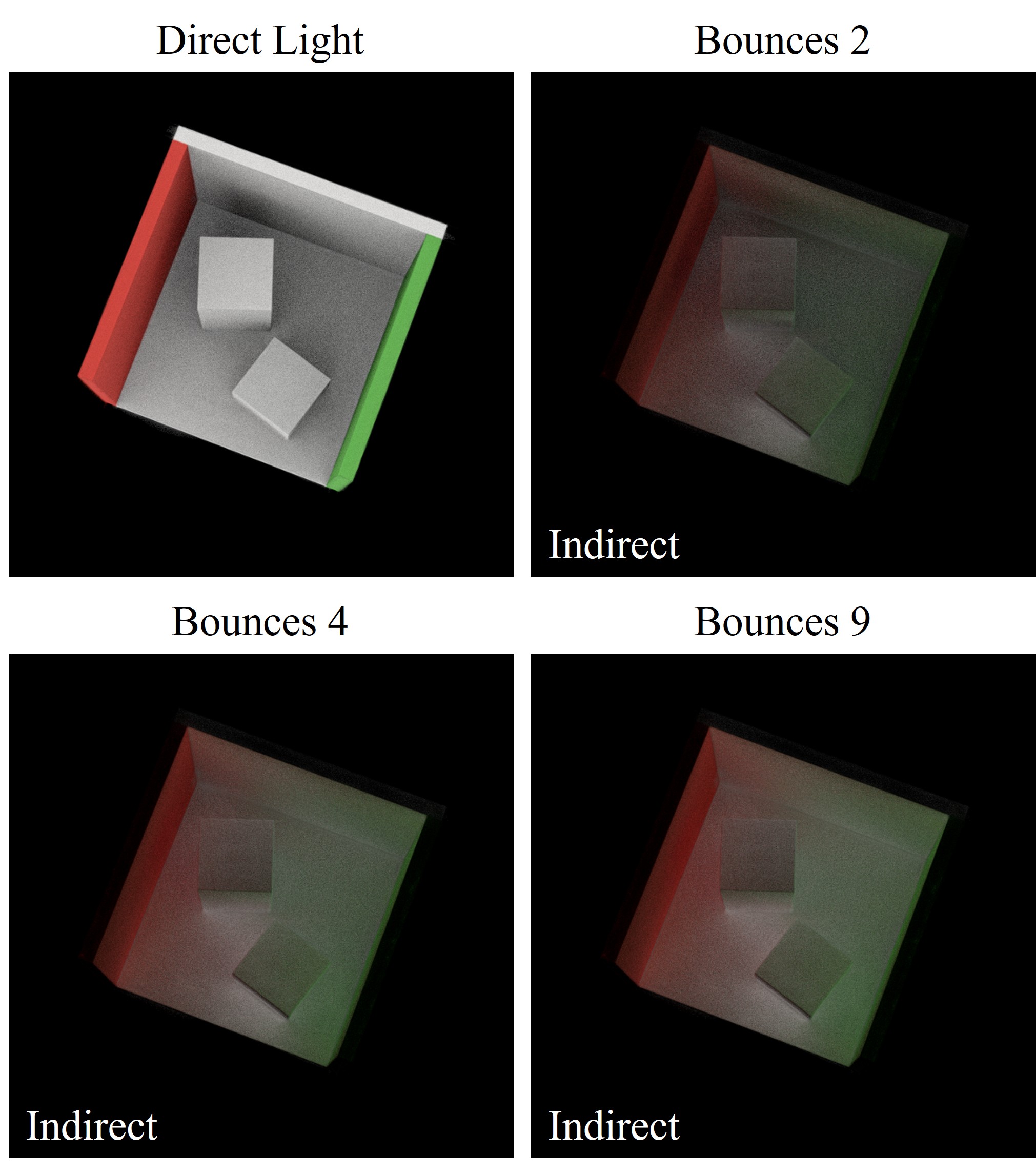}
    \caption{
    Rendering results of the recovered Cornell box under different light transport settings. We compare direct illumination with path-traced rendering using up to 2, 4, and 9 bounces. Increasing the bounce count captures more indirect illumination.
    }
    \label{fig:multi_bounce}
    \vspace{-0.15in}
\end{figure}

\begin{figure}[!htbp]
    \centering
    \includegraphics[width=\linewidth]{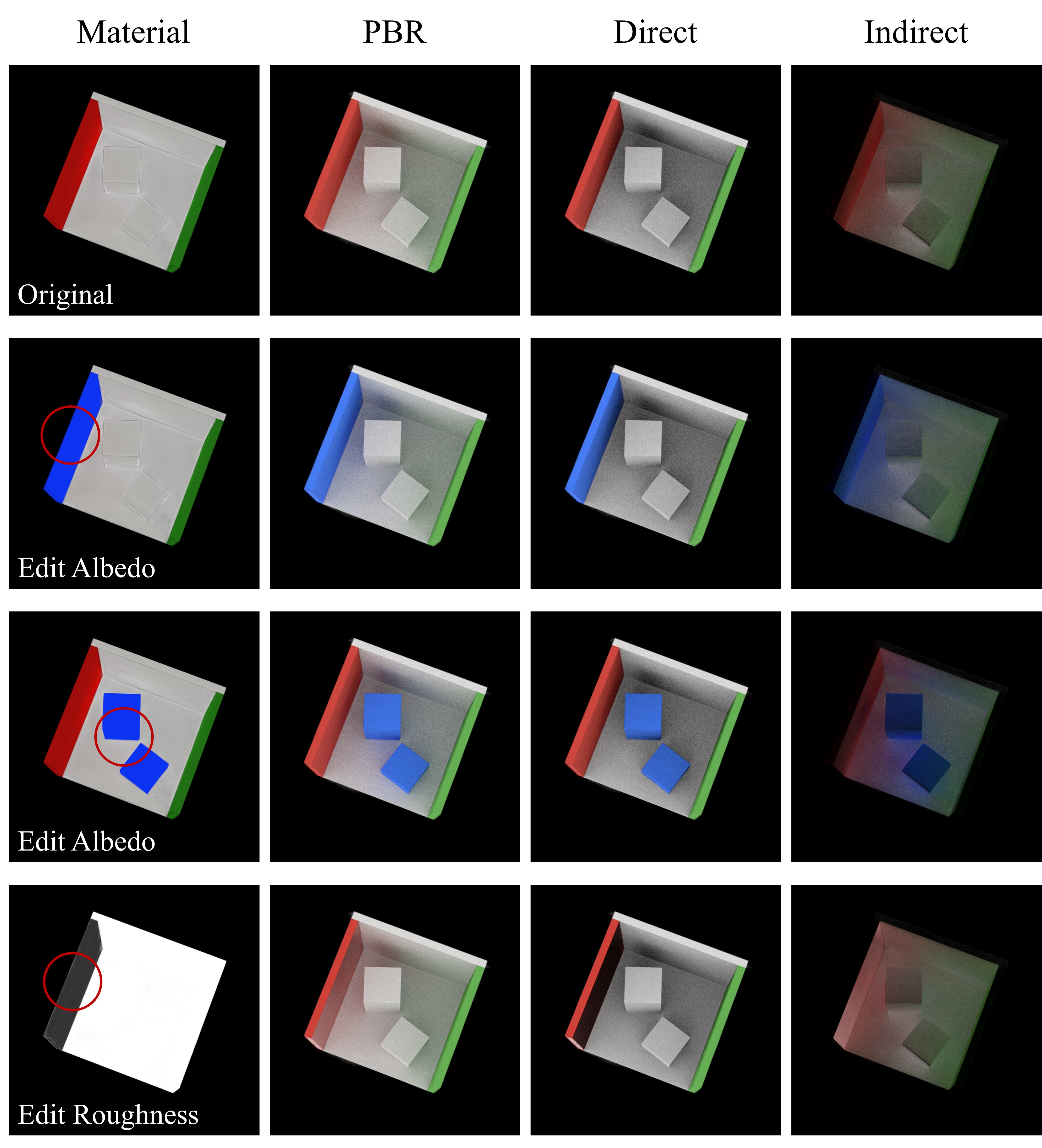}
    \caption{
    Material editing results on the recovered Cornell-box scene. Editing albedo
    and roughness affects both direct appearance and indirect illumination under
    the same ray-traced transport pipeline.
    }
    \label{fig:material_edit}
    \vspace{-0.15in}
\end{figure}

Fig.~\ref{fig:multi_bounce} shows that increasing the maximum number of bounces captures more indirect illumination in the Cornell-box scene. The difference becomes minor beyond 4 bounces, indicating that 4-bounce transport is sufficient for this scene.
Fig.~\ref{fig:material_edit} shows material editing results, where albedo and roughness changes affect both surface appearance and surrounding indirect illumination, demonstrating physically consistent responses under the same path-traced transport model.

\begin{figure}[t]
    \centering
    \includegraphics[width=\linewidth]{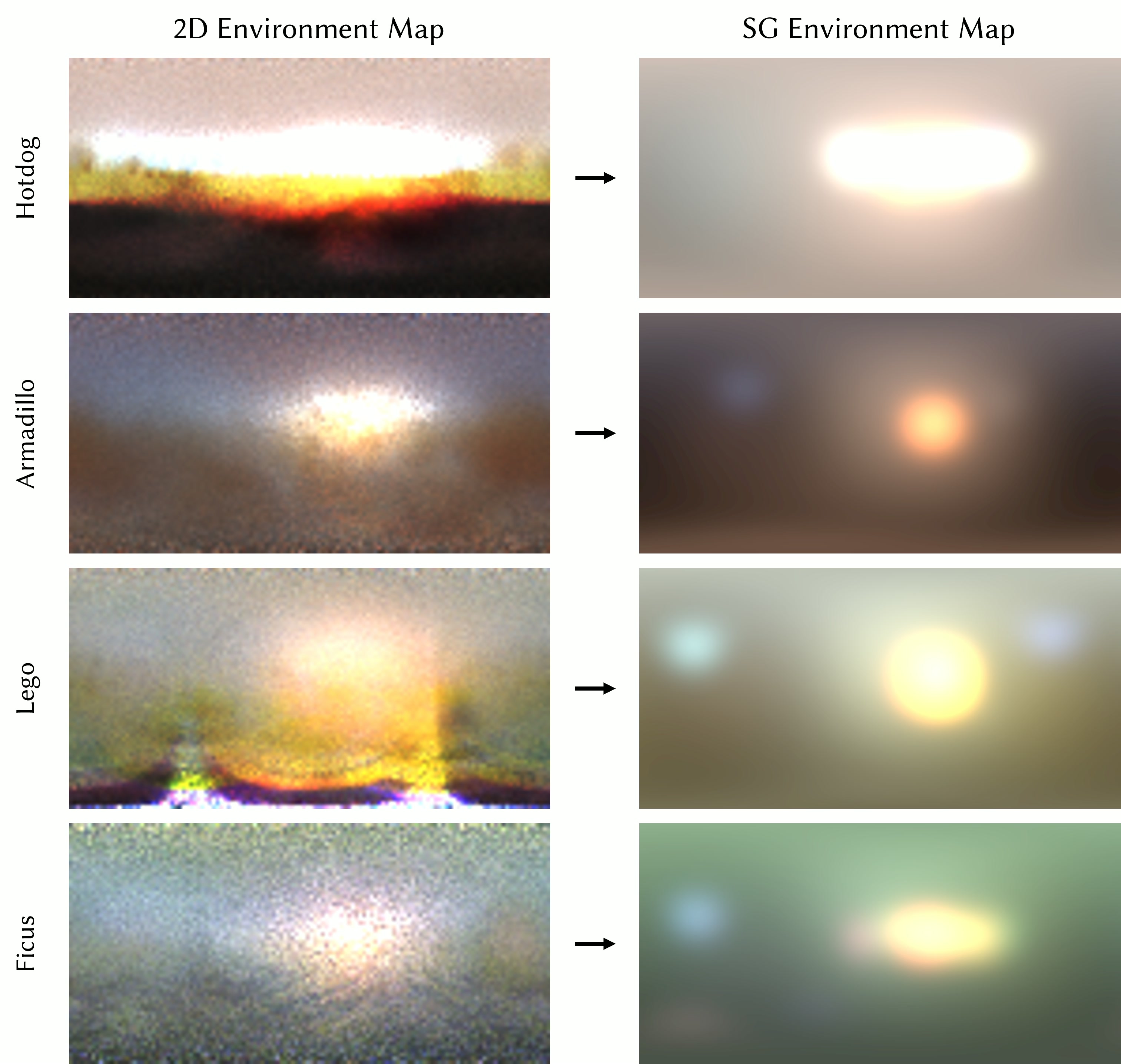}
    \vspace{-0.2in}
    \caption{
    Visualization of the recovered environment illumination. The compact
    Spherical-Gaussian representation accurately captures dominant lighting
    directions while providing a smoother and lower-noise representation in
    low-energy regions.
    }
    \label{fig:envmap_visualization}
\end{figure}

\begin{figure}[!htbp]
    \centering
    \includegraphics[width=\linewidth]{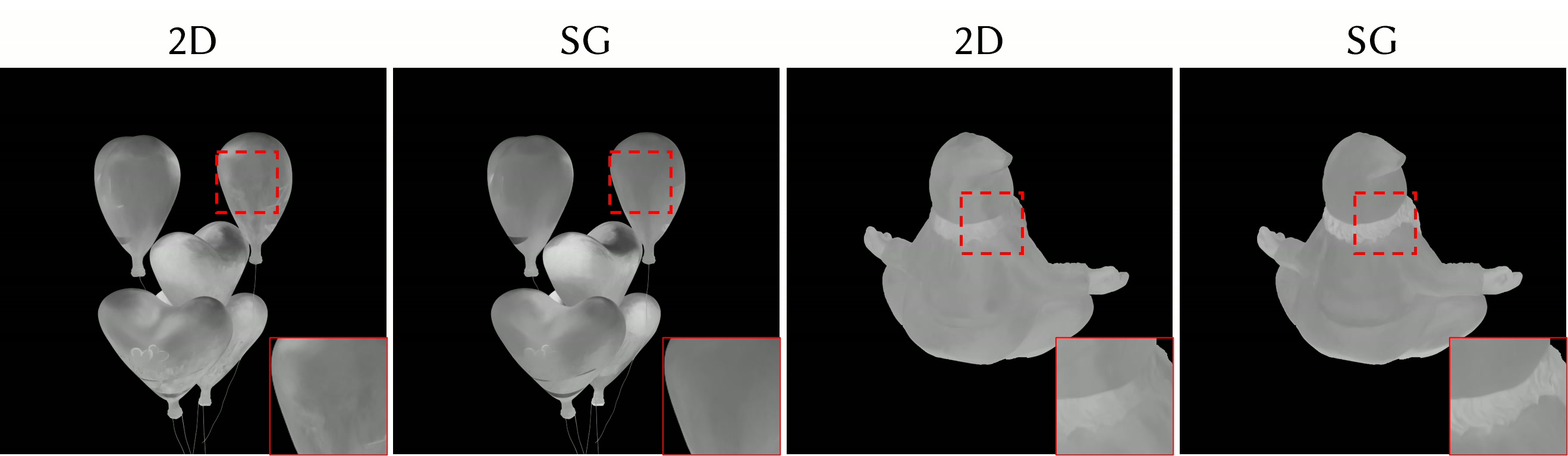}
    \caption{
    Comparison of roughness estimation using the SG and 2D environment-map parameterizations. The SG parameterization generally produces smoother material estimates at the same optimization iteration.
    }
    \label{fig:envmap_ablation}
\end{figure}

\subsection{Heterogeneous Lighting and Geometry}
Our framework resolves visibility and shadowing through ray tracing rather than
precomputed visibility or heuristic approximations. This naturally supports
different lighting configurations, including environment maps, point lights, and
area lights, while allowing multiple geometric representations to be handled
within the same light-transport pipeline. Different geometry types therefore
share the same visibility evaluation and physically based light-transport
formulation, without requiring representation-specific lighting approximations.

\begin{figure*}[t]
  \centering
  \includegraphics[width=\linewidth]{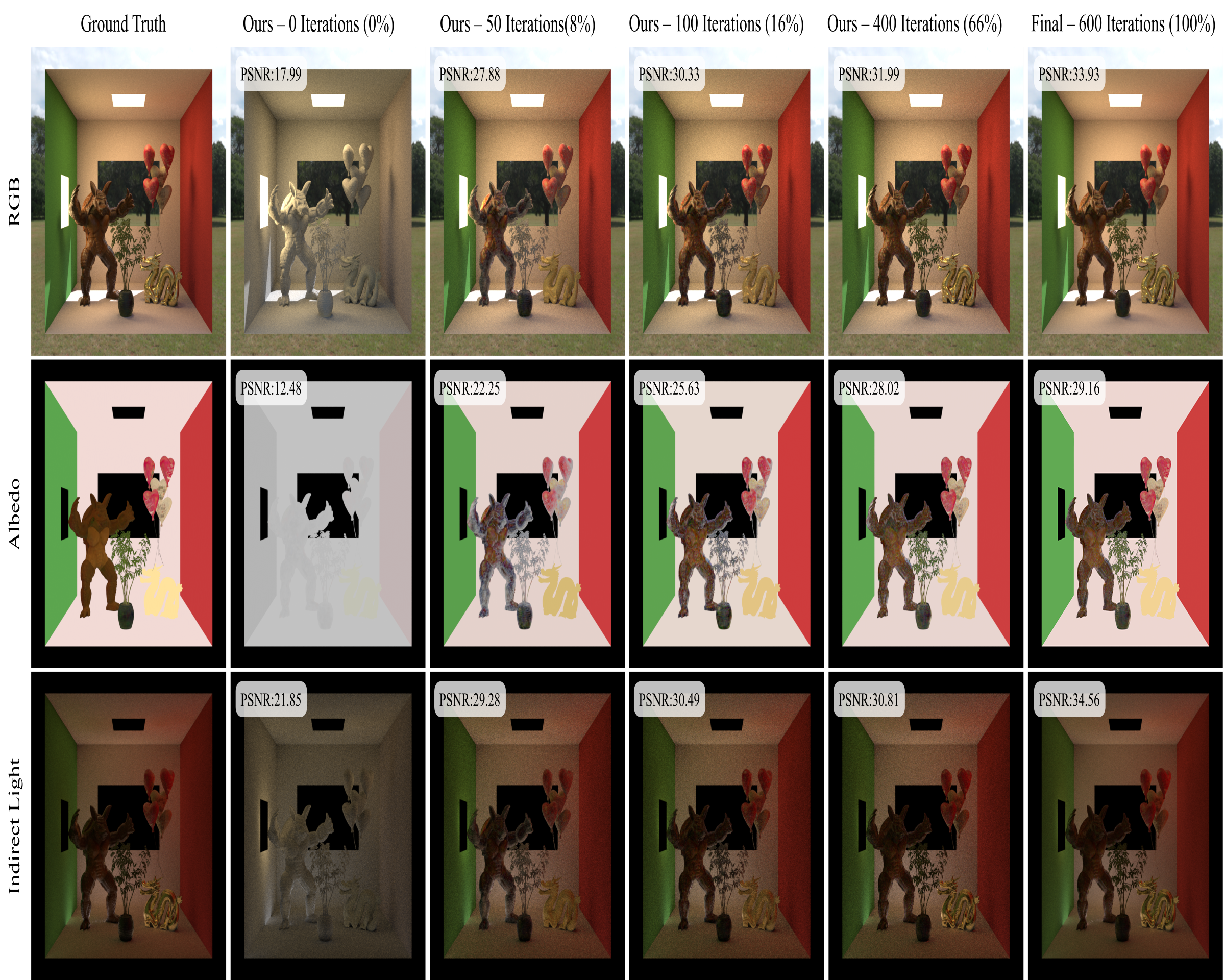}
  \caption{Heterogeneous lighting and geometry. We construct a Cornell-box--style
  scene with mixed geometry: the walls and dragon are represented as triangle
  meshes, while the remaining objects are 3D Gaussian models. Such a hybrid
  representation is practical: regular structures such as walls can be modeled
  compactly with meshes, while objects with more complex local detail can be
  represented more flexibly using Gaussians. The scene is illuminated by two
  area lights and an environment map. Under fixed lighting, we perform
  single-view material estimation for 600 iterations. Our method recovers
  consistent material attributes across mesh and Gaussian representations under
  unified ray-traced light transport.}
  \label{fig:heterogeneous}
\end{figure*}

This flexibility is useful for mixed-material estimation in heterogeneous scenes.
Regular structures such as walls, floors, and ceilings can be represented
efficiently with triangle meshes, while objects with complex geometry and local
detail can use 3D Gaussians. We demonstrate this capability in a scene containing
both representations and multiple material types, illuminated by two area lights
and an environment map. As shown in Fig.~\ref{fig:heterogeneous}, our method
jointly optimizes material properties for different geometry types and recovers
consistent material attributes across the mixed scene.

\subsection{Environment Map}
\label{sec:supp_envmap}
The Spherical Gaussian representation, with substantially fewer parameters, constrains the illumination space, facilitating faster convergence and smoother optimization gradients. Fig.~\ref{fig:envmap_visualization} shows that the compact SG representation accurately captures dominant lighting directions while providing smoother estimates in low-energy regions. Fig.~\ref{fig:envmap_ablation} further shows that the SG parameterization generally yields smoother material estimates.

\begin{figure*}[b]
    \centering
    \includegraphics[width=\linewidth]{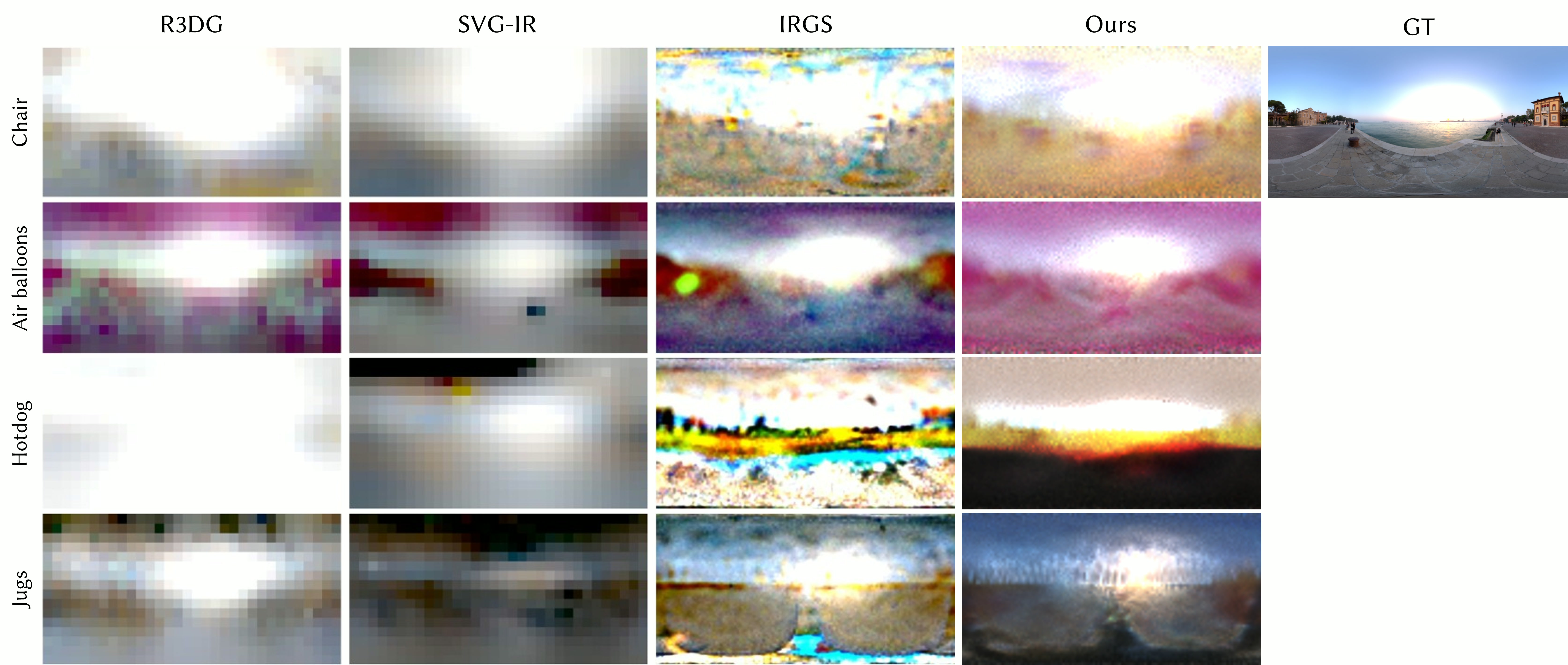}
    \caption{
    Comparison of environment maps recovered by our method and existing inverse-rendering approaches on the Synthetic4Relight dataset. Our method accurately recovers the dominant light sources and their spatial distributions.
    }
    \label{fig:envmap_comparison}
\end{figure*}

For visualization and comparison, we also use the \(128\times64\) 2D environment-map representation as existing methods. Fig.~\ref{fig:envmap_comparison} compares the illumination recovered by our method with that of existing splatting-based inverse-rendering approaches.

\subsection{Extensive Qualitative Results}
\label{sec:extensive_results}
\begin{figure*}[!htbp]
  \centering
  \includegraphics[width=\linewidth]{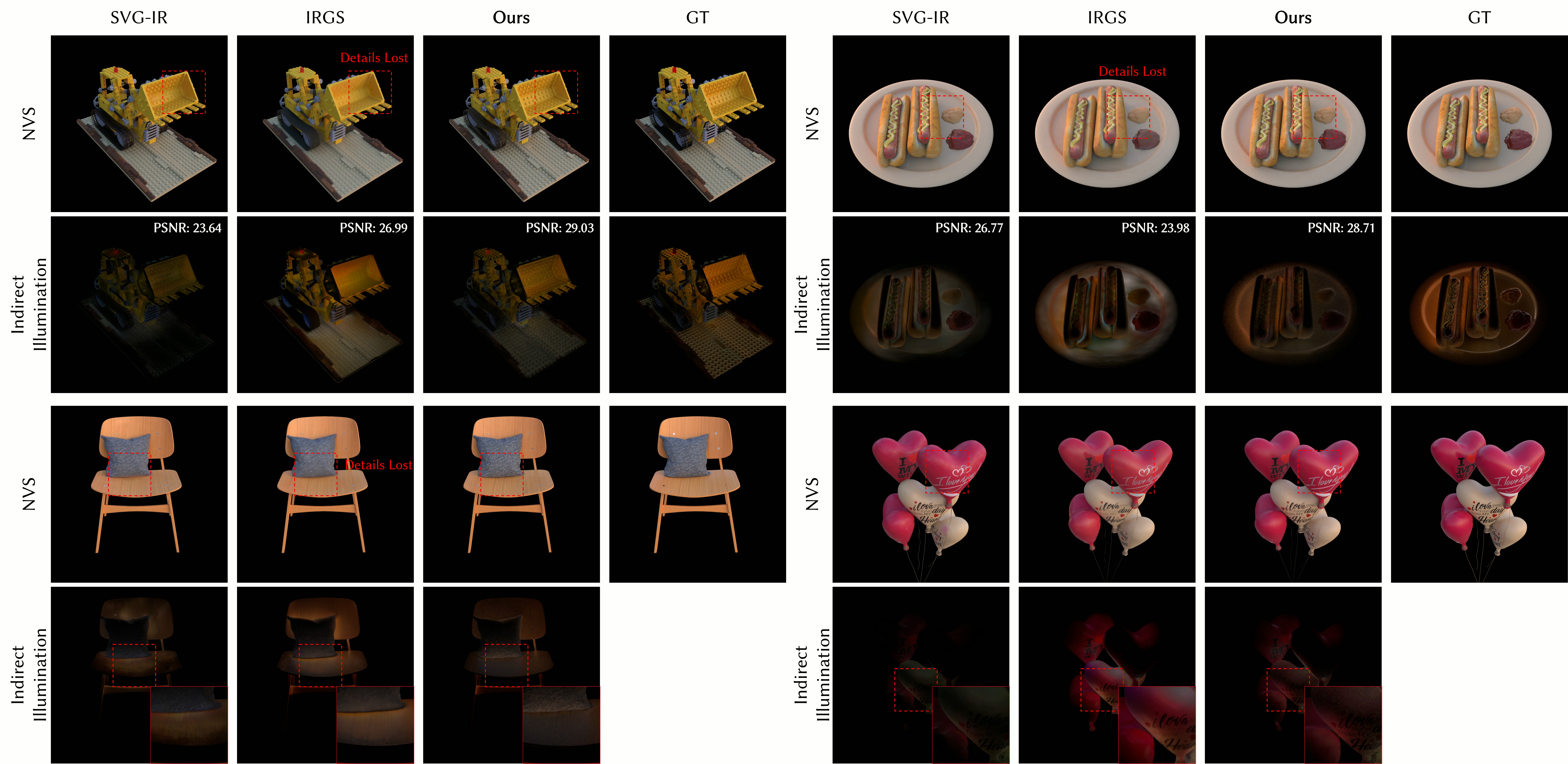}
  \caption{Novel-view synthesis and indirect illumination results. Our framework outperforms IRGS~\citep{IRGS} and SVG-IR~\citep{SVGIR2025} in both novel-view synthesis and indirect illumination modeling.}
  \label{fig:supp_indirect}
\end{figure*}

\begin{figure*}[!htbp]
  \centering
  \includegraphics[width=0.8\linewidth]{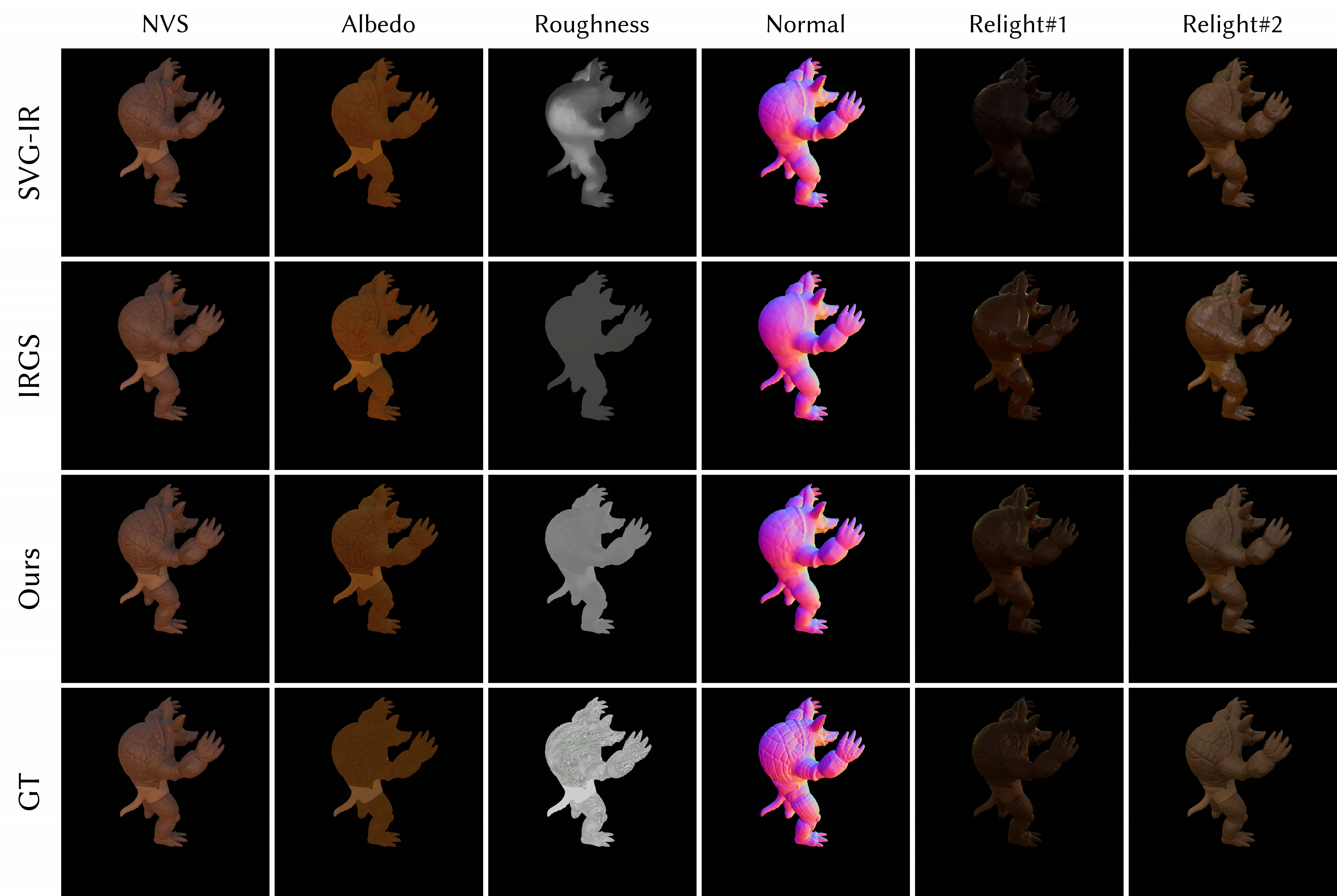}
  \vspace{-0.1in}
  \caption{Qualitative results on the Armadillo scene.}
  \label{fig:armadillo}
\end{figure*}

\begin{figure*}[!htbp]
  \centering
  \includegraphics[width=0.8\linewidth]{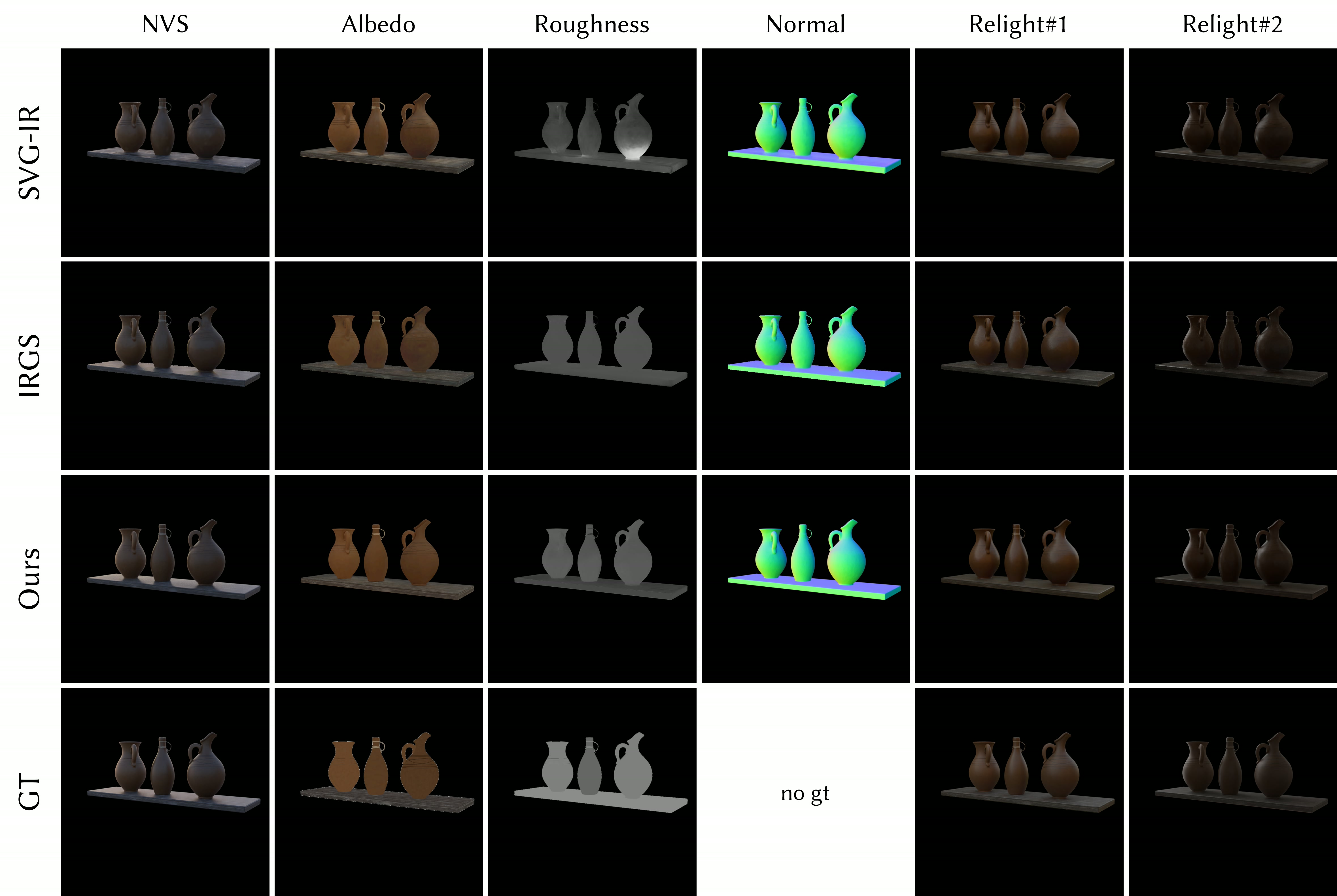}
  \vspace{-0.1in}
  \caption{Qualitative results on the Jugs scene.}
  \label{fig:jugs}
\end{figure*}

We provide additional qualitative results and comparisons in
Figs.~\ref{fig:supp_indirect}--\ref{fig:jugs} to further demonstrate
the effectiveness of our framework.

\begin{figure*}[!htbp]
  \centering
  \includegraphics[width=0.8\linewidth]{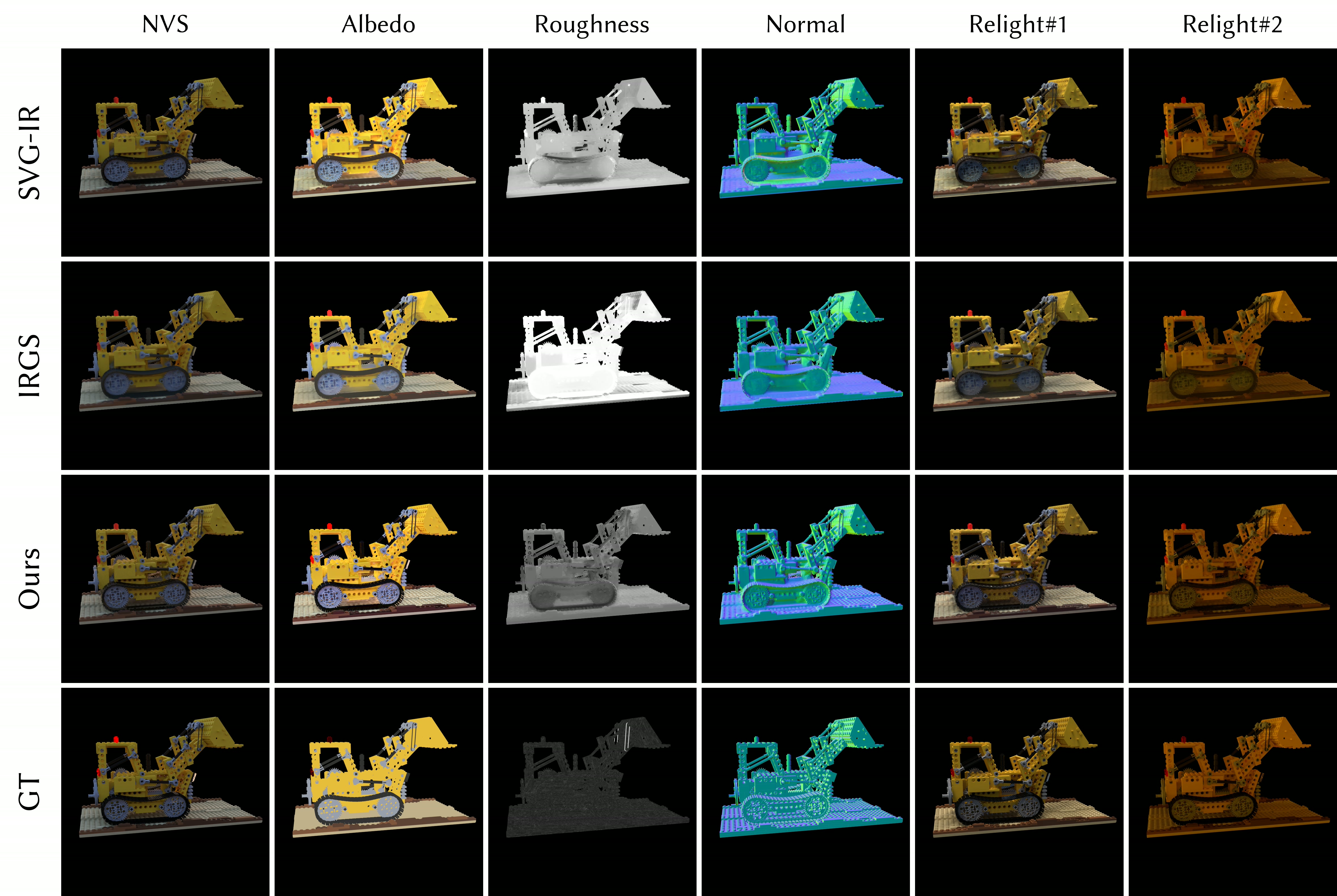}
  \vspace{-0.1in}
  \caption{Qualitative results on the Lego scene.}
\end{figure*}

\begin{figure*}[!htbp]
  \centering
  \includegraphics[width=0.8\linewidth]{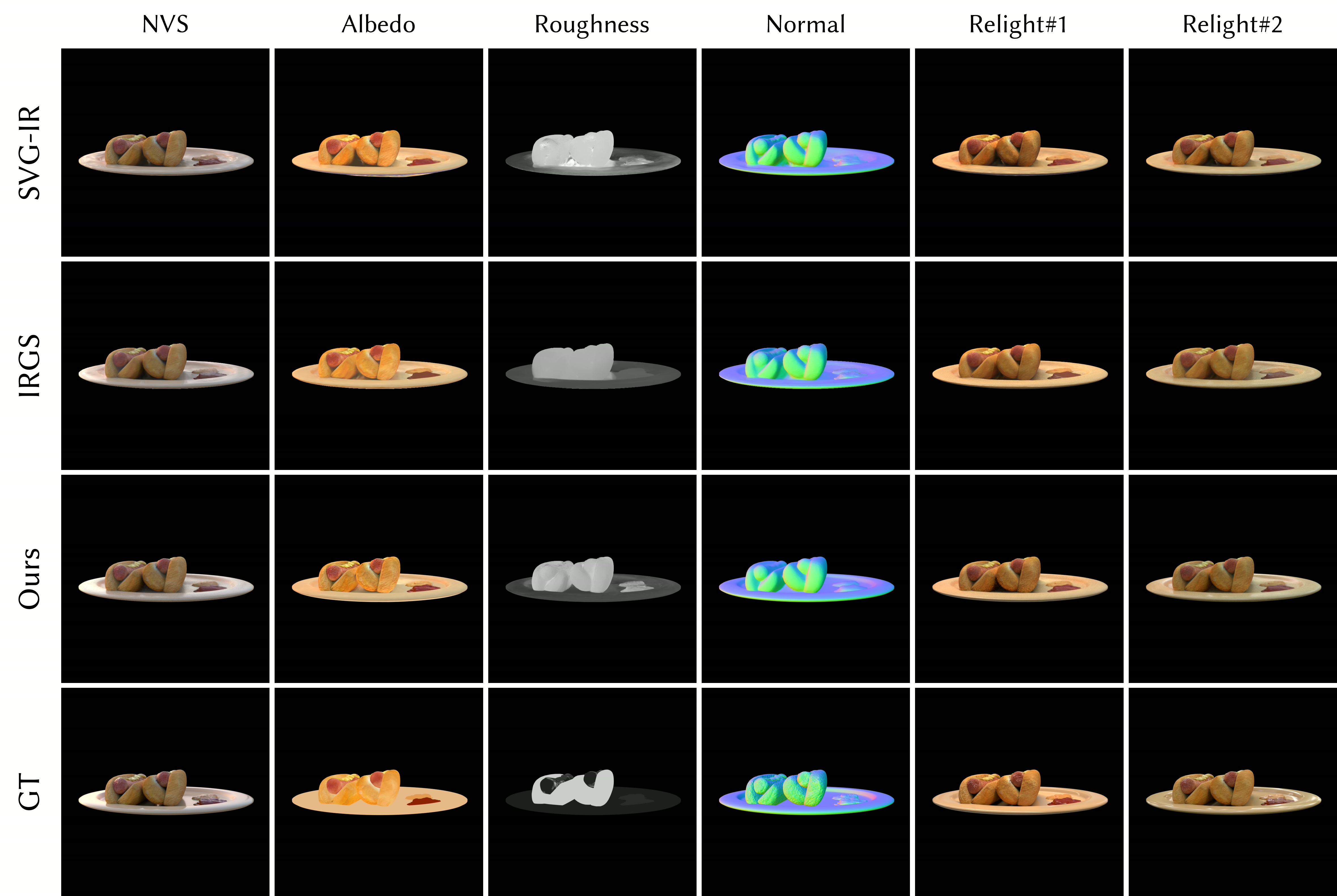}
  \vspace{-0.1in}
  \caption{Qualitative results on the Hotdog scene.}
\end{figure*}

\begin{figure*}[!htbp]
  \centering
  \includegraphics[width=0.8\linewidth]{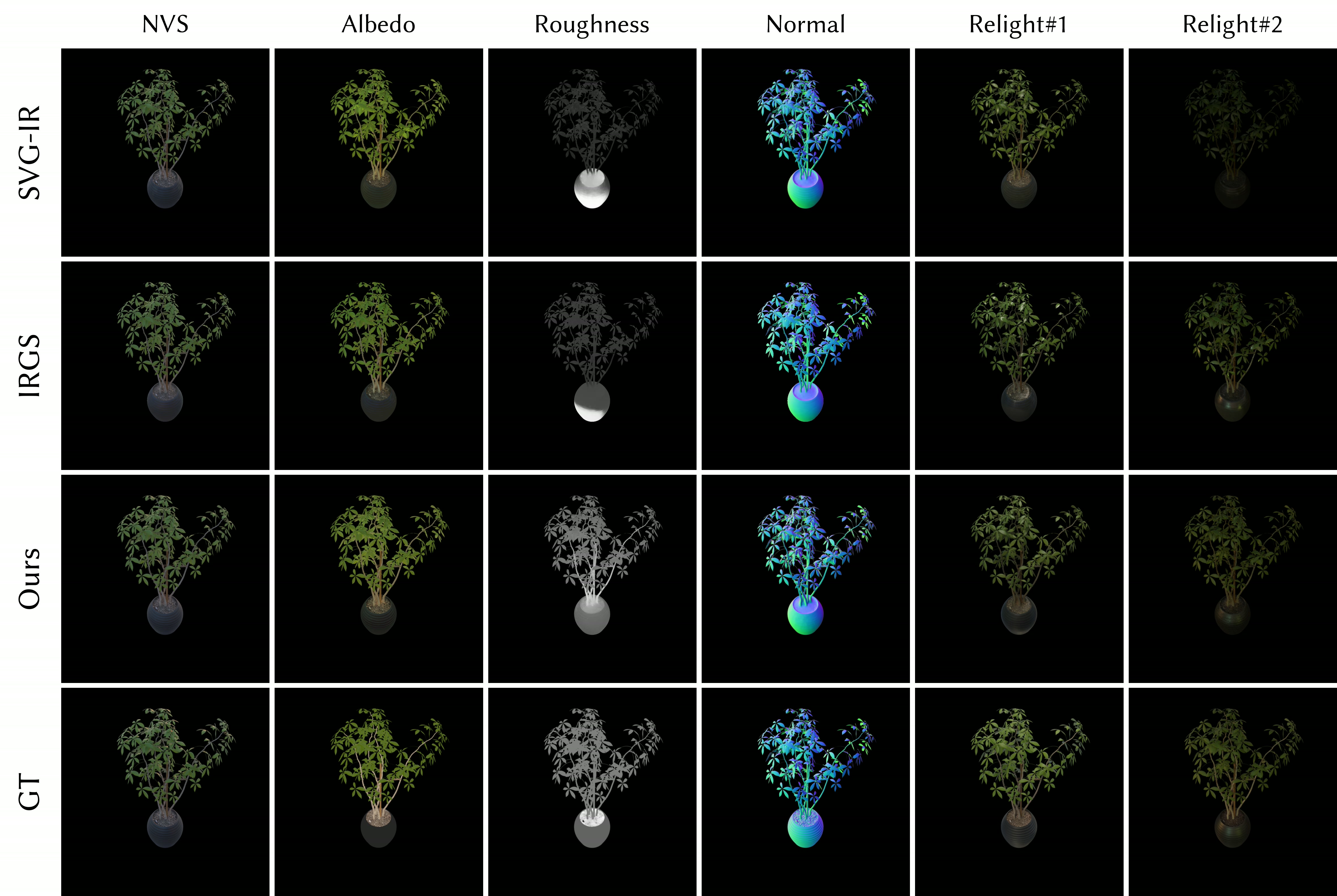}
  \vspace{-0.1in}
  \caption{Qualitative results on the Ficus scene.}
\end{figure*}

\begin{figure*}[!htbp]
  \centering
  \includegraphics[width=0.8\linewidth]{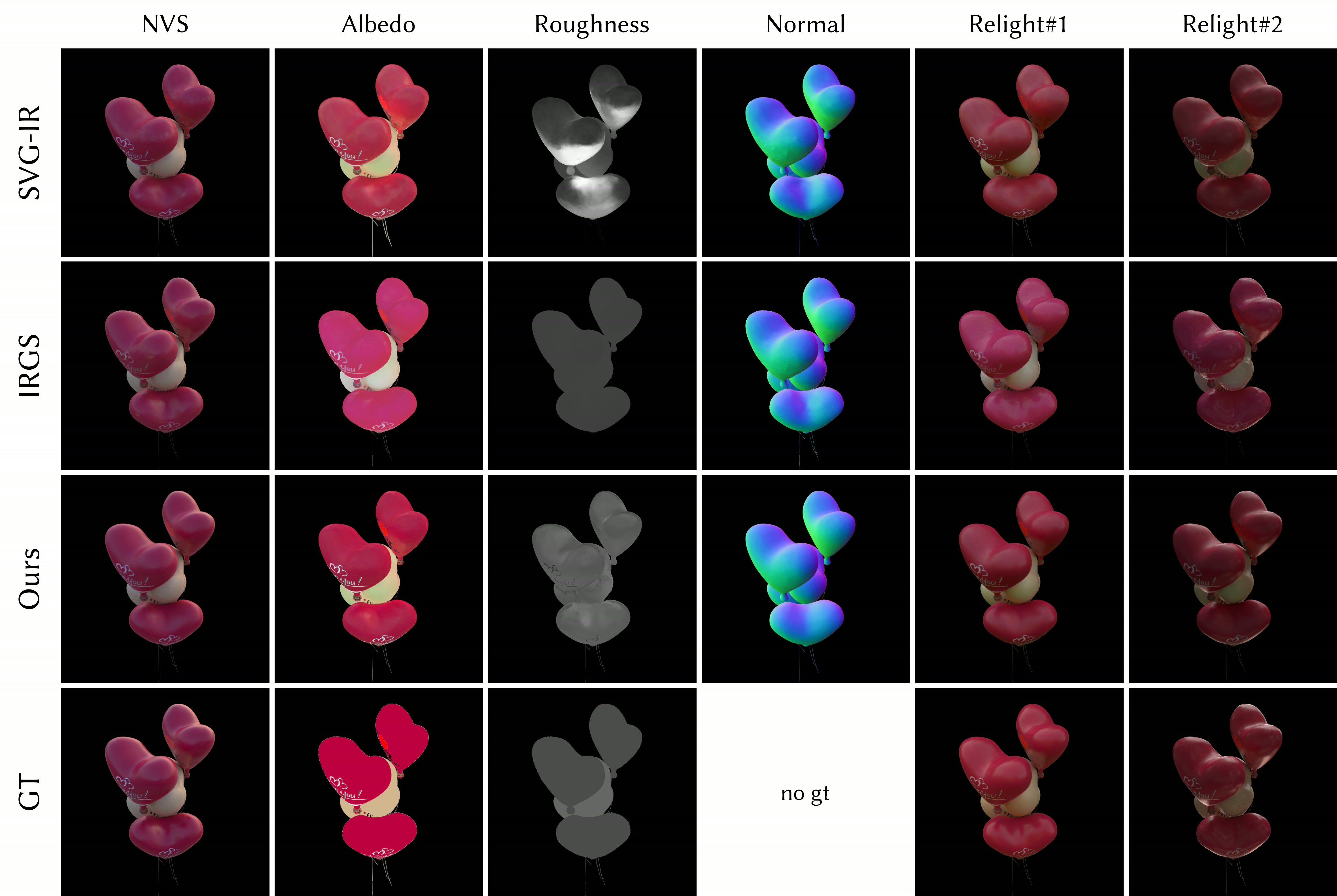}
  \vspace{-0.1in}
  \caption{Qualitative results on the Air Balloons scene.}
\end{figure*}

\begin{figure*}[!htbp]
  \centering
  \includegraphics[width=0.8\linewidth]{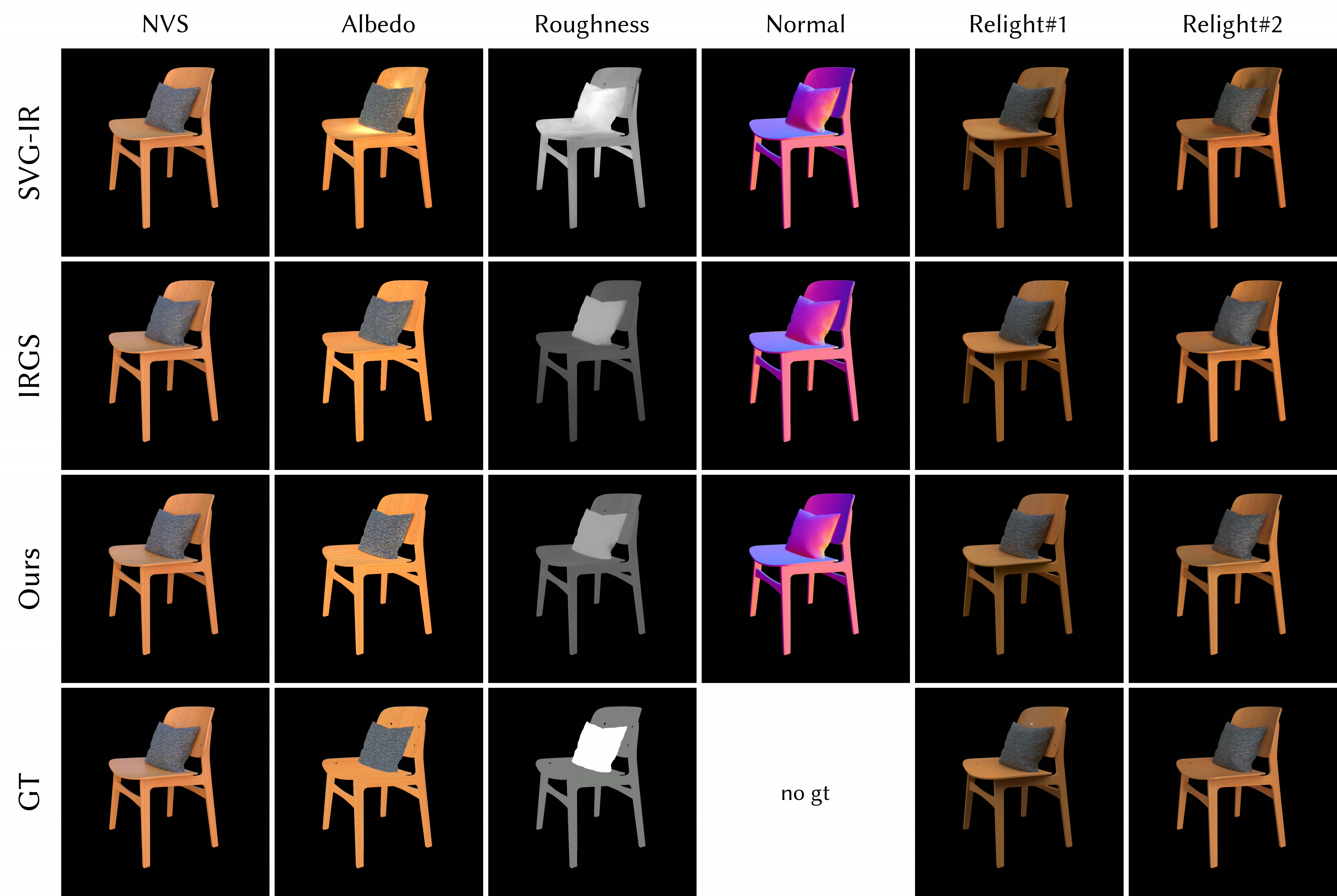}
  \vspace{-0.1in}
  \caption{Qualitative results on the Chair scene.}
\end{figure*}

\clearpage
\putbib[chapters/reference]

\end{bibunit}









\end{document}